\documentclass[reprint, amsmath, amssymb, aps, prx, superscriptaddress]{revtex4-2}

\usepackage{graphicx}
\usepackage{dcolumn}
\usepackage{bm}
\usepackage{enumitem}

\renewcommand{\eqref}[1]{Eq.~(\ref{#1})}
\newcommand{\eqsref}[1]{Eqs.~(\ref{#1})}
\begin{document}

\preprint{APS/123-QED}

\title{Classifying topological floppy modes in the continuum}

\author{Ian Tan}
\affiliation{Department of Engineering, University of Cambridge, Cambridge, CB2 1PZ, UK}
\affiliation{TCM Group, Cavendish Laboratory, University of Cambridge, Cambridge, CB3 0HE, UK}
\author{Anton Souslov}
 \email{as3546@cam.ac.uk}
\affiliation{TCM Group, Cavendish Laboratory, University of Cambridge, Cambridge, CB3 0HE, UK}

\date{\today}

\begin{abstract}
In floppy mechanical lattices, robust edge states and bulk Weyl modes are manifestations of underlying topological invariants.
To explore the universality of these phenomena independent of microscopic detail, we formulate topological mechanics in the continuum.
By augmenting standard linear elasticity with additional fields of soft modes, we define a continuum version of Maxwell counting, which balances degrees of freedom and mechanical constraints.
With one additional field, these augmented elasticity theories can break spatial inversion symmetry and harbor topological edge states. We also show that two additional fields are necessary to harbor Weyl points in two dimensions, and define continuum invariants to classify these states.
In addition to constructing the general form of topological elasticity based on symmetries, we derive the coefficients based on the systematic homogenization of microscopic lattices.
By solving the resulting partial differential equations, we efficiently predict coarse-grained deformations due to topological floppy modes without the need for a detailed lattice-based simulation.
Our discovery formulates novel design principles and efficient computational tools for topological states of matter, and points to their experimental implementation 
in mechanical metamaterials.
\end{abstract}

\maketitle


\section{Introduction}
Topological phenomena occur for a broad range of classical waves. 
Even such disparate phenomena as light in a waveguide~\cite{haldane2008possible, rechtsman2013photonic, silveirinha_chern_2015,  lu_topological_2018, roberts_topological_2022},  waves in fluids~\cite{Khanikaev_Fleury_Mousavi_Alù_2015,  delplace_topological_2017, shankar_topological_2017, souslov_topological_2019, tauber_bulk-interface_2019, venaille_wave_2021}, and vibrations in elastic solids~\cite{Casadei_Delpero_Bergamini_Ermanni_Ruzzene_2012, Wang_Lu_Bertoldi_2015, susstrunk_observation_2015, susstrunk_classification_2016, Matlack_designing_2018, scheibner_non-hermitian_2020, shankar_geometric_2022, fossati_odd_2024} have all been explored using a topological framework. One fundamental mechanical model characterized by integer invariants is a lattice of masses connected by springs~\cite{kane_topological_2014, lo_topology_2021, wang_topological_2023}. For example, mechanical lattices on the edge of stability, known as isostatic or Maxwell lattices~\cite{maxwell1864calculation, calladine_buckminster_1978}, exhibit topological phenomena in the form of Weyl modes in the bulk~\cite{rocklin_mechanical_2016, stenull_topological_2016} and edge modes in a finite system~\cite{kane_topological_2014, lubensky_phonons_2015, mao_maxwell_2018}. The same phenomena appear in other settings, such as kirigami~\cite{chen_topological_2016}, origami~\cite{berry_topological_2020}, and geared metamaterials~\cite{meeussen_geared_2016, ma_nonlinear_2023}, which hints at a topological theory independent of microscopic detail. In all these cases, one edge is significantly softer than the rest of the material due to topological polarization, enabling potential applications from cushioning to vibrational damping~\cite{bilal_intrinsically_2017, bergne_scalable_2022, tang_fully_2024}. 

Recent work~\cite{sun_continuum_2020, saremi_topological_2020, nassar_microtwist_2020} has proposed continuum models that capture some of the rich phenomenology of topological mechanics. This is remarkable, given that the topological winding number in discrete systems, defined using the Brillouin zone, is associated with lattice periodicity. Refs.~\cite{sun_continuum_2020, saremi_topological_2020} propose continuum theories that capture topological polarization by breaking spatial inversion symmetry, using dependence on strain gradients in addition to the usual stress-strain relations. By contrast, Ref.~\cite{nassar_microtwist_2020} considers a weakly-distorted 2D kagome lattice to derive elasticity augmented by one additional degree of freedom. This approach is further adapted to obtain a continuum model for weakly-distorted 3D pyrochlore lattices in Ref.~\cite{xia_microtwist_2021}. None of these previous topological theories in the continuum exhibit Weyl zero modes, which for discrete systems appear generically in both models~\cite{rocklin_mechanical_2016} and experiments~\cite{bilal_intrinsically_2017}. We therefore ask: Is it possible to classify all topological floppy modes based on the continuum theories that exhibit them? 

\begin{figure*}
    \centering
\includegraphics{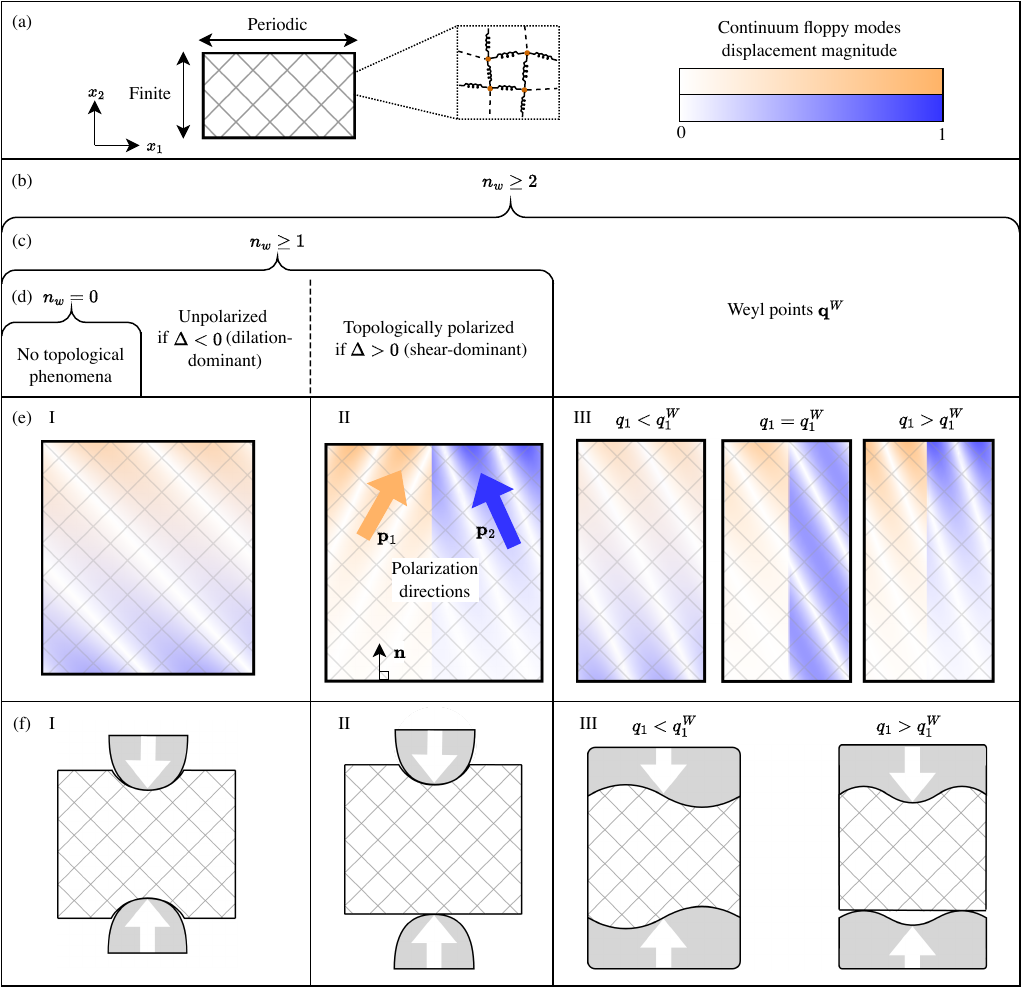}
    \caption{Classification of topological elasticity in two dimensions. (a) We consider a continuum system finite in the $x_2$-direction and periodic along the $x_1$-direction. Such a system could represent the continuum limit of a ball-and-spring network, but our results are independent of the microscopic realization. We indicate the displacement magnitudes of the two continuum edge floppy modes [defined in \eqref{eq:def_continuum_edge_mode}] using color gradients. (b--d) We construct an elasticity theory by adding $n_w$ kinematic fields arising from local soft modes to the displacement field. Then, we classify the resulting phenomenology according to $n_w$. A theory with $n_w = 0$ corresponds to standard linear elasticity, which is unable to capture topological phenomena. For $n_w \geq 1$, topological polarization is present if $\Delta$ [defined in \eqref{eq:Delta_definition}] is positive, and is characterized by the polarization directions $\mathbf{p}_1, \mathbf{p}_2$. Theories with $n_w \geq 2$ can exhibit even richer phenomenology by describing systems with Weyl points. (e) Representative displacement magnitudes associated with the continuum floppy modes are shown in columns~I~--~III. Floppy modes localized on the same edge are displayed side-by-side for clarity. Column II illustrates how both floppy modes are localized on the same edge if the normal $\mathbf{n}$ to the strip edges makes an acute angle with both $\mathbf{p}_1$ and $\mathbf{p}_2$. In column III, as $q_1$ passes through Weyl point coordinate $q_1^W$, the blue floppy mode switches from being localized at the bottom edge to the top edge, becoming a bulk mode at the transition point. (f) The physical phenomena associated with topological mechanics in the continuum are illustrated using indentation tests in columns I -- III. Edges with no localized floppy modes are more rigid. The presence of a Weyl point, shown in column III, results in a wavevector-dependent edge rigidity. The deformations shown here have been exaggerated for clarity.}
    \label{fig:classification}
\end{figure*}

Here we augment standard linear elasticity with additional fields of soft modes to write down the general form of topological mechanics in the continuum, which we show arises naturally from the homogenization of microscopic lattices.
We define a Maxwell criterion and topological invariants in the continuum independent of any underlying lattice.
We demonstrate that topological edge modes can be captured by elasticity augmented by at least one additional field. In two dimensions, we prove that the point at which the material becomes topologically polarized is equivalent to the transition from so-called dilation-dominant to shear-dominant Guest-Hutchinson modes.
We then show that for Weyl points to be present in two-dimensional elasticity, at least two additional fields are necessary.
Thus, we classify topological floppy modes in the continuum using the number of additional soft fields, as summarized in Fig.~\ref{fig:classification}.
We arrive at a continuum approach for topological modeling, where the resulting equations of mechanical equilibrium are efficiently solved without the full lattice structure.
Our fundamental models of topological states at the largest length scale open up a systematic approach toward new topological field theories.

\section{Generalized elasticity and the Maxwell criterion}
\label{sec:generalized_elasticity}
Standard linear elasticity (i.e., Cauchy elasticity) is governed by equations that are symmetric under spatial inversion. Consequently, for any finite strip of material obeying standard linear elasticity, the existence of a floppy mode localized on one edge implies the existence of another floppy mode localized on the opposite edge. Thus, standard linear elasticity is unable to capture the asymmetric distribution of edge modes characteristic of topological polarization. In this section, we construct a generalized elasticity theory that breaks spatial inversion, and explore its phenomenology in later sections. Although we construct this theory without reference to any underlying microscopic structure, we show in Sec.~\ref{sec:homogenization} that our theory arises naturally as the continuum limit of a ball-and-spring lattice. We begin by constructing an elastic energy density and derive equations of motion via the Euler-Lagrange equations. We then define a Maxwell criterion and winding numbers in the continuum, which are our main results in this section.

\subsection{The continuum equations of motion}
Standard elasticity is determined by the displacement field $\mathbf{u}(\mathbf{x},t)$ at position $\mathbf{x}$ and time $t$. In our description, we consider $n_w$ additional continuum fields $\{\varphi_m(\mathbf{x},t)\}_{m = 1}^{n_w}$, which represent internal degrees of freedom, corresponding to local displacements. Although Refs.~\cite{nassar_microtwist_2020,xia_microtwist_2021} have considered specific cases where augmented continuum theories with $n_w = 1$ and $n_w = 3$ model kagome and pyrochlore lattices, respectively, we instead follow a more general approach based on symmetries without considering a microscopic lattice. 

Generically, elasticity results from the energetic costs of spatial variations in the kinematic fields, so the elastic energy density $V$ depends on $\nabla \mathbf{u}$, $\nabla \varphi_1, \ldots, \nabla \varphi_{n_w}$, where $\nabla$ is the spatial gradient. We call these kinematic quantities the \textit{generalized strain measures}. To consider only the longest length scales, we do not consider any higher-order gradients. However, to break spatial inversion symmetry, we need to consider the fields $\varphi_1, \ldots, \varphi_{n_w}$ alongside their gradients in the elastic energy. To show this, we first define a gradient-dependent $\tilde{\Lambda} \equiv [\nabla \mathbf{u} \; \nabla \varphi_1 \ldots \nabla \varphi_{n_w} ]^{\mathrm{T}}$, and consider an elastic energy density 
\begin{equation}
\label{eq:elastic_energy_density_preliminary}
    \tilde{V} = \frac{1}{2} \tilde{\Lambda}^{\mathrm{T}} \mathbf{\breve{K}} \tilde{\Lambda},
\end{equation}
where $\mathbf{\breve{K}}$ is a generalized stiffness matrix. The elasticity based on $\tilde{V}$ cannot capture topological polarization, because it is symmetric under spatial inversion: the parity transformation $(\mathbf{x},\mathbf{u},\varphi_m) \mapsto -(\mathbf{x},\mathbf{u},\varphi_m)$ leaves both $\tilde{\Lambda}$ and $\tilde{V}$ unchanged. 

To construct a linear lowest-order-gradient theory that breaks spatial inversion, we define a vector of generalized strain measures that includes $\varphi_1, \ldots, \varphi_{n_w}$:
\begin{equation}
    \Lambda \equiv [\nabla^s \mathbf{u} \; \nabla \varphi_1 \ldots \nabla \varphi_{n_w} \; \varphi_1 \ldots \varphi_{n_w}]^{\mathrm{T}},
\end{equation}
where we have replaced $\nabla \mathbf{u}$ by its symmetrization $\nabla^s \mathbf{u} = \frac{1}{2}\left(\nabla \mathbf{u} + (\nabla \mathbf{u})^\mathrm{T}\right)$ because the isotropy of space requires that the elastic energy be invariant under infinitesimal rigid-body rotations of the system, represented by the antisymmetric part of $\nabla \mathbf{u}$.
By including dependence on $\varphi_1, \ldots, \varphi_{n_w}$, the elastic energy density is no longer symmetric under spatial inversion due to the presence of terms with only a single gradient:
\begin{equation}
\label{eq:elastic_energy_density_general}
    V = \frac{1}{2} \Lambda^{\mathrm{T}} \mathbf{\hat{K}} \Lambda,
\end{equation}
where $\mathbf{\hat{K}}$ is a generalized stiffness matrix, which is symmetric and positive semi-definite to ensure stability. 
Heuristically, the expression for $V$ includes both parity-even terms from $\tilde{V}$ and parity-odd terms of the form $\varphi_m \nabla \mathbf{u}$ and $\varphi_m \nabla \varphi_k$, which results in an elasticity that generically breaks spatial inversion symmetry.

We introduce a fundamental difference between $\varphi_1, \ldots, \varphi_{n_w}$ and $\mathbf{u}$ as kinematic fields: $\mathbf{u}$ affects $V$ only through its gradients because of the elastic system's uniform translational symmetry, whereas no such symmetries exist for $\varphi_1, \ldots, \varphi_{n_w}$. We show in Sec.~\ref{sec:homogenization} that when the continuum theory is obtained from homogenizing a lattice, the energetic cost of non-zero $\varphi_1, \ldots, \varphi_{n_w}$ is associated with the lattice being gapped at sufficiently small wavenumbers, a necessary condition for the presence of topological polarization. 

To obtain the equations of motion of a continuum system with our elastic energy density, we define a kinetic energy
\begin{equation}
    E_K = \frac{1}{2} (\partial_t \Psi)^{\mathrm{T}} \mathbf{\hat{M}} \, \partial_t \Psi,
\end{equation}
where $\Psi \equiv [\mathbf{u} \; \varphi_1, \ldots, \varphi_{n_w}]^{\mathrm{T}}$ 
are the generalized displacements and $\mathbf{\hat{M}}$ is the generalized mass matrix, which is symmetric and positive-definite. For completeness, we define the total potential energy density of the system to be $V + V_{\mathrm{ext}}$, where $V_{\mathrm{ext}}(\mathbf{u}, \varphi_1, \ldots, \varphi_{n_w})$ accounts for conservative body force density $\mathbf{f}(\mathbf{x},t)$ and generalized body torque density $\tau_k(\mathbf{x},t)$. Applying the Euler-Lagrange equations to the Lagrangian density $\mathcal{L} = E_K - (V + V_{\mathrm{ext}})$ results in the equations of motion
\begin{subequations}
\label{eq:continuum_eq}
    \begin{equation}
	\label{eq:continuum_eq_motion_u}
	\rho \partial_t^2 \mathbf{u}(\mathbf{x},t) + \sum_{m=1}^{n_w} \partial_t^2 \varphi_m(\mathbf{x},t) \mathbf{p}_m  = \mathbf{f}(\mathbf{x},t) + \nabla \cdot \mathbf{T}(\mathbf{x},t)
    \end{equation}
    and for $k = 1, \ldots, n_w$,
    \begin{multline}
    \label{eq:continuum_eq_motion_phi}
        \mathbf{p}_k \cdot \partial_t^2\mathbf{u}(\mathbf{x},t) + \sum_{m = 1}^{n_w} \mu_{km} \partial_t^2 \varphi_m(\mathbf{x},t) \\
        = \tau_k(\mathbf{x},t) - \eta_k(\mathbf{x},t) + \nabla \cdot \mathbf{c}_k(\mathbf{x},t),
    \end{multline}
\end{subequations}
where $\rho$ is the mass density, and $\mathbf{p}_k$, $\mu_{km}$ are the generalized inertia and moment of inertia densities, respectively. Equations~(\ref{eq:continuum_eq}) contain the stress measures 
\begin{equation*}
\mathbf{T}(\mathbf{x},t) \textrm{, }  \{\mathbf{c}_k(\mathbf{x},t)\}_{k=1}^{n_w} \textrm{ and } \{- \eta_k(\mathbf{x},t) \}_{k=1}^{n_w}
\end{equation*}
dual to the strain measures \begin{equation*}
\nabla^s \mathbf{u}(\mathbf{x},t)\textrm{, } \{\nabla \varphi_m(\mathbf{x},t)\}_{m=1}^{n_w} \textrm{ and } \{\varphi_m(\mathbf{x},t)\}_{m=1}^{n_w},
\end{equation*}
respectively. The form of \eqref{eq:continuum_eq_motion_u} shows that $\mathbf{T}(\mathbf{x},t)$ is the well-known Cauchy stress from standard elasticity. Under the assumption of linearity, and for a conservative system, the stress and strain measures are related by the constitutive relations
\begin{subequations}
    \label{eq:constitutive}
    \begin{align}
	   \mathbf{T} &= \mathsf{C}:\nabla^s \mathbf{u} + \sum_{m=1}^{n_w}( \mathfrak{B}_m \cdot \nabla \varphi_m + \mathbf{N}_m \varphi_m ) \\
	   \mathbf{c}_k & = \mathfrak{B}_k^{\mathrm{T}} : \nabla^s \mathbf{u} + \sum_{m = 1}^{n_w}( \mathbf{M}_{km} \cdot \nabla \varphi_m + \mathbf{h}_{km} \varphi_m) \\
	   \eta_k & = \mathbf{N}_k :\nabla^s \mathbf{u} + \sum_{m = 1}^{n_w} \left(\mathbf{h}_{mk} \cdot \nabla \varphi_m + J_{km} \varphi_m \right), 
    \end{align}
\end{subequations}
where $\mathsf{C}$ is the fourth rank
(i.e., order) elastic tensor from standard elasticity, $\mathfrak{B}_k$ and $\mathbf{N}_k$ are third rank and second rank tensors, respectively. 
The coefficients $\mathbf{h}_{km}$ and $J_{km}$ are vectors and scalars, respectively. The product $:$ denotes the double contraction of tensors, so that in index notation with the summation convention, $(\mathsf{C}:\nabla^s \mathbf{u})_{ij} = (\mathsf{C})_{ijkl} (\nabla^s \mathbf{u})_{kl}$, and $(\mathfrak{B}_k^{\mathrm{T}} : \nabla^s \mathbf{u})_{i} = (\mathfrak{B}_k^{\mathrm{T}})_{ijl} (\nabla^s \mathbf{u})_{jl}$. The transpose operation on third rank tensors is defined to satisfy $(\mathfrak{B}^{\mathrm{T}})_{ijk} = (\mathfrak{B})_{kij}$, motivated by thinking of the third rank tensor as a map from the space of vectors to the space of second rank tensors. We refer the coefficients in \eqref{eq:constitutive} as \textit{generalized elastic moduli}. Defining the generalized stresses $\Sigma \equiv [\mathbf{T} \; \mathbf{c}_1 \ldots \mathbf{c}_{n_w} \; \eta_1 \ldots \eta_{n_w}]^{\mathrm{T}}$, these constitutive relations~(\ref{eq:constitutive}) can be expressed compactly as
\begin{equation}
    \label{eq:constitutive_compact}
    \Sigma = \mathbf{\hat{K}} \Lambda.
\end{equation}

\subsection{Maxwell criterion and topological floppy modes}
\label{sec:maxwell_condition}
We have constructed a generalized elasticity theory that breaks spatial inversion symmetry. We proceed to show how our theory enables the definition of topological invariants. In Sec.~\ref{sec:classifying}, we show how these invariants are linked to topological polarization and Weyl modes in the continuum. First, we briefly review topological floppy modes in discrete mechanical lattices~\cite{kane_topological_2014, rocklin_mechanical_2016} to motivate the developments in this section, deferring further details to Sec.~\ref{sec:site_displacements_and_compatibility_matrix}. We consider periodic ball-and-spring lattices in $d$-dimensions with unit cells containing $n_s$ sites connected by $n_b$ bonds modeled as linear-elastic springs. The kinematics of these lattices is described using a compatibility matrix, which is a linear map from the space of site displacements to the space of bond extensions. Periodic lattices are conveniently studied in wavevector space, or $\mathbf{q}$-space, in which we consider quantities varying with spatial position $\mathbf{x}$ according to $e^{i \mathbf{q} \cdot \mathbf{x}}$. The compatibility matrix $\mathbf{C}(\mathbf{q})$ for wavevector $\mathbf{q}$ relates the amplitudes of the unit cell site displacements $\Phi \equiv [\mathbf{u}_1  \ldots  \mathbf{u}_{n_s}]^{\mathrm{T}}$ to those of bond extensions $\Xi(\mathbf{q}) \equiv [ e_1(\mathbf{q}) \ldots e_{n_b}(\mathbf{q}) ]^{\mathrm{T}}$ by $\Xi(\mathbf{q}) = \mathbf{C}(\mathbf{q}) \Phi$. In this context, we focus on zero modes, which correspond to displacements $\Phi$ with zero bond extensions $\Xi$.

The compatibility matrix $\mathbf{C}(\mathbf{q})$ has dimensions $n_b \times n_s d$. We recall that the Maxwell criterion for ball-and-spring lattices is $n_s d = n_b$~\cite{maxwell1864calculation, calladine_buckminster_1978}, which is equivalent to $\mathbf{C}(\mathbf{q})$ being square. When lattices satisfy the Maxwell criterion, we define topological invariants using $\det \mathbf{C}(\mathbf{q})$ via~\cite{kane_topological_2014, rocklin_mechanical_2016}
\begin{equation}
\label{eq:discrete_invariant}
    n(\mathcal{C}) = \frac{1}{2 \pi i} \int_{\mathcal{C}} \frac{\nabla_{\mathbf{q}} \det \mathbf{C}(\mathbf{q})}{\det \mathbf{C}(\mathbf{q})} \cdot \mathrm{d}\mathbf{q},
\end{equation}
where $\mathcal{C}$ is a closed path in $\mathbf{q}$-space on which $\det \mathbf{C}(\mathbf{q}) \neq 0$. The path $\mathcal{C}$ chosen depends on whether the invariant is computed for edge modes (for which $\mathcal{C}$ is a line that spans $\mathbf{q}$-space) or Weyl modes (which are enclosed by a loop $\mathcal{C}$). Topological invariants of this form encode information about zero modes because zero modes at wavevector $\mathbf{q}$ exist if and only if $\det \mathbf{C}(\mathbf{q}) = 0$.

We formulate a Maxwell criterion for the continuum by defining an effective compatibility matrix from our generalized elasticity theory. To begin, we recall that the generalized displacements $\Psi = [\mathbf{u} \; \varphi_1 \ldots \varphi_{n_w}]^{\mathrm{T}}$ represent our $d + n_w$ continuum degrees of freedom. As in the discrete case~\cite{lubensky_phonons_2015} (c.f. Sec.~\ref{sec:site_displacements_and_compatibility_matrix}), we consider generalized displacements of the form $\Psi = \hat{\Psi} e^{i \mathbf{q} \cdot \mathbf{x}}$, where the complex components of $\mathbf{q}$ represent spatially growing and decaying modes and the hat over vectors and scalars indicates that they represent Fourier amplitudes. In this representation, the (complex) generalized strain measures are
\begin{align*}
    \nabla^s \mathbf{u} &= \mathrm{sym}\left(\left(\hat{\mathbf{u}} e^{i \mathbf{q} \cdot \mathbf{x}}\right) \otimes (i \mathbf{q}) \right) \\
    \nabla \varphi_m &= i\mathbf{q} \, \hat{\varphi}_{m} e^{i \mathbf{q} \cdot \mathbf{x}} \\
    \varphi_m &= \hat{\varphi}_{m} e^{i \mathbf{q} \cdot \mathbf{x}},
\end{align*}
where $\mathrm{sym}(\mathbf{A}) = \frac{1}{2}\left(\mathbf{A} + \mathbf{A}^{\mathrm{T}}\right)$ is the symmetrization of $\mathbf{A}$. The strain measures $\Lambda$ are linear in the generalized displacements $\Psi$, i.e., $\Lambda = \mathbf{\hat{C}}(\mathbf{q}) \hat{\Psi} e^{i \mathbf{q} \cdot \mathbf{x}}$ where $\mathbf{\hat{C}}(\mathbf{q})$ is a $\mathbf{q}$-dependent linear map from the generalized displacements $\Psi$ to the generalized strains $\Lambda$. The map $\mathbf{\hat{C}}(\mathbf{q})$ forms a matrix in the orthonormal basis $\{\mathbf{e}_j\}_{j = 1}^d$ of our $d$-dimensional space. The generalized stresses are given by $\Sigma = \mathbf{\hat{K}} \mathbf{\hat{C}}(\mathbf{q}) \hat{\Psi} e^{i \mathbf{q} \cdot \mathbf{x}}$. Let $n_K = \mathrm{rank} \, \mathbf{\hat{K}}$ be the dimension of the range space of $\mathbf{\hat{K}}$, and $y_1, \ldots, y_{n_K}$ be a basis for this range space. The number $n_K$ is the continuum version of the number of bonds $n_b$ in the discrete case, because it represents the number of constraints imposed by the elastic moduli. We define $\mathbf{P}_K$ to be the matrix with columns $y_1, \ldots, y_{n_K}$. 

In a discrete lattice, pre-multiplying the compatibility matrix by the diagonal matrix containing bond spring constants results in a linear map from site displacements to bond forces. We take the matrix $\mathbf{P}_K^{\mathrm{T}} \mathbf{\hat{K}} \mathbf{\hat{C}}(\mathbf{q})$ as a continuum analog to this linear map. Although this choice is not unique, we show that the topological quantities obtained are well-defined and do not depend on the arbitrary choice of gauge. The matrix $\mathbf{P}_K^{\mathrm{T}} \mathbf{\hat{K}} \mathbf{\hat{C}}(\mathbf{q})$ has dimensions $n_K \times (d + n_w)$. A natural definition for the Maxwell criterion in the continuum is
\begin{equation}
    \label{eq:maxwell_continuum}
    n_K = d + n_w,
\end{equation}
which balances the number of constraints $n_K$ against the number of degrees of freedom $d + n_w$ and is equivalent to $\mathbf{P}_K^{\mathrm{T}} \mathbf{\hat{K}} \mathbf{\hat{C}}(\mathbf{q})$ being square. Zero modes at wavevector $\mathbf{q}$ exist if and only if there are generalized displacements $\Lambda$ at $\mathbf{q}$ that result in $V = \frac{1}{2} \Lambda^{\mathrm{T}} \mathbf{\hat{K}} \Lambda = 0$. This condition is equivalent to the existence of a non-trivial null space of $\mathbf{P}_K^{\mathrm{T}} \mathbf{\hat{K}} \mathbf{\hat{C}}(\mathbf{q})$, which given the Maxwell criterion \eqref{eq:maxwell_continuum}, is equivalent to $\det \mathbf{P}_K^{\mathrm{T}} \mathbf{\hat{K}} \mathbf{\hat{C}}(\mathbf{q}) = 0$. Therefore, we take the quantity $\det \mathbf{P}_K^{\mathrm{T}} \mathbf{\hat{K}} \mathbf{\hat{C}}(\mathbf{q})$ as a continuum analog to $\det \mathbf{C}(\mathbf{q})$, which we use to define topological invariants by analogy with \eqref{eq:discrete_invariant}. Since these topological invariants depend on $\det \mathbf{P}_K^{\mathrm{T}} \mathbf{\hat{K}} \mathbf{\hat{C}}(\mathbf{q})$ only via the ratio 
\begin{equation*}
    \frac{\nabla_{\mathbf{q}} \det \mathbf{P}_K^{\mathrm{T}} \mathbf{\hat{K}} \mathbf{\hat{C}}(\mathbf{q})}{\det \mathbf{P}_K^{\mathrm{T}} \mathbf{\hat{K}} \mathbf{\hat{C}}(\mathbf{q})},
\end{equation*}
any matrix whose determinant is proportional to $\det \mathbf{P}_K^{\mathrm{T}} \mathbf{\hat{K}} \mathbf{\hat{C}}(\mathbf{q})$ could be chosen to define the topological invariant. We exploit this gauge choice to define an effective compatibility matrix in the continuum, $\mathbf{P}_K^{\mathrm{T}} \mathbf{\hat{C}}(\mathbf{q})$, which we use to compute a topological invariant via $\det \mathbf{P}_K^{\mathrm{T}} \mathbf{\hat{C}}(\mathbf{q})$. To see that 
\begin{equation*}
\det \mathbf{P}_K^{\mathrm{T}} \mathbf{\hat{K}} \mathbf{\hat{C}}(\mathbf{q}) = \det \mathbf{P}_K^{\mathrm{T}} \mathbf{\hat{K}} \mathbf{P}_K \det \mathbf{P}_K^{\mathrm{T}}\mathbf{\hat{C}}(\mathbf{q}), 
\end{equation*}
note that $\mathbf{\hat{K}}$ is symmetric and $\det \mathbf{P}_K^{\mathrm{T}} \mathbf{\hat{K}} \mathbf{P}_K > 0$ is a constant, so that $\mathbf{P}_K^{\mathrm{T}} \mathbf{\hat{K}} \mathbf{\hat{C}}(\mathbf{q}) = \mathbf{P}_K^{\mathrm{T}} \mathbf{\hat{K}} \mathbf{P}_K \mathbf{P}_K^{\mathrm{T}}\mathbf{\hat{C}}(\mathbf{q})$, where $\mathbf{P}_K \mathbf{P}_K^{\mathrm{T}}$ projects onto the orthogonal complement of the null space of $\mathbf{\hat{K}}$. Thus, we define our topological invariants via
\begin{equation}
    \label{eq:continuum_topological_invariants}
    n[\mathbf{q}(\tau)] = \frac{1}{2 \pi i} \int_{\tau_0}^{\tau_1} \frac{\frac{\mathrm{d}}{\mathrm{d} \tau} \det \mathbf{P}_K^{\mathrm{T}} \mathbf{\hat{C}}(\mathbf{q}(\tau))}{\det \mathbf{P}_K^{\mathrm{T}} \mathbf{\hat{C}}(\mathbf{q}(\tau))} \, \mathrm{d}\tau,
\end{equation}
where $\mathbf{q}(\tau)$ represents a parametrization of a path in $\mathbf{q}$-space by the parameter $\tau \in [\tau_0, \tau_1]$. For a closed path, $\mathbf{q}(\tau_0) = \mathbf{q}(\tau_1)$. 

When considering topological edge modes in Sec.~\ref{sec:edge_modes}, we modify this form of invariant to account for the absence of a Brillouin zone in the continuum, adapting an approach applied to higher strain gradient theories in Ref.~\cite{saremi_topological_2020}. We consider non-closed paths $\mathbf{q}(\tau)$ for which $\mathbf{q}$ is real to retain the bulk-edge correspondence seen in discrete topological mechanics: the invariant is computed using bulk modes but contains information about the edges. The modified invariant counts only edge modes that are visible on large length scales. To ensure this, consider a mode localized on an edge normal to $\mathbf{e}_k$ (and decaying along $\mathbf{e}_k$) with complex wavevector $\mathbf{q} = \sum_{j = 1}^d q_j \mathbf{e}_j$ parametrized by expressing the complex component $q_k$ normal to the edge as a function of the real components $\{q_j\}_{j \neq k}$ along the edge, i.e., $q_k(\{q_j\}_{j \neq k})$. In the continuum, we take the limit $q_{j(\neq k)} \to 0$ and require that
\begin{equation}
\label{eq:def_continuum_edge_mode}
    \lim_{q_{j(\neq k)} \to 0} q_k(\{q_j\}_{j \neq k}) = 0,
\end{equation}
i.e., both the wavevector along the edge and the inverse penetration depth of the mode tend to zero.
We refer to edge modes satisfying this property as \textit{continuum edge modes}.

The quantity $\det \mathbf{P}_K^{\mathrm{T}} \mathbf{\hat{C}}(\mathbf{q})$ depends on $\mathbf{q}$ only via $i \mathbf{q}$, and has the polynomial form 
\begin{equation}
    \label{eq:general_form_continuum_det_C}
    \det \mathbf{P}_K^{\mathrm{T}} \mathbf{\hat{C}}(\mathbf{q}) = \sum_{m = d}^{d + n_w} \frac{i^m}{m!} P_m(\mathbf{q}),
\end{equation}
where $P_m(\mathbf{q})$ is a homogeneous polynomial with real coefficients of degree $m$ in the components of $\mathbf{q}$. The highest degree present is $d + n_w$ because $\mathbf{P}_K^{\mathrm{T}} \mathbf{\hat{C}}(\mathbf{q})$ has dimensions $(d + n_w) \times (d + n_w)$ and each element in the matrix is at most $\mathcal{O}(q_j)$.
Since $\mathbf{P}_K^{\mathrm{T}} \mathbf{\hat{C}}(\mathbf{q})$ has a null space of at least $d$ dimensions, corresponding to the $d$ uniform translations, we conclude that $P_m(\mathbf{q}) = 0$ for $0 \leq m < d$.

We have defined an effective compatibility matrix, from which we will compute topological invariants for edge modes and Weyl points in Sec.~\ref{sec:classifying}. Importantly, our results are completely independent of microscopic detail and arise from the continuum model. Although we have used periodic lattices to motivate our definitions, none of our results require the existence of an underlying lattice, only that the generalized elastic moduli satisfy our Maxwell criterion, \eqref{eq:maxwell_continuum}. Therefore, our approach might also apply to non-periodic topological systems, which is a topic of recent interest in areas as diverse as gyroscopic metamaterials~\cite{nash_topological_2015, mitchell_amorphous_2018, mitchell_real-space_2021}, fiber networks~\cite{zhou_topological_2018}, quasicrystals~\cite{zhou_topological_2019}, and in formulating model-free topological mechanics~\cite{guzman_model-free_2024}.

\subsection{Relation to higher strain gradient theories}
Here we show that continuum theories involving the gradient of the linearized strain, $\nabla (\nabla^s \mathbf{u})$, such as those in Refs.~\cite{sun_continuum_2020, saremi_topological_2020}, can be obtained as special cases of our generalized elasticity theory. Following an approach similar to Ref.~\cite{Forest_Sab_2020}, we constrain the additional fields $\varphi_1, \ldots, \varphi_{n_w}$ so that they are linear functions of the linearized strain $\nabla^s \mathbf{u}$. To do this, we
set $n_w = \frac{1}{2}d (d + 1)$, which is the number of independent components of $\nabla^s \mathbf{u}$, and impose the constraint
\begin{equation}
\label{eq:additional_fields_constrained}
    \varphi_m = \mathbf{D}_m : \nabla^s \mathbf{u},
\end{equation}
for $m = 1, \ldots, n_w$, where $\mathbf{D}_m$ is a second rank tensor, and in this case, the double contraction $:$ is equivalent to the trace of the matrix product. Using \eqref{eq:additional_fields_constrained}, the elastic energy density $V$ becomes dependent on the gradient of linearized strain $\nabla (\nabla^s \mathbf{u})$ via the generalized strain measures $\nabla \varphi_m$, which are linear combinations of the components of $\nabla (\nabla^s \mathbf{u})$. Continuum theories involving higher gradients of linearized strain can be similarly recovered by incorporating more additional fields $\varphi_m$ and expressing these fields in terms of higher strain gradients.

Our continuum theory and those in Refs.~\cite{sun_continuum_2020, saremi_topological_2020} require more coefficients than standard linear elasticity to capture topological phenomena. We argue that, to capture the same phenomena, our formulation requires fewer additional coefficients than theories involving strain gradients such as $\nabla (\nabla^s \mathbf{u})$. The number of these coefficients is determined by the square of the dimensionality of the space of generalized strains in the theory. For example, in strain-gradient linear elasticity involving both $\nabla^s \mathbf{u}$ and $\nabla (\nabla^s \mathbf{u})$, the number of these coefficients is $\frac{1}{4}d^2(d+1)^4$. By contrast, in our generalized elasticity theory, the number of coefficients is $(d/2 + n_w)^2(d + 1)^2$. These two quantities are equal when
\begin{equation}
\label{eq:minimality_n_w}
    n_w = \frac{1}{2}d^2.
\end{equation}
Substituting $d = 2$ and $3$ into \eqref{eq:minimality_n_w}, we see that elasticity theories with $n_w = 1$ have fewer coefficients than lowest-order strain gradient theories. 
Both of these approaches capture topological polarization (see Refs.~\cite{sun_continuum_2020, saremi_topological_2020}, Fig.~\ref{fig:classification}, and Sec.~\ref{sec:classifying}), but augmented elasticity with $n_w = 1$ captures the same phenomenology with fewer coefficients. Additionally, Weyl points are found in two-dimensional elasticity theories with $n_w = 2$ (Sec.~\ref{sec:weyl_modes}), so we capture a class of topological phenomena not previously seen in higher strain gradient theories.

\section{Continuum limit of a ball-and-spring lattice}
\label{sec:obtaining_continuum_theory}
We have constructed the form of a generalized elasticity theory to capture topological phenomena, based on the requirement that the theory be able to break spatial inversion symmetry. Here we take the continuum limit of a generic ball-and-spring lattice and show that our continuum theory emerges naturally. This homogenization procedure shows that the additional fields $\varphi_1, \ldots, \varphi_{n_w}$ correspond to local soft modes in the lattice. This procedure also links the generalized elastic moduli to the compatibility matrix of the underlying lattice.

Generically, our continuum theory describes physical phenomena on a macroscopic length scale $L$ in the limit of a small size $a$ for the lattice unit cell (see Fig.~\ref{fig:epsilon_scaling}).
To take this limit, we define a small parameter $\delta \equiv a/L$. We require a homogenization procedure to obtain continuum equations of motion systematically from the microscopic equations in the limit $\delta \to 0$. Such a procedure should have the following features:
\begin{enumerate}
    \item \label{list:generic} The lattice geometry enters continuum elasticity only through a few parameters, the generalized elastic moduli.
    \item \label{list:meaningful} The continuum degrees of freedom, including the displacement field, correspond to physical displacements of the lattice.
    \item \label{list:asymptotic} In the continuum limit $\delta \to 0$, all terms are retained up to the lowest order in $\delta$. 
\end{enumerate}
\subsection{Lattice site displacements and the compatibility matrix}
\label{sec:site_displacements_and_compatibility_matrix}
In this section, we introduce our notation for exploring the continuum limit of ball-and-spring lattices. Throughout this work, we consider a generic lattice in $d$-dimensions whose unit cell contains $n_s$ sites labeled by $h = 1, \ldots, n_s$. We define the diagonal mass matrix $\mathbf{M}_a$ with dimensions $n_s d \times n_s d$ having the masses of the particles at the corresponding sites along its diagonal. These sites are connected by $n_b$ bonds, modeled as Hookean springs with spring constants that are the elements of the diagonal spring constant matrix $\mathbf{K}_a$. 

We now consider how the masses $\mathbf{M}_a$ and spring constants $\mathbf{K}_a$ scale with the lattice constant $a$. In the continuum limit, the mass density of the system should be preserved, so $\mathbf{M}_a = a^d \mathbf{M}$, where $\mathbf{M}$ is independent of $a$. The scaling of spring constants $\mathbf{K}_a$ with lattice constant $a$ depends on the nature of the elastic bonds connecting the sites. For $d = 3$, we model the bonds as linearly elastic rods. The spring constant for the extension of a rod is given by $K = EA/L_0$, where $A$ is the cross-sectional area, $L_0$ is its rest length, and $E$ is the Young's modulus. We consider the limit $a \to 0$ while preserving the shape and material of the rods, so that $A \varpropto a^2$ and $L_0 \varpropto a$, and the spring constant scales as $K \varpropto a$. In 2D, we treat the bonds as linearly elastic ribbons, for which the spring constants are independent of $a$. For a physical quasi-two-dimensional system, this scaling corresponds to our intuition: for a constant out-of-plane thickness and in-plane dimensions scaling as $a$, $A \varpropto a$, and $K$ is independent of $a$. These two cases we consider are summarized together as $\mathbf{K}_a = a^{d-2} \mathbf{K}$, where $\mathbf{K}$ is independent of the microscopic length scale $a$.

In order to go from site displacements to continuum fields, we first define $\Phi \equiv [\mathbf{u}_1  \ldots  \mathbf{u}_{n_s}]^{\mathrm{T}}$ to be the $n_sd$-dimensional vector containing all of the site displacements for a single unit cell. We choose the primitive lattice vectors $\mathbf{a}_1, \ldots, \mathbf{a}_{d}$, and index the unit cells by integers $m_1, \ldots, m_d$. With this definition, the position of the center of each unit cell is 
\begin{equation*}
 \mathbf{x} = m_1 \mathbf{a}_1 + \ldots + m_d \mathbf{a}_d   
\end{equation*}
 and the site displacements are specified by 
 \begin{equation*}
  \Phi^{(m_1, \ldots, m_d)} = [\mathbf{u}_1^{(m_1, \ldots, m_d)}  \quad \ldots \quad \mathbf{u}_{n_s}^{(m_1, \ldots, m_d)} ]^{\mathrm{T}}.   
 \end{equation*}
 To specify the continuum degrees of freedom $\Phi(\mathbf{x},t)$,  we replace $\mathbf{u}_h^{(m_1, \ldots, m_d)}$ by smooth fields $\mathbf{u}_h$ that depend on position $\mathbf{x}$ and time $t$ using the relation $\mathbf{u}_h^{(m_1, \ldots, m_d)}(t) = \mathbf{u}_h(\mathbf{x}, t)$.

To lowest order in displacements, the extension of the $k$th bond ($k = 1, \ldots, n_b$) is given by
\begin{equation}
\label{eq:extensions}
        e_k^{(m_1, \ldots, m_d)} = \mathbf{\hat{s}}_k \cdot \left(\mathbf{u}_{h_2(k)}^{(m_1', \ldots, m_d')} - \mathbf{u}_{h_1(k)}^{(m_1, \ldots, m_d)} \right),
\end{equation}
where $h_1(k)$ and $h_2(k)$ are the indices of the sites connected by the bond. In this expression, $\mathbf{\hat{s}}_k$ is a unit vector pointing along the bond from site $h_1(k)$ to site $h_2(k)$, $(m_1, \ldots, m_d)$ indexes the unit cell for the $k$th bond and the site $h_1(k)$, and $(m_1', \ldots, m_d')$ indexes the unit cell for $h_2(k)$. 
The compatibility matrix is a linear map from the space of site displacements $\Phi$ to the space of bond extensions $\Xi \equiv [ e_1 \ldots e_{n_b} ]^{\mathrm{T}}$ that captures the relationship in \eqref{eq:extensions}. 

We now focus on the continuum description for periodic lattices, which are conveniently studied in wavevector space. To briefly review this approach, the wavevector $\mathbf{q} = \frac{1}{2 \pi}(\bar{q}_1 \mathbf{b}_1 + \ldots + \bar{q}_d \mathbf{b}_d)$ can be expressed in terms of the primitive reciprocal lattice vectors $\mathbf{b}_1, \ldots, \mathbf{b}_d$, defined from the lattice vectors $\mathbf{a}_k$ via $\mathbf{b}_{j} \cdot \mathbf{a}_k = 2 \pi \delta_{jk}$. In our convention, $\bar{q}_r$ are dimensionless and $\mathbf{b}_j$ have units of inverse length. The overbar $\bar{\cdot}$ is used to distinguish these dimensionless components $\bar{q}_r$ from the components $q_j$ in $\mathbf{q} = q_1 \mathbf{e}_1 + \ldots + q_d \mathbf{e}_d$, where $\{\mathbf{e}_j\}_{j = 1}^d$ is an orthonormal basis.
We represent lattice displacements using the form
\begin{equation}
\label{eq:harmonic_displacements}
\begin{aligned}
    \mathbf{u}_{h}^{(m_1, \ldots, m_d)} (\mathbf{q}) &= \mathbf{u}_h \exp(i \mathbf{q} \cdot (m_1 \mathbf{a}_1 + \ldots + m_d \mathbf{a}_d)) \\
    & = \mathbf{u}_h \exp(i (\bar{q}_1 m_1 + \ldots + \bar{q}_d m_d)),
\end{aligned}
\end{equation}
which is a convenient representation of a Fourier transform.
Substituting \eqref{eq:harmonic_displacements}  into \eqref{eq:extensions} gives rise to bond extensions
\begin{displaymath}
    e_k^{(m_1, \ldots, m_d)}(\mathbf{q}) = e_k(\mathbf{q}) e^{i \mathbf{q} \cdot (m_1 \mathbf{a}_1 + \ldots + m_d \mathbf{a}_d)},
\end{displaymath}
where the extension amplitudes are given by
\begin{equation}
\label{eq:extensions_wavevector}
    e_k(\mathbf{q}) = \mathbf{\hat{s}}_k \cdot \left(\mathbf{u}_{h_2(k)}e^{i \mathbf{q} \cdot \mathbf{n}_k} - \mathbf{u}_{h_1(k)} \right),  
\end{equation}
with $\mathbf{n}_k \equiv (m_1'-m_1)\mathbf{a}_1 + \ldots + (m_d' - m_d) \mathbf{a}_d$ pointing between the unit cells. In wavevector space, the compatibility matrix $\mathbf{C}(\mathbf{q})$ relates the amplitudes of site displacements $\Phi$ to those of bond extensions $\Xi(\mathbf{q}) \equiv [e_1(\mathbf{q}) \ldots e_{n_b}(\mathbf{q})]^\mathrm{T}$ via
\begin{equation}
\label{eq:compatibility_matrix_wavevector}
    \Xi(\mathbf{q}) = \mathbf{C}(\mathbf{q}) \Phi,
\end{equation}
the $k$th component of which is \eqref{eq:extensions_wavevector}.
Significantly, this expression for the matrix $\mathbf{C}(\mathbf{q})$ applies to complex-valued $\bar{q}_r$, which represent exponentially localized modes.

To obtain a continuum theory, we re-express the compatibility matrix in terms of the displacement gradients. The extension of each bond is given to first order by
\begin{align}
    \label{eq:continuum_extensions}
        e_k (\mathbf{x}, t)
        &= \mathbf{\hat{s}}_k \cdot (\mathbf{u}_{h_2(k)}(\mathbf{x} + \mathbf{n}_k, t) - \mathbf{u}_{h_1(k)} (\mathbf{x}, t)) \\
	& \approx \mathbf{\hat{s}}_k \cdot ((1 + \mathbf{n}_k \cdot \nabla)\mathbf{u}_{h_2(k)}(\mathbf{x}, t) - \mathbf{u}_{h_1(k)} (\mathbf{x}, t)), \nonumber
\end{align}
where $\nabla$ represents differentiation with respect to $\mathbf{x}$, and we take the continuum limit where $|\mathbf{n}_k| \sim \mathcal{O}(a) \sim \mathcal{O}(\delta) L$. To obtain a continuum version of \eqref{eq:compatibility_matrix_wavevector}, we re-write \eqref{eq:continuum_extensions} as
\begin{equation}
\label{eq:continuum_compatibility_preliminary}
    \Xi(\mathbf{x},t) = (\mathbf{C} + \mathbf{C}_{\mathbf{x}}) \Phi(\mathbf{x},t),
\end{equation}
where $\Xi(\mathbf{x}, t) \equiv [e_1(\mathbf{x}, t) \ldots e_{n_b}(\mathbf{x}, t)]^{\mathrm{T}}$, and $(\mathbf{C} + \mathbf{C}_{\mathbf{x}})$ is the continuum compatibility matrix. We have decomposed the compatibility matrix into the matrix $\mathbf{C}$ that does not contain differential operators $\nabla$, and the differential operator $\mathbf{C}_{\mathbf{x}}$, which is  $\mathcal{O}(\delta)$. We compute these terms in wavevector space to be $\mathbf{C} = \mathbf{C}(\mathbf{q})|_{\mathbf{q} = \mathbf{0}}$ and 
\begin{equation*}
    \mathbf{C}_{\mathbf{x}} = \sum_{j,r = 1}^d (-i) (\partial_{\bar{q}_r} \mathbf{C}(\mathbf{q})|_{\mathbf{q} = \mathbf{0}} )(\mathbf{a}_r \cdot \mathbf{e}_j) \partial_{x_j}.
\end{equation*}
To show that $\mathbf{C}_{\mathbf{x}} \Phi$ is  $\mathcal{O}(\delta)$, we non-dimensionalize the spatial variable by $L$, giving $\mathbf{x}^* = L^{-1} \mathbf{x}$. By our assumption that $L$ is the length scale over which spatial variations in $\Phi$ occur, the derivatives $\partial_{x_j^*} \Phi$ are of order 1. We find 
\begin{equation*}
\mathbf{C}_{\mathbf{x}} \Phi = \sum_{j,r = 1}^d (-i) (\partial_{\bar{q}_r} \mathbf{C}(\mathbf{q})|_{\mathbf{q} = \mathbf{0}} )(L^{-1} \mathbf{a}_r \cdot \mathbf{e}_j) \partial_{x_j^*} \Phi    
\end{equation*}
is $\mathcal{O}(\delta)$, because $|\mathbf{a}_r| \varpropto a$, giving $|L^{-1} \mathbf{a}_r| \sim \mathcal{O}(\delta)$, and all the other factors are of order 1. This decomposition of the compatibility matrix shows how spatial gradients enter the continuum equations of motion for $\Phi(\mathbf{x}, t)$.

\subsection{The continuum displacement field and other kinematic fields}
\label{sec:displacement_and_other}

The continuum field $\Phi(\mathbf{x}, t)$ contains all of the lattice degrees of freedom, but we now decompose this field into soft and high-frequency modes, where the soft degrees of freedom include the displacement field $\mathbf{u}(\mathbf{x},t)$. 
We define $\mathbf{u}(\mathbf{x},t)$ to be the average lattice site displacement:
\begin{equation}
\label{eq:average_unit_cell_displacement}
	\mathbf{u}(\mathbf{x},t) = \frac{1}{n_s} \sum_{h = 1}^{n_s} \mathbf{u}_h(\mathbf{x}, t).
\end{equation}
To decompose the rest of the degrees of freedom into zero modes and high-frequency modes, we diagonalize the dynamical matrix $\mathbf{C}^{\mathrm{T}} \mathbf{K} \mathbf{C}$.

The eigenvectors corresponding to zero eigenvalues can be split into the displacements $\{u_n\}_{n =1}^d$ uniform on each site and the (non-translational) floppy modes $\{w_m\}_{m = 1}^{n_w}$, while those corresponding to the non-zero eigenvalues are the the high-frequency modes $\{v_k\}_{k = 1}^{n_s d - (d + n_w)}$, with corresponding projection operators $\mathbf{P}_u$, $\mathbf{P}_w$, and $\mathbf{P}_v$. The null space of $\mathbf{C}$ is spanned by the eigenvectors for the zero modes, which are orthogonal to the high-frequency modes. We then decompose the continuum degrees of freedom into the subspaces spanned by these three sets of eigenvectors:
\begin{equation}
\label{eq:phi_decomposition}
\begin{aligned}
    \Phi(\mathbf{x}, t) &= (\mathbf{P}_u + \mathbf{P}_w + \mathbf{P}_v) \Phi(\mathbf{x}, t) \\
    &\equiv \Phi_u(\mathbf{x}, t) + \Phi_w(\mathbf{x}, t) + \Phi_v(\mathbf{x}, t).
\end{aligned}
\end{equation}
We show in Sec.~\ref{sec:homogenization} that the degrees of freedom associated with the high-frequency modes do not appear explicitly in the continuum equations of motion, having been ``integrated out.''
We then identify 
\begin{equation}
\label{eq:additional_fields}
\varphi_m(\mathbf{x}, t) = w_m^{\mathrm{T}} \Phi(\mathbf{x}, t) = w_m^{\mathrm{T}} \Phi_w(\mathbf{x}, t)
\end{equation}
as the continuum soft modes that we use to augment standard linear elasticity.  This definition justifies our use of $n_w$ to denote both the number of additional fields in Sec.~\ref{sec:generalized_elasticity} and the number of floppy modes (i.e., non-translational zero modes) in this section.

\subsection{Perturbing the compatibility matrix}
\label{sec:perturbing_lattice}
Even though topologically polarized lattices are not spatial-inversion symmetric, this broken symmetry might not be visible on large length scales. We demonstrate that perturbing a lattice configuration about a gapless state ensures that the homogenized theory preserves the symmetry breaking terms. Thus, we generalize the geometrical perturbations employed in Refs.~\cite{nassar_microtwist_2020, xia_microtwist_2021} from specific lattices to a generic lattice, and justify the form of the perturbation.

The compatibility matrix $\mathbf{C}(\mathbf{q})$ of a generic lattice depends on the wavevector components $\bar{q}_r$ ($r = 1, \ldots, d$) via complex phases of the form $e^{i \bar{q}_r}$, as shown in \eqref{eq:extensions_wavevector}. Therefore, a generic form for the determinant is
\begin{equation}
\label{eq:detCexpansion}
    \det \mathbf{C}(\mathbf{q}) = \sum_{\ell} c_{\ell} \exp (i \mathbf{n}_{\ell} \cdot \mathbf{q}) = \sum_{m = 0}^{\infty} \frac{i^m}{m!} P_m (\mathbf{q}),
\end{equation}
where the sum over $\ell$ is a finite sum and $P_m(\mathbf{q}) \equiv \sum_{\ell} c_{\ell} ( \mathbf{n}_{\ell} \cdot \mathbf{q})^m$ is a homogeneous polynomial of degree $m$ in the components $\bar{q}_r$, with real coefficients. We obtain the last equality in \eqref{eq:detCexpansion} using the Taylor series for the exponential function. In our notation, $\bar{q}_r^{-1}$ is the characteristic number of unit cells over which spatial variations occur. For these variations to be visible in the continuum, $\bar{q}_r^{-1}$ should scale as $L / a$, and therefore, $\bar{q}_r \sim \mathcal{O}(\delta)$. This observation tells us which (lowest-order) terms in \eqref{eq:detCexpansion} enter the continuum theory.

The continuum limit reproduces topological mechanics only if the coefficients $c_{\ell}$ depend on $\delta$. We show this by contradiction: otherwise, $\det \mathbf{C}(\mathbf{q}) = \frac{i^m_0}{m_0!} P_{m_0} (\mathbf{q})+ \mathcal{O}(\delta^{m_0+1})$ in the continuum limit, where the terms lowest order in $\delta$ come from just one $P_m$. Therefore, the zero modes in this limit correspond to the solutions of $P_{m_0} (\mathbf{q}) = 0$, a polynomial with real coefficients. For a given wavevector in the surface Brillouin zone for the edges of a strip, specified by real values of $\bar{q}_1, \ldots, \bar{q}_{d-1}$, the values of $\bar{q}_d$ that satisfy this zero-mode equation are either real (indicating bulk modes) or occur in complex conjugate pairs (corresponding to edge modes symmetrically distributed between opposite edges). This implies that topological polarization, if present in the discrete lattice, will not be visible in the continuum limit unless the coefficients $c_{\ell}$ depend on $\delta$.  

To capture topological polarization at large length scales, the polynomial equation satisfied by $\bar{q}_d$ must have complex coefficients. To achieve this, we begin with a gapless lattice, and perturb the positions of the sites to arrive at a gapped configuration. This perturbation turns the floppy modes of the gapless lattice into soft modes. The wavevector-space compatibility matrix becomes
\begin{equation}
\label{eq:perturbed_compatibility_matrix}
    \mathbf{C}_p(\mathbf{q}; \varepsilon) = \mathbf{C}(\mathbf{q}) + \varepsilon \mathbf{C}_w(\mathbf{q}) + \mathcal{O}(\varepsilon^2),
\end{equation}
where $\varepsilon$ is a small perturbation parameter. The matrix $\mathbf{C}(\mathbf{0})$ has a $(d + n_w)$-dimensional null space, corresponding to the presence of $n_w$ floppy modes at $\mathbf{q} = \mathbf{0}$. The perturbations to site positions are chosen so that $\mathbf{C} + \varepsilon \mathbf{C}_w$ has null space of dimension $d$, consisting of only the translational zero modes. By the same reasoning used in Sec.~\ref{sec:site_displacements_and_compatibility_matrix} to obtain \eqref{eq:continuum_compatibility_preliminary}, we deduce that the continuum counterpart to $\mathbf{C}_p(\mathbf{q}; \varepsilon)$ is
\begin{equation}
\label{eq:continuum_compatibility_matrix_homogenization}
    \mathbf{C}_{p, \mathbf{x}} \equiv \mathbf{C} + \mathbf{C}_{\mathbf{x}} + \varepsilon (\mathbf{C}_w + \mathbf{C}_{\mathbf{x},w}) + \mathcal{O}(\varepsilon^2),
\end{equation}
where we use the decomposition $\mathbf{C}_w = \mathbf{C}_w(\mathbf{0})$ and 
\begin{equation*}
{\mathbf{C}_{\mathbf{x}, w} = \sum_{j,r = 1}^d (-i) (\partial_{\bar{q}_r} \mathbf{C}_w(\mathbf{0}) )(\mathbf{a}_r \cdot \mathbf{e}_j) \partial_{x_j}},
\end{equation*}
following our decomposition of the unperturbed compatibility matrix into $\mathbf{C} + \mathbf{C}_{\mathbf{x}}$ in \eqref{eq:continuum_compatibility_preliminary}. As in Sec.~\ref{sec:site_displacements_and_compatibility_matrix}, $\mathbf{C}_{\mathbf{x}, w} \Phi \sim \mathcal{O}(\delta)$.

\begin{figure}
    \centering \includegraphics[width=0.45\textwidth]{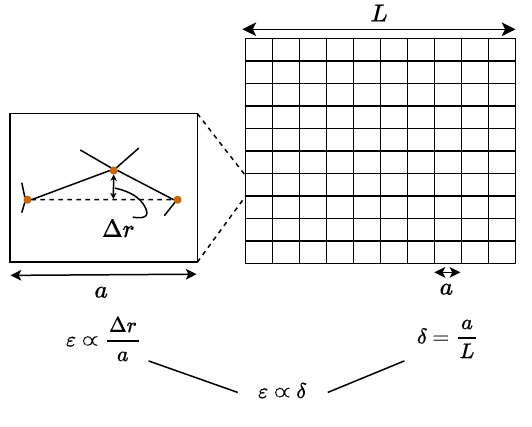}
    \caption{Physical interpretation of the scaling relationship $\varepsilon \sim \mathcal{O}(\delta)$. This choice of scaling relationship links two ratios. The grid on the right represents the many unit cells present in a system with length scale $L$. The box on the left represents a single unit cell and the perturbation $\Delta r$ applied to the site positions in order to obtain a gapped lattice from a gapless configuration. The relationship $\varepsilon \sim \mathcal{O}(\delta)$ connects a length scale associated with the geometry of a single unit cell to the characteristic number of unit cells over which spatial variations occur in the continuum. In the continuum theory, this scaling relationship ensures that we consider elastic terms of the same order in $\delta$, leading to elasticity with broken spatial inversion symmetry.}
    \label{fig:epsilon_scaling}
\end{figure}

Suppose
\begin{equation}
\label{eq:epsilon_scaling}
    \varepsilon \sim \mathcal{O}(\delta),
\end{equation}
which is illustrated in Fig.~\ref{fig:epsilon_scaling}. Physically, this choice of scaling for $\varepsilon$ means that the geometric perturbation to the site positions \emph{within} a unit cell relative to unit cell size $a$ has magnitude proportional to the inverse of the number of unit cells $\delta^{-1} = L / a$ over which spatial variations need to occur to be visible at the continuum level. In Appendix~\ref{sec:elastic_energy}, we show that this scaling relation ensures that elastic energy density remains bounded in the continuum limit.

In a lattice $\varepsilon$-perturbed away from a configuration with $n_w$ floppy modes, the coefficients $c_{\ell}(\varepsilon)$ of $q_r$ in the expansion of $\det \mathbf{C}_p(\mathbf{q}; \varepsilon)$ are polynomials in $\varepsilon$. The terms that are of order $d$ in $q_r$ (the lowest possible order of $q_r$) contain a factor of $\varepsilon^{n_w}$, raising their overall order in $\delta$ to $d + n_w$. In general, as shown rigorously in Appendix~\ref{sec:detC_perturbed_derivation},
\begin{align}
\label{eq:detCperturbed}
        & \det \mathbf{C}_p(\mathbf{q}; \varepsilon) \\
        & \qquad = \sum_{m = d}^{d + n_w} \frac{i^m}{m!}\varepsilon^{d+n_w-m} \bar{P}_m(\bar{q}_1, \ldots, \bar{q}_d) + \mathcal{O}(\delta^{d + n_w + 1}), \nonumber
\end{align}
where 
\begin{equation*}
\bar{P}_m(\bar{q}_1, \ldots, \bar{q}_d) = \sum_{k_1 + \ldots + k_d = m} \bar{A}_{k_1, \ldots, k_d} \bar{q}_1^{k_1} \ldots \bar{q}_d^{k_d}
\end{equation*}
for real coefficients $\bar{A}_{k_1, \ldots, k_d}$. Terms of order $m$ in $\bar{q}_r$ have order $d + n_w - m$ in $\varepsilon$, so the lowest order terms in $\delta$ have order $d + n_w$. These terms are shown in the sum over $m = d, \ldots, d+n_w$ in \eqref{eq:detCperturbed}. Given real values of $\bar{q}_1, \ldots, \bar{q}_{d-1}$, solving for zero modes using the terms that dominate in the continuum limit is equivalent to obtaining roots of a polynomial in $\bar{q}_d$ with complex coefficients. Therefore, an asymmetric distribution of floppy modes across opposite edges is possible, because complex roots are not constrained to occur in complex conjugate pairs. This motivates the study of $\varepsilon$-perturbed lattices to observe topological phenomena in the continuum limit. 

We have shown that choosing $\varepsilon \sim \mathcal{O}(\delta)$ is sufficient to capture topological polarization at small wavenumbers $\bar{q}_r$, when the polarization is present in the discrete lattice. To show that \eqref{eq:epsilon_scaling} is necessary for modeling topological mechanics in the continuum, we write $\varepsilon \sim \mathcal{O}(\delta^{\nu})$, where $\nu \geq 0$ because we require $\varepsilon$ to remain bounded in the continuum limit. The possibility that $\varepsilon$ is independent of $\delta$ is accounted for by the case $\nu = 0$. We show in Appendix~\ref{sec:necessity} that $\varepsilon \sim \mathcal{O}(\delta)$ is the only choice that captures topological polarization because $\nu \neq 1$ results in continuum equations of motion that are spatial inversion symmetric.

\subsection{Universal dependence of continuum floppy modes on lattice perturbations}

\begin{figure*}
    \centering \includegraphics{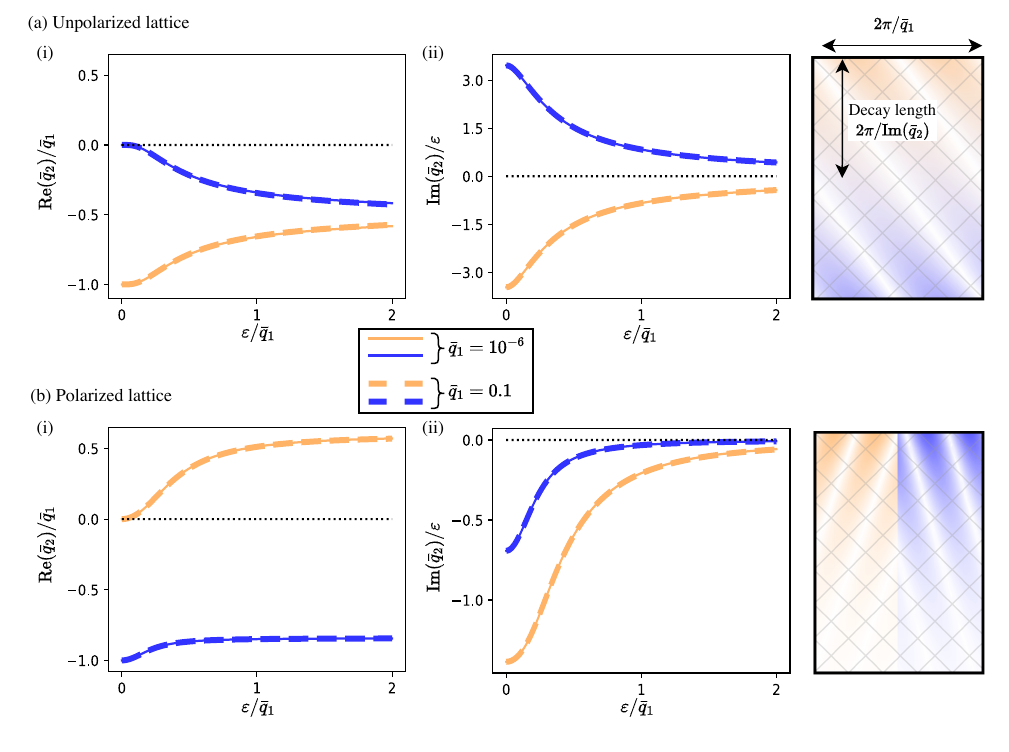}
    \caption{Universal dependence of continuum floppy modes on lattice perturbations. We compare analytical predictions with numerical results from strips of $\varepsilon$-perturbed kagome lattices (defined in Fig.~\ref{fig:distorted_kagome}) periodic along the $\mathbf{b}_1$-direction and finite along the $\mathbf{b}_2$-direction. We investigate how the wavenumber component $\mathrm{Re}(\bar{q}_2)$ and inverse decay length $\mathrm{Im}(\bar{q}_2)$ normal to the strip edge depend on the geometrical perturbation $\varepsilon$ to the lattice and on the wavenumber component $q_1$ parallel to the edge. For both the (a) unpolarized lattice and the (b) polarized lattice, we find that (i) $\mathrm{Re}(\bar{q}_2) / \bar{q}_1$ and (ii) $\mathrm{Im}(\bar{q}_2) / \varepsilon$ depend only on the ratio $\varepsilon / \bar{q}_1$, provided $\varepsilon$ and $\bar{q}_1$ are sufficiently small. The plots show two sets of curves: the thin solid curves correspond to $\bar{q}_1 = 10^{-6}$ and the thick dashed curves correspond to $\bar{q}_1 = 0.1$. Despite the large difference between the $\bar{q}_1$ values, the solid and dashed curves are nearly indistinguishable. The curves are colored according to the floppy modes they represent. In (a)(ii), the inverse decay lengths $\mathrm{Im}(\bar{q}_2)$ have opposite signs, corresponding to floppy modes localized on opposite edges, while (b)(ii) exhibits continuum floppy modes with $\mathrm{Im}(\bar{q}_2)$ of the same sign. The $\bar{q}_r$ values shown are dimensionless, as defined in Sec.~\ref{sec:site_displacements_and_compatibility_matrix}. The rectangles on the right are strips showing representative displacement magnitudes for the continuum floppy modes.}
    \label{fig:perturbation_graphs}
\end{figure*}
\begin{figure}
    \centering \includegraphics{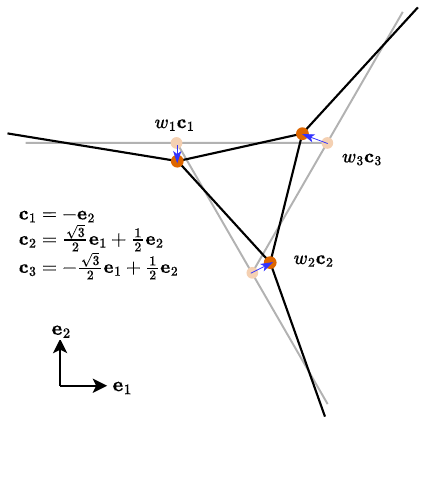}
    \caption{Constructing distorted kagome lattices. The positions of sites in the kagome unit cell are shifted along directions parallel to unit vectors $\mathbf{c}_1, \mathbf{c}_2, \mathbf{c}_3$ by distances $w_1, w_2, w_3$ respectively. The blue arrows represent the changes in site position due to the geometrical distortion. The undistorted unit cell has three sites placed at the vertices of an equilateral triangle. The distorted unit cell shown corresponds to $w_1 = 1, w_2 = 1.5, w_3 = 1.75$.}
    \label{fig:distorted_kagome}
\end{figure}
We now show that considering lattices $\varepsilon$-perturbed away from a gapless configuration enables us to characterize the universal behaviors of their floppy modes. We consider strips of two-dimensional lattices, which are periodic along the $\mathbf{b}_1$-direction and finite along the $\mathbf{b}_2$-direction, where $\{\mathbf{b}_1, \mathbf{b}_2\}$ are reciprocal lattice vectors with $\mathbf{b}_2$ normal to the edge of the strip. Thus, we require that the wavevector component $\bar{q}_1$ be real, but allow $\bar{q}_2$ to be complex, to account for the exponential localization of floppy modes on the strip edges. We retain only the terms lowest order in $\delta$, so solving for floppy modes using \eqref{eq:detCperturbed} is equivalent to solving
\begin{equation}
\label{eq:universal_zero_modes}
    \sum_{m = d}^{d + n_w} \frac{i^m}{m!} \bar{P}_m\left(\frac{\bar{q}_1}{\varepsilon}, \frac{\bar{q}_2}{\varepsilon}\right) = 0,
\end{equation}
obtained by dividing \eqref{eq:detCperturbed} through by $\varepsilon^{d + n_w}$. The form of this equation guarantees that the floppy mode takes the form
\begin{equation}
\label{eq:universal_scaling_over_epsilon}
    \frac{\bar{q}_2}{\varepsilon} = F\left(\frac{\bar{q}_1}{\varepsilon}\right).
\end{equation}
This equation implies that $\bar{q}_2 / \bar{q}_1 = (\varepsilon/ \bar{q}_1) F(\bar{q}_1/ \varepsilon)$. Thus, the ratio $\varepsilon/\bar{q}_1$ determines the values of both $\bar{q}_2/\varepsilon$ and $\bar{q}_2 / \bar{q}_1$ at sufficiently small $\varepsilon$ and $\bar{q}_r$. 

In Fig.~\ref{fig:perturbation_graphs}, we plot two key quantities as a function of $\varepsilon / \bar{q}_1$, which demonstrates the universality of these solutions: $\mathrm{Re}(\bar{q}_2) / \bar{q}_1$ and $\mathrm{Im}(\bar{q}_2) / \varepsilon$. We plot $\varepsilon/\bar{q}_1$ on the horizontal axis because $\varepsilon$ parametrizes the perturbation away from the gapless configuration.
The quantity $\mathrm{Re}(\bar{q}_2) / \bar{q}_1$ determines the direction of the wavefronts associated with a localized floppy mode.
In general, the displacements due to a floppy mode vary sinusoidally with wavefronts normal to $\bar{q}_1 \mathbf{b}_1 + \mathrm{Re}(\bar{q}_2) \mathbf{b}_2$, and with an amplitude exponentially decaying into the bulk. The quantity that characterizes this decay is the inverse decay length $\mathrm{Im}(\bar{q}_2)$, so the ratio $\mathrm{Im}(\bar{q}_2) / \varepsilon$ characterizes how the perturbation $\varepsilon$ sets this decay length scale.

We demonstrate this universal relation numerically by computing the floppy modes of two distorted kagome lattices: unpolarized and polarized (Fig.~\ref{fig:perturbation_graphs}). The geometrical configurations of these lattices are obtained from the standard kagome lattice via perturbations parameterized by $w_1, w_2, w_3$, shown in Fig.~\ref{fig:distorted_kagome}. The unpolarized lattice corresponds to $w_1 = w_2 = w_3 = \varepsilon$, and the polarized lattice corresponds to $-w_1 = w_2 = -w_3 = \varepsilon$. In Fig.~\ref{fig:perturbation_graphs}, we plot $\mathrm{Re}(\bar{q}_2) / \bar{q}_1$ and $\mathrm{Im}(\bar{q}_2) / \varepsilon$ against $\varepsilon / \bar{q}_1$ for values of $\bar{q}_1 = 10^{-6}$ and $0.1$.  Despite the two values of $\bar{q}_1$ differing by several orders of magnitude, the rescaled solutions fall on universal curves over the range $0 \leq \varepsilon / \bar{q}_1 \leq 2$, demonstrating the analytical prediction \eqref{eq:universal_scaling_over_epsilon}. 

The asymptotic behaviors of the curves in the limits $\varepsilon/\bar{q}_1 \to 0$ and $\varepsilon/\bar{q}_1 \to +\infty$ are a general consequence of the form of \eqref{eq:detCperturbed}. In the limit $\varepsilon/\bar{q}_1 \to 0$, we set $\varepsilon = 0$ in \eqref{eq:detCperturbed} and retain terms lowest order in $\delta$. Then, $\bar{q}_2 / \bar{q}_1$ satisfies $\bar{P}_{d + n_w}(1, \bar{q}_2/ \bar{q}_1) = 0$, which is satisfied by the bulk floppy modes of the gapless configuration.  These bulk floppy modes correspond to real $\bar{q}_2$, so that for the kagome lattice, $\bar{q}_2/\bar{q}_1 = \mathrm{Re}(\bar{q}_2) / \bar{q}_1 \rightarrow 0$ or $-1$, consistent with the numerical limits on the left-hand side of Figs.~\ref{fig:perturbation_graphs}a(i),b(i). 

For $\varepsilon/\bar{q}_1 \to +\infty$, we use \eqref{eq:universal_zero_modes} and bring out a factor of $(\bar{q}_1/\varepsilon)^m$ from each homogeneous polynomial $\bar{P}_m\left(\frac{\bar{q}_1}{\varepsilon}, \frac{\bar{q}_2}{\varepsilon}\right)$ to obtain
\begin{equation}
    \sum_{m = d}^{d + n_w} \left(\frac{\bar{q}_1}{\varepsilon}\right)^m \frac{i^m}{m!} \bar{P}_m\left(1, \frac{\bar{q}_2}{\bar{q}_1}\right) = 0.
\end{equation}
In the limit $\varepsilon/\bar{q}_1 \to +\infty$, only the $m = d$ term remains, so that $\bar{q}_2 / \bar{q}_1$ satisfies $\bar{P}_{d}(1, \bar{q}_2/\bar{q}_1) = 0$. Since these values of $\bar{q}_2/\bar{q}_1$ satisfy a polynomial equation with real coefficients, they occur either (a) in complex conjugate pairs or (b) as real numbers. These two cases correspond to the (a) unpolarized and (b) polarized distorted kagome lattices in Fig.~\ref{fig:perturbation_graphs}. The two floppy modes of the unpolarized lattice have $\mathrm{Re}(\bar{q}_2) / \bar{q}_1$ curves that asymptotically approach the same value [right-hand side of Fig.~\ref{fig:perturbation_graphs}a(i)], while those of the polarized lattice converge to distinct values [right-hand side of Fig.~\ref{fig:perturbation_graphs}b(i)]. In Sec.~\ref{sec:polarization_2D}, we generalize this result to any continuum topological mechanics in two dimensions. 

For the asymptotics of $\mathrm{Im}(\bar{q}_2) / \varepsilon$ [Fig.~\ref{fig:perturbation_graphs}a(ii),b(ii)], we show that $\mathrm{Im}(\bar{q}_2) / \varepsilon$ scales as $(\varepsilon/\bar{q}_1)^{-\chi}$ in the limit $\varepsilon/\bar{q}_1 \to +\infty$, where $\chi = 1$ for unpolarized lattices and $\chi = 2$ for polarized lattices. To begin, we expand $F(\bar{q}_1/\varepsilon)$ in \eqref{eq:universal_scaling_over_epsilon} as a power series for small $\bar{q}_1/ \varepsilon$,
\begin{equation}
\label{eq:power_series_F}
    \frac{\bar{q}_2}{\varepsilon} = F'(0) \frac{\bar{q}_1}{\varepsilon} + \frac{1}{2} F''(0) \left(\frac{\bar{q}_1}{\varepsilon}\right)^2 + \mathcal{O}\left(\left(\frac{\bar{q}_1}{\varepsilon}\right)^3\right),
\end{equation}
where $F(0) = 0$ because we are considering only continuum edge modes defined by \eqref{eq:def_continuum_edge_mode}, which satisfy $\bar{q}_2 \to 0$ when $\bar{q}_1 \to 0$. We show in Appendix~\ref{sec:analyticity} that $F$ is analytic on a neighborhood of zero, so this expansion is valid. Substituting \eqref{eq:universal_scaling_over_epsilon} into \eqref{eq:universal_zero_modes} and differentiating with respect to $\bar{q}_1/\varepsilon$ shows that $F'(0)$ satisfies $\bar{P}_d(1, F'(0)) = 0$. As in the previous paragraph, the unpolarized lattice has $F'(0)$ occurring in complex conjugate pairs, while the polarized lattice has real values of $F'(0)$. When taking the limit $\varepsilon / \bar{q}_1 \to +\infty$, we discard the higher-order terms in the power series \eqref{eq:power_series_F} because $\bar{q}_1 / \varepsilon \to 0$. For the unpolarized lattice, 
\begin{equation*}
    \mathrm{Im}(\bar{q}_2) / \varepsilon \approx (\varepsilon / \bar{q}_1)^{-1} \mathrm{Im}\,F'(0) ,
\end{equation*}
because $\mathrm{Im}\,F'(0) \neq 0$. For the polarized lattice, 
\begin{equation*}
\mathrm{Im}(\bar{q}_2) / \varepsilon \approx \frac{1}{2}(\varepsilon / \bar{q}_1)^{-2}\mathrm{Im}\,F''(0) ,
\end{equation*}
because $\mathrm{Im}\,F'(0) = 0$. These analytical results correspond to the different asymptotics in Figs.~\ref{fig:perturbation_graphs}a(ii),b(ii) for $\varepsilon/\bar{q}_1 \to +\infty$.

In summary, we have characterized how the continuum floppy modes of $\varepsilon$-perturbed lattices depend on the perturbation parameter $\varepsilon$. We find different scaling relationships for the unpolarized and polarized cases. We use the specific case of distorted kagome lattices to illustrate our universal analytical predictions in the continuum.

\begin{figure*}
    \centering
    \includegraphics{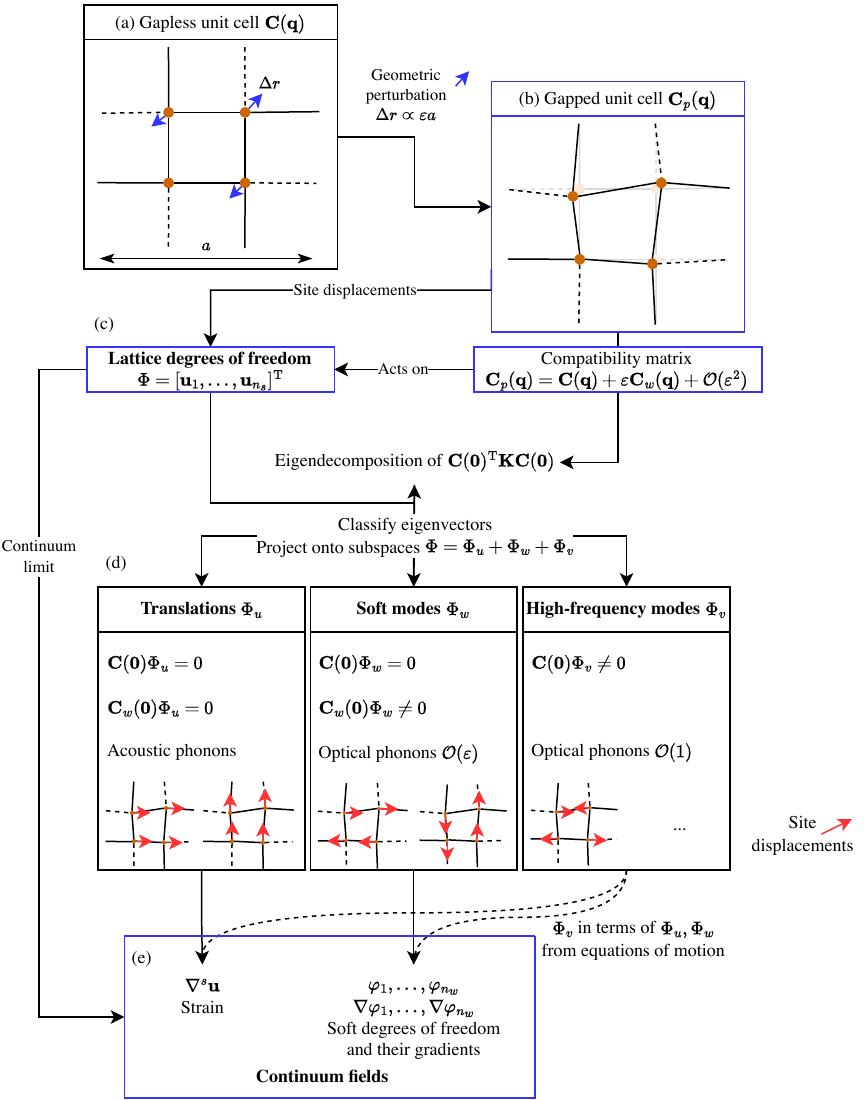}
    \caption{How to obtain continuum fields from a discrete lattice. (a) A gapless lattice is geometrically perturbed, shifting the site positions by distances proportional to $\varepsilon a$, where $a$ is the unit-cell size. (b) The resulting gapped lattice has a compatibility matrix given by \eqref{eq:perturbed_compatibility_matrix}, which acts on the site displacements to give the bond extensions. (c) The lattice degrees of freedom are the $n_s d$ site displacements, which are projected onto three subspaces, as shown in \eqref{eq:phi_decomposition}. (d) The subspaces are spanned by the eigenvectors of $\mathbf{C}(\mathbf{0})^{\mathrm{T}}\mathbf{K}\mathbf{C}(\mathbf{0})$, classified according to whether they are in the null spaces of both $\mathbf{C}(\mathbf{0})$ and $\mathbf{C}_w(\mathbf{0})$ [translations $\Phi_u$], in only the null space of $\mathbf{C}(\mathbf{0})$ [soft modes $\Phi_w$], or in neither of these null spaces [high-frequency modes $\Phi_v$]. These eigenvectors are natural ways to express the lattice degrees of freedom. The equations of motion are used to express $\Phi_v$ in terms of $\Phi_u$ and $\Phi_w$, given by \eqref{eq:pseudo-inversion_Phi_v}. (e) The continuum fields arise naturally from the degrees of freedom in (d). The translations $\Phi_u$ give rise to strain, and the soft modes $\Phi_w$ give rise to scalar fields and their gradients. These continuum fields satisfy the equations of motion \eqsref{eq:continuum_eq} and the constitutive relations \eqref{eq:constitutive}. In taking the continuum limit, the $n_s d$ lattice degrees of freedom, where $d$ is the number of spatial dimensions, have been reduced to $d + n_w$ continuum fields, where $n_w$ is the number of local soft modes.}
    \label{fig:eigendecomposition}
\end{figure*}
\subsection{The homogenization procedure}
\label{sec:homogenization}
In this section, we show how to obtain continuum equations of motion, \eqref{eq:continuum_eq}, systematically from the discrete lattice equations of motion, with additional details in Appendix~\ref{sec:projecting_subspaces}. Figure~\ref{fig:eigendecomposition} outlines the overall approach, bringing together concepts introduced in the previous subsections. We begin with the equations of motion for the degrees of freedom of the discrete lattice $\Phi^{(m_1, \ldots, m_d)}$, which we replace with the continuum fields $\Phi(\mathbf{x},t)$. 
These $n_s d$ fields obey Newton's second law, which takes the form:
\begin{align}
	\label{eq:full_eqs_motion}
	\mathbf{M}_a \partial_t^2 \Phi(\mathbf{x}, t) = -\mathbf{Q}_{p, \mathbf{x}} \mathbf{K}_a \mathbf{C}_{p,\mathbf{x}} \Phi(\mathbf{x}, t) + F_a(\mathbf{x},t), 
\end{align}
where $F_a$ is a vector of external forces, $\mathbf{C}_{p,\mathbf{x}}$ is the continuum compatibility matrix defined in \eqref{eq:continuum_compatibility_matrix_homogenization}, and
\begin{displaymath}
    \mathbf{Q}_{p, \mathbf{x}} \equiv (\mathbf{C} + \varepsilon \mathbf{C}_w - \mathbf{C}_{\mathbf{x}} - \varepsilon \mathbf{C}_{\mathbf{x},w} + \mathcal{O}(\varepsilon^2))^{\mathrm{T}}
\end{displaymath}
is the continuum equilibrium matrix obtained from $\mathbf{C}_{p,\mathbf{x}}$ by replacing all spatial derivatives $\partial_{x_j}$ with $-\partial_{x_j}$ and taking the matrix transpose. This equilibrium matrix maps bond tensions to the resultant forces on lattice sites due to the bonds and lets us define the continuum dynamical matrix, $\mathbf{Q}_{p, \mathbf{x}} \mathbf{K}_a \mathbf{C}_{p,\mathbf{x}}$. 

To reduce the number of degrees of freedom in \eqref{eq:full_eqs_motion}, we use the decomposition in \eqref{eq:phi_decomposition} by projecting \eqref{eq:full_eqs_motion} onto the high-frequency modes using the operator $\mathbf{P}_v$, uniform displacements using $\mathbf{P}_u$, and floppy modes using $\mathbf{P}_w$. We use the $\mathbf{P}_v$ projection to eliminate the high-frequency degrees of freedom $\Phi_v$. Retaining only terms lowest order in $\delta$ (i.e., neglecting $\mathcal{O}(\delta^2)$ terms) in the $\mathbf{P}_v$ projection, we find
\begin{align}
	\label{eq:pre-pseudo-inversion_Phi_v}
    &\mathbf{P}_v \mathbf{C}^{\mathrm{T}}\mathbf{K} \mathbf{C} \mathbf{P}_v \Phi_v(\mathbf{x},t) \\
    &\approx -\mathbf{P}_v \mathbf{C}^{\mathrm{T}} \mathbf{K} (\mathbf{C}_{\mathbf{x}} (\Phi_u(\mathbf{x}, t) + (\mathbf{C}_{\mathbf{x}} + \varepsilon \mathbf{C}_w) \Phi_w(\mathbf{x}, t)), \nonumber
\end{align}
where, significantly, the terms corresponding to the inertial dynamics are neglected due to being higher order in $\delta$, so that \eqref{eq:pre-pseudo-inversion_Phi_v} is a constraint that uniquely determines $\Phi_v$.

We now solve \eqref{eq:pre-pseudo-inversion_Phi_v} for $\Phi_v$ and substitute the result into the other projections of the equations of motion.
We define $\mathbf{P}_v \mathbf{C}^{\mathrm{T}}\mathbf{K} \mathbf{C} \mathbf{P}_v |_{U_v}$ to be the restriction of $\mathbf{P}_v \mathbf{C}^{\mathrm{T}}\mathbf{K} \mathbf{C} \mathbf{P}_v$ to the subspace $U_v$ spanned by the high-frequency modes. 
 Since $\Phi_v(\mathbf{x},t)$ is a linear combination of eigenvectors of $\mathbf{C}^{\mathrm{T}}\mathbf{K} \mathbf{C}$ with non-zero eigenvalues, this restriction is invertible.
 We then define 
\begin{equation}
\label{eq:def_B}
    \mathbf{B} = \left( \mathbf{P}_v \mathbf{C}^{\mathrm{T}}\mathbf{K} \mathbf{C} \mathbf{P}_v |_{U_v} \right)^{-1} \mathbf{P}_v,
\end{equation}
 and solve \eqref{eq:pre-pseudo-inversion_Phi_v} for $\Phi_v$ in terms of $\Phi_u, \Phi_w$:
\begin{equation}
	\label{eq:pseudo-inversion_Phi_v}
	\Phi_v \approx -\mathbf{B} \mathbf{C}^{\mathrm{T}} \mathbf{K} (\mathbf{C}_{\mathbf{x}} \Phi_u + (\mathbf{C}_{\mathbf{x}} + \varepsilon \mathbf{C}_w) \Phi_w).
\end{equation}
This result shows that $|\Phi_v(\mathbf{x},t)|$ is $ \mathcal{O}(\delta)$. The matrix $\mathbf{B}$ is also known as the Moore-Penrose pseudoinverse of $\mathbf{C}^{\mathrm{T}}\mathbf{K} \mathbf{C}$~\cite{axler_linear_2024}.

We now use \eqref{eq:pseudo-inversion_Phi_v} to eliminate $\Phi_v(\mathbf{x},t)$ from the equations of motion. In Appendix~\ref{sec:projecting_subspaces}, we derive the projections of \eqref{eq:full_eqs_motion} onto the low-frequency subspaces using $\mathbf{P}_u$ and $\mathbf{P}_w$ to arrive at \eqref{eq:eq_motion_projection_Px} and \eqref{eq:eq_motion_projection_Pw}, respectively. Substituting \eqref{eq:pseudo-inversion_Phi_v} results in the continuum equations of motion
\begin{equation}
	\label{eq:continuum_eqs_preliminary}
	\begin{aligned}
		& a^2 \mathbf{P}_u (\mathbf{M} \partial_t^2 ( \Phi_u + \Phi_w) - F) \\ 
        &\qquad = \mathbf{P}_u \mathbf{C}_{\mathbf{x}}^{\mathrm{T}} \mathbf{\tilde{K}} (\varepsilon \mathbf{C}_w \Phi_w + \mathbf{C}_{\mathbf{x}}( \Phi_u + \Phi_w))\\
		& a^2 \mathbf{P}_w (\mathbf{M} \partial_t^2 ( \Phi_u + \Phi_w) - F) \\ 
        &\qquad =\mathbf{P}_w \mathbf{C}_{\mathbf{x}}^{\mathrm{T}} \mathbf{\tilde{K}} (\varepsilon \mathbf{C}_w \Phi_w + \mathbf{C}_{\mathbf{x}}( \Phi_u + \Phi_w)) \\
        & \qquad \qquad - \varepsilon \mathbf{P}_w \mathbf{C}_w^{\mathrm{T}} \mathbf{\tilde{K}} (\varepsilon \mathbf{C}_w \Phi_w + \mathbf{C}_{\mathbf{x}}( \Phi_u + \Phi_w)),
	\end{aligned}
\end{equation}
where we introduce the force density $F \equiv F_a / a^d$ in the continuum limit and
\begin{equation}
\label{eq:def_tilde_K}
    \mathbf{\tilde{K}} = \mathbf{K} - \mathbf{K} \mathbf{C} \mathbf{P}_v \mathbf{B} \mathbf{P}_v \mathbf{C}^{\mathrm{T}} \mathbf{K}
\end{equation}
is the effective spring constant matrix that takes into account the relaxation due to ``integrating out'' the high-frequency modes $\Phi_v$. Equation~(\ref{eq:def_tilde_K}) shows that the high-frequency modes reduce the spring constant matrix from $\mathbf{K}$ to $\mathbf{\tilde{K}}$ by accounting for the relaxation in the ball-and-spring network (see Appendix~\ref{sec:symmetries} for details), which is also observed in amorphous solids~\cite{Zaccone_Scossa-Romano_2011,liu_local_2022}. 
In Appendix~\ref{sec:gen_elastic_moduli}, we derive the mapping between Eqs.~(\ref{eq:continuum_eqs_preliminary}) and the continuum equations of motion Eqs.~(\ref{eq:continuum_eq}) with constitutive relations \eqref{eq:constitutive}, previously introduced based on symmetries. Appendix~\ref{sec:gen_elastic_moduli} also derives the generalized elastic moduli in terms of the compatibility and spring constant matrices of the underlying discrete lattice, \eqref{eq:elastic_tensors}. In Appendix~\ref{sec:symmetries}, we use the compatibility matrix of the microscopic lattice to derive the symmetry relations \eqref{eq:relations} satisfied by these moduli. These relations are consistent with the \textit{a priori} symmetries used in Sec.~\ref{sec:generalized_elasticity} to write down the augmented continuum theory.

A key feature of the homogenization procedure resulting in Eqs.~(\ref{eq:continuum_eq}--\ref{eq:constitutive}) is that all the terms retained in the equations of motion have the same order in $\delta$ as $\delta \to 0$. This ensures that any topological effects remain visible at large length scales. The passage from the discrete lattice to a continuum model involves a reduction in degrees of freedom, from the $n_s d$ components in \eqref{eq:full_eqs_motion} to the $d + n_w$ components in \eqsref{eq:continuum_eq}.
Since our homogenization procedure involves retaining terms of lowest order in $\delta$ in the equations of motion, we expect that solving for zero modes using the homogenized continuum theory is equivalent to solving for zero modes using the lowest-order terms in $\delta$ in $\det \mathbf{C}_p(\mathbf{q}; \varepsilon)$. This equivalence is proven in Appendix~\ref{sec:recovering_determinant}, where we show that $\det \mathbf{P}_K^{\mathrm{T}} \mathbf{\hat{C}}(\mathbf{q})$ of the continuum theory is proportional to the lowest-order terms in $\det \mathbf{C}_p(\mathbf{q}; \varepsilon)$ when both the discrete and the continuum Maxwell criteria are satisfied.

The relations $\varepsilon \sim \mathcal{O}(\delta)$ and $|\Phi_v| \sim \mathcal{O}(\delta)$ in our homogenization procedure are physically significant because they ensure that the elastic energy density remains bounded as $\delta \to 0$, as shown in Appendix~\ref{sec:elastic_energy}. We therefore interpret these scaling relations as follows: high-frequency degrees of freedom $\Phi_v$ have small amplitudes which scale as $\delta$, and the soft degrees of freedom $\Phi_w$ do not scale with $\delta$ when $\varepsilon \sim \mathcal{O}(\delta)$.

This homogenization procedure links our generalized continuum theory Eqs.~(\ref{eq:continuum_eq}, \ref{eq:constitutive}) to microscopic lattice-based realizations. With this connection, we see that the continuum displacement field $\mathbf{u}$ represents the average site displacement within a unit cell and the additional fields $\varphi_1, \ldots, \varphi_{n_w}$ represent local soft modes of the lattice. The mass density $\rho$ corresponds to the mass density of the lattice, and the body force density $\mathbf{f}$ is the total force density acting over the sites of a unit cell. Gapping the lattice by perturbing its configuration away from a gapless state gives rise to dependence of the elastic energy density on the fields $\varphi_1, \ldots, \varphi_{n_w}$ in addition to their gradients. Our results therefore provide design principles for realizing topological phenomena in the continuum using mechanical metamaterials. These principles dictate how the underlying lattice geometry gives rise to the desired number of local soft modes $n_w$ and the desired generalized elastic moduli.

\vspace*{-\baselineskip}
\section{Classifying topological phenomena}
\label{sec:classifying}
We classify topological floppy modes in the continuum according to invariants given by \eqref{eq:continuum_topological_invariants}, and the number $n_w$ of additional fields necessary to capture them.
\vspace*{-\baselineskip}
\subsection{Topological edge modes}
\label{sec:edge_modes}
Setting $n_w = 0$ in the continuum equations of motion Eq.~(\ref{eq:continuum_eq}, \ref{eq:constitutive}) gives standard linear elasticity. For this case, the equations are symmetric under spatial inversion and cannot capture topological polarization. Here we focus on the case $n_w \geq 1$ and use \eqref{eq:continuum_topological_invariants} to define an invariant for edge modes in the continuum. We also show that in two dimensions, the presence of shear-dominant Guest-Hutchinson modes~\cite{rocklin_transformable_2017} is sufficient for topological polarization in the continuum.

\subsubsection{Topological invariant for edge modes}
For edge modes, the path of integration in \eqref{eq:continuum_topological_invariants} is a straight line. 
To define this path, we consider floppy modes localized on an edge normal to $\mathbf{e}_k$, where $\{\mathbf{e}_j\}_{j = 1}^d$ is an orthonormal basis. To count the continuum edge modes, which we previously defined in \eqref{eq:def_continuum_edge_mode}, we use the numbers $N_{+,k}$ and $N_{-,k}$ of such modes on opposite sides, i.e., decaying along the $+\mathbf{e}_k$ and $-\mathbf{e}_k$ directions respectively. Using the approach of Ref.~\cite{saremi_topological_2020} to take the continuum limit, we define the invariant by the difference 
\begin{equation}
\label{eq:edge_modes_invariant}
    N_{+, k} - N_{-,k} = \lim_{\alpha \to 0^{+}} \frac{1}{\pi i} \int_{-\alpha^c}^{\alpha^c} \frac{\frac{\mathrm{d}}{\mathrm{d} \tau} \det \mathbf{P}_K^{\mathrm{T}} \mathbf{\hat{C}}(\mathbf{q}_{\alpha}^k(\tau))}{\det \mathbf{P}_K^{\mathrm{T}} \mathbf{\hat{C}}(\mathbf{q}_{\alpha}^k(\tau))} \, \mathrm{d}\tau,
\end{equation}
where the integration path is given by
\begin{displaymath}
    \mathbf{q}_{\alpha}^k(\tau) = \tau \mathbf{e}_k + \alpha \sum_{\substack{j = 1, j \neq k}}^d \mathbf{e}_j
\end{displaymath}
for real $\tau$ and exponent $c$ satisfying $0 < c < 1/(d + n_w)$, i.e., along this path $\tau$ ranges over the $q_k$ component and all other components of $\mathbf{q}$ are set to the constant $\alpha$. We prove this version of the topological bulk-edge correspondence in the continuum in Appendix~\ref{sec:counting_edge_modes} and summarize below.

The path of integration in \eqref{eq:edge_modes_invariant} is not a closed path because there is no Brillouin zone in the continuum, i.e., there are no non-contractible loops in $\mathbf{q}$-space for which the components of $\mathbf{q}$ are real. However, we can choose loops in the complex plane that enclose the complex roots $\tau = q_k$ of the polynomial $\det \mathbf{P}_K^{\mathrm{T}} \mathbf{\hat{C}}(\mathbf{q}_{\alpha}^k(\tau))$. Cauchy's argument principle (see e.g., Ref.~\cite{asmar2018complex}) applied to such loops counts the difference $N_{+, k} - N_{-,k}$. However, when we use loops involving complex components of $\mathbf{q}$, this topological invariant is not calculated solely from bulk information. To define a bulk invariant, we take the limit $\alpha \to 0^{+}$ in \eqref{eq:edge_modes_invariant}, for which contributions from complex $\mathbf{q}$ components are negligible, see Appendix~\ref{sec:counting_edge_modes} for a proof. Taking the limit $\alpha \to 0^{+}$ also ensures that $N_{+, k} - N_{-,k}$ counts only floppy modes that satisfy the defining property of continuum edge modes, \eqref{eq:def_continuum_edge_mode}. We conclude that this continuum version of the topological invariant relies only on information within the continuum compatibility matrix $\det \mathbf{P}_K^{\mathrm{T}} \mathbf{\hat{C}}(\mathbf{q})$ at small real values of $q_j$, as reflected in \eqref{eq:edge_modes_invariant}.

The requirement that the exponent $c$ satisfy the inequality ${c < 1/(d + n_w)}$ ensures that the domain of integration $[-\alpha^c, \alpha^c]$ is sufficiently large to capture the roots $\tau = q_k$ of $\det \mathbf{P}_K^{\mathrm{T}} \mathbf{\hat{C}}(\mathbf{q}_{\alpha}^k(\tau))$ corresponding to the continuum edge modes. We show in Appendix~\ref{sec:counting_edge_modes} that this requirement can be relaxed to $0 < c < 1$ when the components $q_k$ for all of the edge modes are analytic functions of $\alpha$.

\subsubsection{Characterizing topological polarization in 2D elasticity}
\label{sec:polarization_2D}
\begin{figure*}
    \centering
    \includegraphics{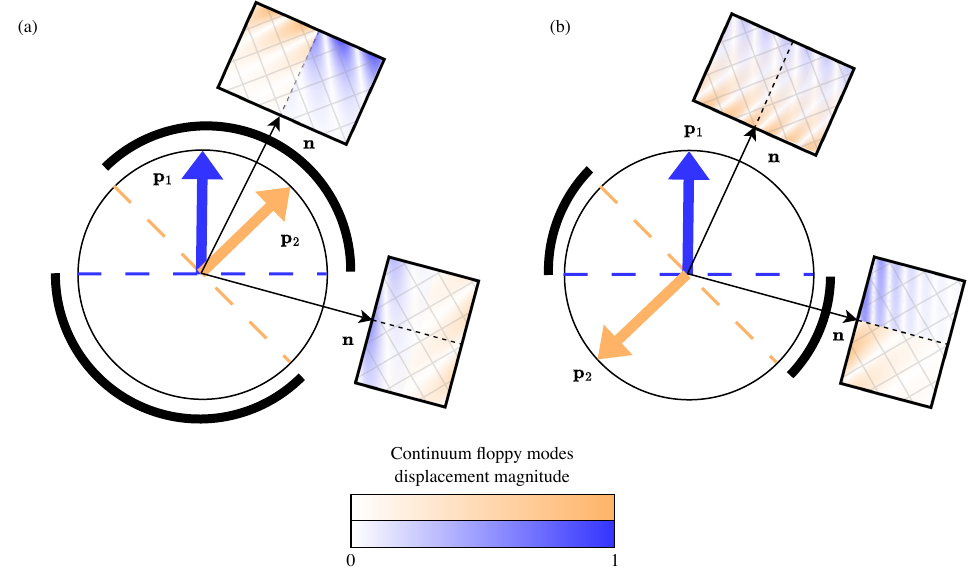}
    \caption{Shear-dominant Guest-Hutchinson modes (i.e., $\Delta > 0$) imply topological polarization in the continuum. Each of the two continuum floppy modes is associated with a polarization direction that can be computed from the generalized elastic moduli \eqref{eq:polarization_direction} when $\Delta > 0$. The continuum floppy mode distribution between the edges of a strip is a function of the normal $\mathbf{n}$ to the strip edges. This distribution is completely determined by the polarization directions $\mathbf{p}_1$, $\mathbf{p}_2$. The rectangles represent strips of different orientations and the displacement magnitudes of the two localized floppy modes are indicated in the legend. The thick black arcs indicate normal directions for which both floppy modes are localized on the same edge. These directions make acute angles with both polarization directions. The soft directions are indicated by the dashed lines. 
    For the polarization $\mathbf{p}_1$ ($\mathbf{p}_2$),
    the vector arrow, the line for soft directions, and the corresponding floppy modes are colored in blue (orange).
   (a) and (b) have the same soft directions but opposite signs for $\mathbf{p}_2$, resulting in different floppy mode distributions for the same strip orientation.}
    \label{fig:2Dpolarisation}
\end{figure*}

In this subsection, we show how the generalized elastic moduli in~\eqref{eq:constitutive} fully characterize topological polarization in two dimensions. In this case, there are exactly two continuum edge modes, and the material is topologically polarized if a strip can be cut such that both floppy modes are localized on the same edge. By contrast, in the unpolarized case, every strip has floppy modes localized on opposite edges. We show that topological polarization is determined by the sign of a discriminant $\Delta$. When $\Delta > 0$, we define polarization directions $\mathbf{p}_1$ and $\mathbf{p}_2$ that characterize the edge orientations $\mathbf{n}$ for which  floppy modes are asymmetrically localized, see Fig.~\ref{fig:2Dpolarisation}. These polarization directions play a role in the continuum analogous to the role that the polarization lattice vector $\mathbf{R}_{\mathrm{T}}$, introduced in Ref.~\cite{kane_topological_2014}, plays for discrete lattices. Our key result is that systems with the discriminant $\Delta > 0$ are topologically polarized, whereas those with $\Delta < 0$ are not. These two cases correspond to the shear-dominant and dilation-dominant lattices studied in Ref.~\cite{rocklin_transformable_2017}, which showed that the Guest-Hutchinson mode (i.e., a zero mode with uniform strain) in a 2D topologically polarized Maxwell lattice is necessarily shear-dominant. Here we show that the converse is true in the continuum: the condition that two-dimensional elasticity has a shear-dominant Guest-Hutchinson mode is \emph{sufficient} for topological polarization.

To define the discriminant $\Delta$, we first consider the strip geometry. Consider an orthonormal basis $\{\mathbf{e}_1, \mathbf{e}_2\}$ and a strip with edges parallel to $\mathbf{e}_1$. To find the zero modes, we use the continuum analog of the compatibility matrix $\mathbf{P}_K^{\mathrm{T}} \mathbf{\hat{C}}(\mathbf{q})$, which depends only on the generalized elastic moduli, as explained in Sec.~\ref{sec:maxwell_condition}. We consider only systems with no bulk floppy modes, so there are no purely real non-zero solutions $\mathbf{q}$ to $\det \mathbf{P}_K^{\mathrm{T}} \mathbf{\hat{C}}(\mathbf{q}) = 0$. For a zero mode, the wavevector $\mathbf{q} = q_1 \mathbf{e}_1 + q_2 \mathbf{e}_2$ is given by $q_2 = F_j(q_1)$, where $j = 1, \ldots, 2 + n_w$ indexes the zero modes for a given value of $q_1$. Here, each $F_j$ is a function that satisfies $\det \mathbf{P}_K^{\mathrm{T}} \mathbf{\hat{C}}(q_1 \mathbf{e}_1 + F_j(q_1) \mathbf{e}_2) = 0$ for all $q_1$ on a sufficiently small neighborhood of zero.
To study the properties of these zero modes, we write
\begin{equation}
    P_m(\mathbf{q}) = P_m(q_1\mathbf{e}_1 + q_2 \mathbf{e}_2) = \sum_{j = 0}^m A_{j,m-j} q_1^j q_2^{m-j}
\end{equation}
in \eqref{eq:general_form_continuum_det_C}, where the polynomial coefficients $A_{j,k}$ are real. We consider the case $m = 2$, for which $P_2(\mathbf{q})$ is a quadratic form,
\begin{align}
\label{eq:quadratic_form}
    P_2 (q_1\mathbf{e}_1 + q_2 \mathbf{e}_2) &= \begin{bmatrix}
        q_1 \\ q_2
    \end{bmatrix}^{\mathrm{T}}
    \begin{bmatrix}
        A_{2,0} & \frac{1}{2} A_{1,1} \\
        \frac{1}{2} A_{1,1} & A_{0,2}
    \end{bmatrix}
    \begin{bmatrix}
        q_1 \\ q_2
    \end{bmatrix} \\
    &\equiv \begin{bmatrix}
        q_1 \\ q_2
    \end{bmatrix}^{\mathrm{T}}
    \mathbf{A}_2
    \begin{bmatrix}
        q_1 \\ q_2
    \end{bmatrix}, \nonumber
\end{align}
where we defined $\mathbf{A}_2$.
We then define the discriminant,
\begin{equation}
\label{eq:Delta_definition}
    \Delta = A_{1,1}^2 - 4 A_{2,0} A_{0,2} = -4 \det \mathbf{A}_2,
\end{equation}
which is independent of our choice of basis $\{\mathbf{e}_1, \mathbf{e}_2\}$, because under a change of basis, $\mathbf{A}_2$ transforms to $\mathbf{Q} \mathbf{A}_2 \mathbf{Q}^{\mathrm{T}}$ with orthogonal $\mathbf{Q}$, i.e., $\mathbf{Q} \mathbf{Q}^{\mathrm{T}} = \mathbf{I}$. The sign of $\Delta$ indicates the nature of the Guest-Hutchinson mode~\cite{guest_determinacy_2003, rocklin_transformable_2017, mao_maxwell_2018} associated with the elasticity theory. We show in Appendix~\ref{sec:guest-hutchinson} that $\Delta > 0$ and $\Delta < 0$ are equivalent to the Guest-Hutchinson mode being shear-dominant and dilation-dominant, respectively. The discriminant $\Delta$ characterizes topological polarization in the continuum, as we proceed to show.

As a first step, we define the \emph{soft directions} in the bulk. These directions correspond to solutions of $P_2(\mathbf{q}) = 0$, i.e., solutions of $\det \mathbf{P}_K^{\mathrm{T}} \mathbf{\hat{C}}(\mathbf{q}) = 0$ to lowest order in $q_1$ and $q_2$.
Because $P_2$ is a quadratic form, these solutions lie along lines in $\mathbf{q}$-space with two soft directions for the case $\Delta > 0$, as introduced in Ref.~\cite{saremi_topological_2020}. Without loss of generality, we assume that $q_1 = 0$ is not a soft direction, so that we have $A_{0,2} \neq 0$.

We now show that there are two continuum edge modes, which by definition satisfy $\lim_{q_1 \to 0} F_j(q_1) = 0$, i.e., \eqref{eq:def_continuum_edge_mode} in two dimensions. Setting $q_1 = 0$ in $\det \mathbf{P}_K^{\mathrm{T}} \mathbf{\hat{C}}(\mathbf{q}) = 0$ results in 
\begin{equation*}
    q_2^2 \left(A_{0,2} + \sum_{m = 3}^{2 + n_w} \!\! A_{0,m} q_2^{m-2}\right) = 0,
\end{equation*}
which has $q_2 = 0$ as a solution with multiplicity 2 when $A_{0,2} \neq 0$. Therefore, there are exactly two continuum edge modes when $A_{0,2} \neq 0$, and we denote these modes by $F_1$ and $F_2$. Using the theory of Puiseux series~\cite{knopp1996theory}, we show in Appendix~\ref{sec:analyticity} that $F_1$ and $F_2$ are analytic functions on a neighborhood of zero. Therefore, the continuum floppy modes can be studied using their power series expansions 
\begin{equation}
\label{eq:power_series}
    F_j(q_1) = \sum_{k = 0}^{\infty} \frac{1}{k!}F_j^{(k)}(0) q_1^k. 
\end{equation}
Setting $q_2 = F_j(q_1)$ in $\det \mathbf{P}_K^{\mathrm{T}} \mathbf{\hat{C}}(\mathbf{q}) = 0$ and differentiating with respect to $q_1$ twice shows that the first derivatives evaluated at zero, $F_j'(0)$, satisfy $P_2\!\left(1, F_j'(0)\right) = 0$.

Since $\Delta$ is the discriminant for the quadratic equation satisfied by $F_j'(0)$, for the case $\Delta < 0$, $F_1'(0)$ and $F_2'(0)$ are a complex conjugate pair. Setting $\beta_1 = \mathrm{Im}\,F_1(0) = - \mathrm{Im}\,F_2(0)$, we see from \eqref{eq:power_series} that the decay constants are $\mathrm{Im}\,F_1(q_1) = \beta_1 q_1 + \mathcal{O}(q_1^2)$ and $\mathrm{Im}\,F_2(q_1) = -\beta_1 q_1 + \mathcal{O}(q_1^2)$. The number of modes localized on each edge must remain the same as $q_1$ is varied, otherwise there would be bulk zero modes corresponding to $q_1$ values at which edge modes switch sides. Thus, the number of modes localized on each edge is completely determined by only the terms lowest order in $q_1$ present in $\mathrm{Im}\,F_j(q_1)$. Therefore, $\Delta < 0$ corresponds to the unpolarized case in which floppy modes are localized on opposite edges, because the decay constants have opposite signs. 

When $\Delta > 0$, $F_1'(0)$ and $F_2'(0)$ are both real. 
Below, we prove the result that the continuum is topologically polarized in this case. This proof can be summarized by the following steps:
\begin{enumerate}
    \item Polarization directions $\mathbf{p}_1$ and $\mathbf{p}_2$ are constructed as vectors orthogonal to the soft directions. We show that these polarization directions indicate the edge on which the continuum floppy modes are localized.
    \item Strips with edges normal to $\mathbf{n}$ have both floppy modes localized on the edge to which $\mathbf{n}$ points if and only if $\mathbf{n}$ makes an acute angle with both polarization directions. This is illustrated in Fig.~\ref{fig:2Dpolarisation}.
    \item Because $\Delta > 0$, the soft directions are distinct. Therefore, there always exist normals $\mathbf{n}$ such that the corresponding strips are topologically polarized (Fig.~\ref{fig:2Dpolarisation}).
\end{enumerate}

We represent the two soft directions $S_1$ and $S_2$ using the unit vectors 
\begin{equation}
    \label{eq:soft_direction_definition}
    \mathbf{e}_1^{S_j} = \frac{\mathbf{e}_1 + F_j'(0) \mathbf{e}_2}{|\mathbf{e}_1 + F_j'(0) \mathbf{e}_2|},
\end{equation}
for $j = 1,2$. Then, we construct two distinct orthonormal bases $\{\mathbf{e}_1^{S_1},\mathbf{e}_2^{S_1}\}$ and $\{\mathbf{e}_1^{S_2},\mathbf{e}_2^{S_2}\}$. Each polarization direction $\mathbf{p}_j$ is given by $\pm\mathbf{e}_2^{S_j}$, where the sign is chosen such that $\mathbf{p}_j$ points toward the soft edge. As we argued previously, the edges on which floppy modes are localized are indicated by the signs of the terms lowest order in $q_1$ present in $\mathrm{Im}\,F_j(q_1)$. When $\Delta > 0$, these terms are evaluated to be 
\begin{equation*}
    \mathrm{Im}\,F_j(q_1) = \frac{1}{2}\mathrm{Im}\,F''_j(0) q_1^2 + \mathcal{O}(q_1^3),
\end{equation*}
so we compute $F''_j(0)$. Setting $q_2 = F_j(q_1)$ in $\det \mathbf{P}_K^{\mathrm{T}} \mathbf{\hat{C}}(\mathbf{q}) = 0$ and differentiating with respect to $\tilde{q}_1$ three times leads to
\begin{equation}
\label{eq:F_double_prime}
    F_j''(0) = \frac{2(-1)^j i \sum_{k = 0}^3 A_{3-k,k} \left(F_j'(0)\right)^k}{3 \sqrt{\Delta}}.
\end{equation}
Because $F_j'(0)$ is real, $F_j''(0)$ is purely imaginary. 
We now show that each continuum floppy mode can be associated with a soft direction. The continuum floppy modes take the form
\begin{align}
\label{eq:zero_mode_wavevector}
    \mathbf{q} &= q_1\!\!\left(\mathbf{e}_1 + F_j'(0) \mathbf{e}_2\right) + i \frac{1}{2} \!\left(\mathrm{Im}\, F_j''(0)\right)\! q_1^2 \mathbf{e}_2 + \mathcal{O}(q_1^3)\! \\
    &= q_1 \sqrt{1 + \left(F_j'(0)\right)^2} \mathbf{e}_1^{S_j} + i \frac{1}{2}\!\left(\mathrm{Im} \,F_j''(0) \right) q_1^2 \mathbf{e}_2 + \mathcal{O}(q_1^3), \nonumber
\end{align}
using the expansion \eqref{eq:power_series} and the definition, \eqref{eq:soft_direction_definition}.
For sufficiently small $q_1$, each continuum edge floppy mode therefore has sinusoidal wavefronts normal to an associated soft direction $\mathbf{e}_1^{S_j}$. 

We identify the polarization direction $\mathbf{p}_j$ by considering the case $\mathbf{e}_1 = \mathbf{e}_1^{S_j}$, i.e., when the strip edges are parallel to a soft direction. In this basis, we adopt the notation $q_2 = F_{S_j}(q_1)$, where $F_{S_j}'(0) = 0$, so the wavevector of the associated floppy mode is $\mathbf{q} = q_1 \mathbf{e}_1^{S_j} + F_{S_j}(q_1) \mathbf{e}_2^{S_j}$. Using \eqref{eq:F_double_prime}, we find $F_{S_j}''(0) = -i \frac{2}{3} \frac{A_{3,0}}{A_{1,1}}$, where $A_{j,k}$ here are computed with respect to the basis $\{\mathbf{e}_1^{S_j}, \mathbf{e}_2^{S_j}\}$. For $j = 1,2$, we define the polarization directions
\begin{equation}
\label{eq:polarization_direction}
    \mathbf{p}_j = -\mathrm{sgn}(\mathrm{Im} \, F_{S_j}''(0)) \mathbf{e}_2^{S_j} = \mathrm{sgn}\left(\frac{A_{3,0}}{A_{1,1}}\right) \mathbf{e}_2^{S_j},
\end{equation}
where $\mathrm{sgn}(x)$ is the sign of $x$, so that $\mathbf{p}_j$ points toward the edge on which the floppy mode is localized. To derive these results, we assumed the generic case $A_{0,2} \neq 0$, which implies that $F_{S_j}$ is analytic. Requiring that $A_{0,2} \neq 0$ is also equivalent to requiring that the normal $\mathbf{e}_2^{S_j}$ not be parallel to a soft direction. Therefore, expression~(\ref{eq:polarization_direction}) is valid whenever the soft directions are not orthogonal. In Appendix~\ref{sec:orthogonal_soft_directions}, we generalize this procedure to include the special case of orthogonal soft directions, provided $1 \leq n_w \leq 2$. 

These polarization directions $\mathbf{p}_j$ determine the floppy mode localization for edges of any orientation not normal to a soft direction.
To show this, we consider a strip with edges parallel to $\mathbf{e}_1$ and normal to $\mathbf{e}_2$. Provided $\mathbf{e}_2 \cdot \mathbf{p}_j \neq 0$, the floppy mode associated with the soft direction $\mathbf{e}_1^{S_j}$ has wavevector
\begin{equation}
    \label{eq:continuum_zero_mode_soft_directions}
    \mathbf{q} = -\frac{\mathrm{sgn}(\mathrm{Im}\,F_{S_j}''(0))}{\mathbf{e}_2 \cdot \mathbf{p}_j} q_1 \mathbf{e}_1^{S_j} - i \frac{|\mathrm{Im}\,F_{S_j}''(0)|}{2(\mathbf{e}_2 \cdot \mathbf{p}_j)^3} q_1^2 \mathbf{e}_2 + \mathcal{O}(q_1^3),
\end{equation}
which we derive in Appendix~\ref{sec:zero_modes_polarization_directions}.
A version of \eqref{eq:continuum_zero_mode_soft_directions} has previously been derived as Eq.~(26) of Ref.~\cite{saremi_topological_2020} without using the polarization directions we introduce. Fig.~\ref{fig:2Dpolarisation} shows the importance of considering polarization directions in addition to soft directions. The two systems shown in Fig.~\ref{fig:2Dpolarisation}(a,b) have the same soft directions but have the opposite polarization vectors $\mathbf{p}_2$, leading to significantly different edge localizations.

The continuum edge mode associated with soft direction $\mathbf{e}_1^{S_j}$ is localized on the edge to which $\mathbf{e}_2$ points if and only if $\mathbf{e}_2$ makes an acute angle with $\mathbf{p}_j$. 
This is the case because $\mathbf{e}_2 \cdot \mathbf{p}_j > 0$ in \eqref{eq:continuum_zero_mode_soft_directions} is equivalent to $\mathrm{Im}(\mathbf{q} \cdot \mathbf{e}_2) < 0$, so the mode grows exponentially as $\exp(i \mathbf{q} \cdot \mathbf{x})$ in the $\mathbf{e}_2$-direction. Therefore, the polarization directions fully determine the distribution of localized floppy modes between the edges of any strip, provided the edges are not orthogonal to a soft direction. When the Guest-Hutchinson mode is shear-dominant (i.e., $\Delta > 0$), it is always possible to choose orientations for which the same edge hosts both floppy modes (i.e., to choose $\mathbf{e}_2$ that makes an acute angle with both polarization directions, see Fig.~\ref{fig:2Dpolarisation}). This reasoning enables us to compute the topological invariant \eqref{eq:edge_modes_invariant} in terms of the polarization directions as
\begin{equation}
    N_{+,2} - N_{-,2} = -\mathrm{sgn}(\mathbf{e}_2 \cdot \mathbf{p}_1) -\mathrm{sgn}(\mathbf{e}_2 \cdot \mathbf{p}_2).
\end{equation}

\begin{figure*}
    \centering \includegraphics{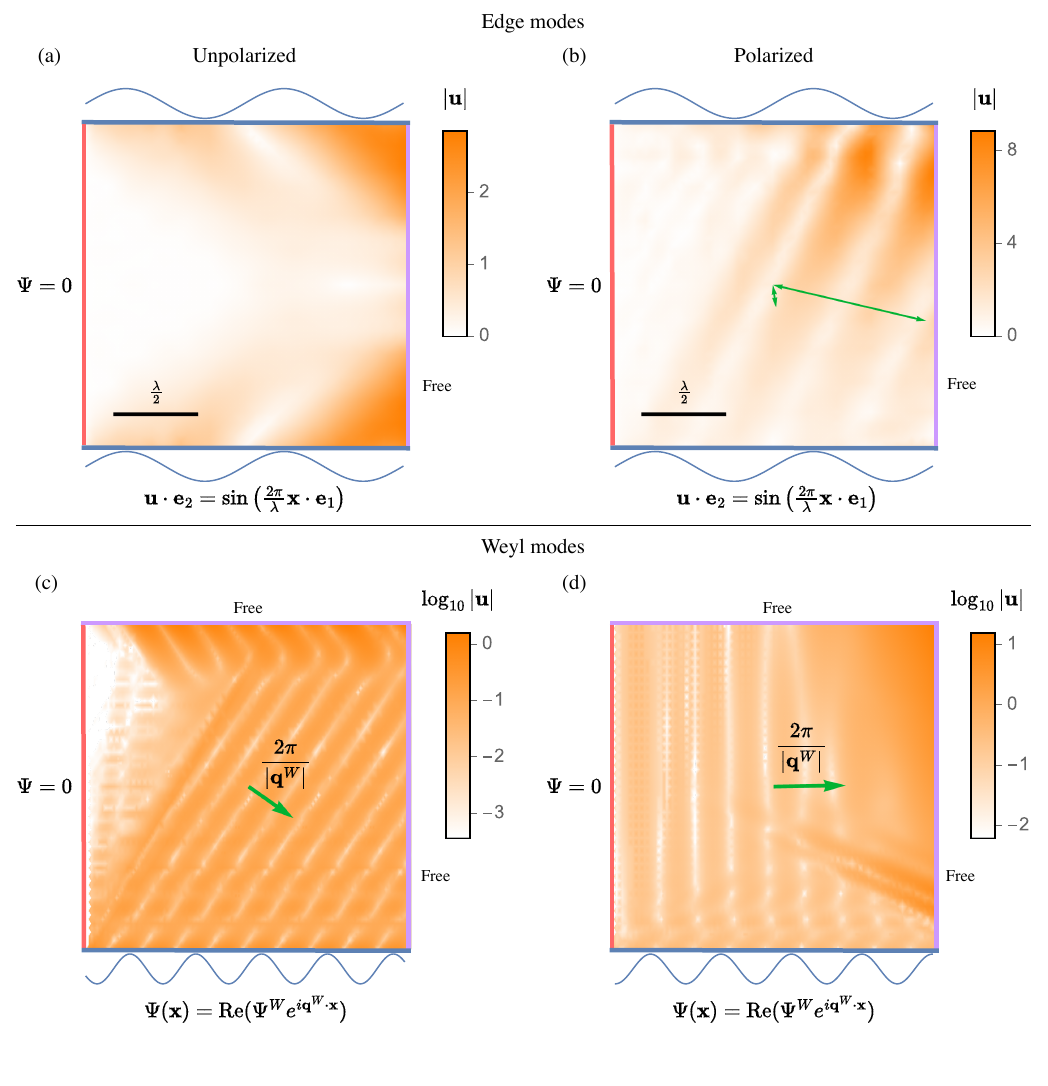}
    \caption{Efficient computation for floppy modes in the continuum. We find floppy modes by solving \eqref{eq:constitutive} with left-hand side stress measures set to zero, $(\mathbf{T}, \mathbf{c}_k, \eta_k) = (\mathbf{0},\mathbf{0},0)$. We specify either Dirichlet boundary conditions on the (generalized) displacements or free boundary conditions if no displacements are specified. Details of the boundary conditions and solution procedure are given in Appendix~\ref{sec:numerical}. (a--b) Edge modes are shown using the magnitude of the displacement field $|\mathbf{u}(\mathbf{x}, t)|$ for the continuum limits of distorted kagome lattices: (a) unpolarized and (b) polarized. In (a), the displacements decay into the bulk from both sides, whereas in (b), topological polarization leads to soft edge modes only on the top side. The length scale (shown by scale bars) in both cases is set by the prescribed displacements $\mathbf{u} \cdot \mathbf{e}_2 = \sin\left(\frac{2 \pi}{\lambda} \mathbf{x} \cdot \mathbf{e}_1 \right)$ on the two horizontal boundaries. The two green double-headed arrows in (b) are parallel to the soft directions, \eqref{eq:soft_direction_definition}, with length computed using \eqref{eq:continuum_zero_mode_soft_directions}. (c--d) Bulk Weyl modes are shown using the logarithm of the displacement field magnitude $\log_{10} |\mathbf{u}(\mathbf{x}, t)|$ for the continuum limits of two different geometrical configurations of distorted double kagome lattices. The wavevectors $\mathbf{q}^W$ corresponding to Weyl points are computed using \eqref{eq:weyl_coordinates}. The green arrows are parallel to $\mathbf{q}^W$, with length equal to the wavelength of the Weyl mode $2 \pi / |\mathbf{q}^W|$, see Appendix~\ref{sec:numerical}.}
    \label{fig:numerical_solutions}
\end{figure*}
We illustrate our analytical results by solving the partial differential equations, \eqref{eq:constitutive}, with the stress measures $\Sigma = 0$ to obtain continuum floppy modes. Provided the continuum Maxwell criterion \eqref{eq:maxwell_continuum} is satisfied, the system of equations~(\ref{eq:constitutive}) consists of $d + n_w$ independent equations for $d + n_w$ dependent variables. In  Fig.~\ref{fig:numerical_solutions}(a) and (b), we show the solutions with $n_w = 1$ for unpolarized and polarized cases, respectively, with details of the boundary conditions in Appendix~\ref{sec:numerical}. In the unpolarized case, we find a symmetric distribution of the edge modes. In contrast, the edge mode distribution is asymmetric in the polarized case, and we show using green arrows that the wavefronts are consistent with the approximation \eqref{eq:continuum_zero_mode_soft_directions}.

This completes our characterization of topological polarization for 2D continua, which is independent of any underlying lattice. We compute the soft and polarization directions from the generalized elastic moduli for any $n_w \geq 1$ via $A_{j,k}$.
Our characterization can be applied to discrete lattices with two edge floppy modes in the continuum, including the distorted kagome lattice~\cite{kane_topological_2014, rocklin_transformable_2017} at small distortions. As demonstrated by Ref.~\cite{rocklin_transformable_2017}, the distorted kagome lattice with shear-dominant Guest-Hutchinson modes can have polarization lattice vector $\mathbf{R}_{\mathrm{T}} = \mathbf{0}$. In Appendix~\ref{sec:polarization_discrete}, we show that this is consistent with our continuum results by demonstrating that the presence of a shear-dominant Guest-Hutchinson mode implies that we can always choose a set of primitive lattice vectors such that $\mathbf{R}_{\mathrm{T}} \neq \mathbf{0}$.

\subsection{Weyl zero modes}
\label{sec:weyl_modes}
\begin{figure*}
    \centering
    \includegraphics{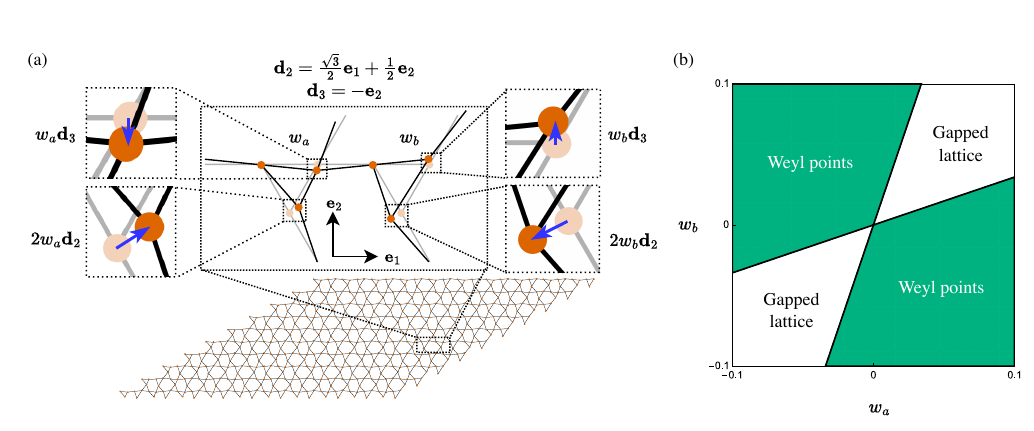}
    \caption{Constructing a lattice with continuum Weyl points. (a) The distorted double-kagome lattice unit cell is based on two kagome unit cells distorted according to the parametrization shown. The unit cell geometry is specified by the parameters $w_a$ and $w_b$, which determine the magnitudes of a geometrical perturbation, shown using blue arrows, for each of the two sub-cells. The case $w_a = w_b$ reduces to the distorted kagome lattice with three sites per primitive unit cell. (b) Phase diagram for the presence of Weyl points in the lattice in part (a), calculated by requiring the expression under the square root in \eqref{eq:weyl_coordinates} to be positive. The green regions correspond to values of $w_a, w_b$ for which Weyl points with $n_{\mathrm{Weyl}} = \pm 1$ can be observed in the continuum.}
    \label{fig:double_kagome_phase}
\end{figure*}
In the context of mechanical lattices of the type studied in Refs.~\cite{kane_topological_2014, rocklin_mechanical_2016}, Weyl points are bulk floppy modes that are topologically protected by a winding number. Although a distinct class of Weyl points has been studied in finite-frequency topological acoustics~\cite{shi_elastic_2019, ganti_weyl_2020, Xiao_Chen_He_Chan_2015}, no previous continuum theories for topological mechanics have considered zero-frequency Weyl modes. Here we show that Eqs.~(\ref{eq:continuum_eq}, \ref{eq:constitutive}) can harbor Weyl points provided that $n_w \geq 2$. For the simplest case $n_w = 2$, we derive an analytical expression for the existence and location of Weyl modes, which we use to construct the phase diagram in Fig.~\ref{fig:double_kagome_phase}.

First, we show that $n_w \geq 2$ is necessary for the existence of Weyl points. These bulk floppy modes necessarily occur at $\mathbf{q}$ values with real components that satisfy $\det \mathbf{P}_K^{\mathrm{T}} \mathbf{\hat{C}}(\mathbf{q}) = 0$. Therefore, the Weyl wavevectors $\mathbf{q}$ must satisfy two simultaneous constraints: 
\begin{equation}
    \mathrm{Re} \det \mathbf{P}_K^{\mathrm{T}} \mathbf{\hat{C}}(\mathbf{q}) = 0 \quad \textrm{ and } \quad \mathrm{Im} \det \mathbf{P}_K^{\mathrm{T}} \mathbf{\hat{C}}(\mathbf{q}) = 0.
    \label{eq:reimcond}
\end{equation}
For $n_w = 1$, these conditions applied to \eqref{eq:general_form_continuum_det_C} give
\begin{equation*}
        P_2(\mathbf{q}) = 0 \quad \textrm{ and } \quad
        P_3(\mathbf{q}) = 0
\end{equation*}
from the real and imaginary parts respectively. Since each $P_m(\mathbf{q})$ is a homogeneous polynomial with real coefficients, the solutions to each equation lie on lines passing through $\mathbf{q} = \mathbf{0}$. Weyl modes would only exist if these lines intersect at $\mathbf{q} \neq \mathbf{0}$, demonstrating that $n_w \geq 2$ is necessary for Weyl points in two dimensions.

We now consider the simplest case $n_w = 2$ and derive conditions that the coordinates $\left(q_1^W, q_2^W\right)$ of these bulk Weyl modes must satisfy. 
Setting the real and imaginary parts to zero, as in \eqref{eq:reimcond}, gives
\begin{subequations}
\label{eq:continuum_weyl_zero_re_im}
    \begin{align}
        -\frac{1}{2!}P_2(\mathbf{q}) + \frac{1}{4!} P_4(\mathbf{q}) &= 0, \label{eq:continuum_weyl_zero_re}\\
        P_3(\mathbf{q}) &= 0. \label{eq:continuum_weyl_zero_im}
    \end{align}
\end{subequations}
Because $P_3(\mathbf{q})$ is a homogeneous polynomial, the coordinates of the Weyl points are located on lines in $\mathbf{q}$-space. These lines can be represented as $q_2 = \kappa q_1$, except when the line is given by $q_1 = 0$ (i.e., the line is horizontal). Under the assumption $q_2 = \kappa q_1$, $\kappa$ satisfies 
\begin{equation}
\label{eq:P_3_weyl}
    P_3(\mathbf{e}_1 + \kappa \mathbf{e}_2) = \sum_{j = 0}^3 A_{j,3-j} \kappa^{3-j} = 0.
\end{equation}
If $A_{0,3} \neq 0$, \eqref{eq:P_3_weyl} is cubic and we obtain up to three real solutions for $\kappa$. We then substitute these $\kappa$ values into  $q_2 = \kappa q_1$ to use in \eqref{eq:continuum_weyl_zero_re}. We arrive at the result that if a bulk mode exists in the continuum, its $\mathbf{q}^W\!$-coordinates $\left(q_1^W, q_2^W\right)$ must satisfy
\begin{equation}
\label{eq:weyl_coordinates}
    q_1^W = \pm \sqrt{12\frac{P_2(\mathbf{e}_1 + \kappa \mathbf{e}_2)}{P_4(\mathbf{e}_1 + \kappa \mathbf{e}_2)}}, \quad q_2^W = \kappa q_1^W.
\end{equation}

The case of a horizontal line $q_1 = 0$ that satisfies \eqref{eq:continuum_weyl_zero_im} needs to be treated separately.
This solution only occurs if $A_{0,3} = 0$. In this case, there are still three cubic solutions, where one of the solutions corresponds to $q_1 = 0$ and has the form:
\begin{equation}
\label{eq:weyl_coordinates_q1_zero}
    q_1^W = 0, \quad q_2^W = \pm \sqrt{12\frac{P_2(\mathbf{e}_2)}{P_4(\mathbf{e}_2)}} = \pm \sqrt{12\frac{A_{0,2}}{A_{0,4}}}.
\end{equation}
In addition, there are two solutions given by \eqref{eq:weyl_coordinates}, corresponding to lines $q_2 = \kappa q_1$. For these two lines, $\kappa$ now satisfies the quadratic equation $ A_{1,2} \kappa^{2} + A_{2,1} \kappa + A_{3,0} = 0$.
Any bulk floppy mode has to satisfy either \eqref{eq:weyl_coordinates} or \eqref{eq:weyl_coordinates_q1_zero}, and both $q_1^W$ and $q_2^W$ must be real.

These bulk floppy modes are only Weyl points if they satisfy the extra condition that the topological invariant is non-zero for an enclosing contour integral in $\mathbf{q}$-space. To compute this winding number, we use \eqref{eq:continuum_topological_invariants} re-expressed as
\begin{equation}
\label{eq:weyl_winding_number}
    n_{\mathrm{Weyl}} = \frac{1}{2 \pi i} \int_{\tau_0}^{\tau_1} \frac{\frac{\mathrm{d}}{\mathrm{d} \tau} \det \mathbf{P}_K^{\mathrm{T}} \mathbf{\hat{C}}(\mathbf{q}(\tau))}{\det \mathbf{P}_K^{\mathrm{T}} \mathbf{\hat{C}}(\mathbf{q}(\tau))} \, \mathrm{d}\tau,
\end{equation}
where $\mathbf{q}(\tau)$ is a parametrization of a closed curve in real $\mathbf{q}$-space that encloses the point  $\left(q_1^W, q_2^W\right)$. We find Weyl points by finding bulk floppy modes with the topologically non-trivial value $n_{\mathrm{Weyl}} \neq 0$. 

Figure~\ref{fig:classification}(e)III illustrates the physical consequences of a Weyl mode in the bulk spectrum.
The blue mode for $q_1 < q_1^W$ is on the bottom of the material sample. 
When $q_1 = q_1^W$, we find that this blue mode has a diverging penetration depth, corresponding to a bulk mode. This phenomenon is in contrast to the topologically gapped case, where no bulk floppy modes exist. For the case $q_1 > q_1^W$, the blue mode again becomes an edge mode, but is now localized on the top edge of the material sample. Unlike a topologically polarized material, a Weyl material has edge modes that switch sides as the wavenumber is varied.
Significantly, unlike the Weyl winding number in Ref.~\cite{rocklin_mechanical_2016}, the continuum definition of $n_{\mathrm{Weyl}}$ in \eqref{eq:weyl_winding_number} does not rely on the existence of a Brillouin zone.

To construct a lattice with Weyl points in the continuum,
we start with a kagome lattice supercell consisting of two kagome unit cells, as shown in Fig.~\ref{fig:double_kagome_phase}(a).
These two kagome unit cells have a total of two local floppy modes at $\mathbf{q} = \mathbf{0}$ (i.e., one mode for each kagome unit cell).
We then apply a perturbation to each of the cells, parametrized by $w_a$ and $w_b$ as shown in Fig.~\ref{fig:double_kagome_phase}(a), which results in what we term the distorted double-kagome (DDK) lattice.
For this lattice, $n_w = 2$, and we now show that the perturbed lattice harbors Weyl modes.
In Figure~\ref{fig:double_kagome_phase}(b), we present a phase diagram of the DDK lattice for the parameters $w_a$ and $w_b$. The green regions correspond to lattice geometries that exhibit Weyl points in the continuum. We plot this phase diagram by finding real non-zero solutions of the form  \eqref{eq:weyl_coordinates}, in combination with winding number $n_{\mathrm{Weyl}} = \pm 1$ computed using \eqref{eq:weyl_winding_number}. 

Unlike lattice-based approaches, our continuum theory exhibits Weyl phenomenology in the continuum solutions to the partial differential equations \eqref{eq:constitutive}. We find these solutions numerically with left-hand side stress measures set to zero, $(\mathbf{T}, \mathbf{c}_k, \eta_k) = (\mathbf{0},\mathbf{0},0)$, and plot the results in Fig.~\ref{fig:numerical_solutions}(c,d), with details of the boundary conditions given in Appendix~\ref{sec:numerical}. The bulk floppy modes associated with the Weyl points are visible as wavefronts that span the bulk of the sample, with direction and wavelength matching the Weyl wavevectors computed using \eqref{eq:weyl_coordinates}.

In summary, we have shown two advantages to our continuum approach to Weyl modes: the analytically obtained Weyl phase diagram in Fig.~\ref{fig:double_kagome_phase}, and efficient solutions for Weyl phenomenology in a finite material sample in Fig.~\ref{fig:numerical_solutions}(c,d).

\section{Conclusions and outlook}
We have presented the general form of a topological continuum theory in Eqs.~(\ref{eq:continuum_eq}, \ref{eq:constitutive}). Our theory of linear elasticity augmented with additional fields captures both topological polarization and Weyl modes independent of any microscopic detail. We showed that these phenomena can be classified in two-dimensional elasticity using the number of additional fields needed to capture them: topological edge states require at least one additional field, and Weyl points require at least two. Like other studies of topological physics in the continuum~\cite{silveirinha_chern_2015, silveirinha_bulk-edge_2016, silveirinha_proof_2019, delplace_topological_2017, souslov_topological_2019, scheibner_non-hermitian_2020, bal_topological_2022, fossati_odd_2024}, we do not rely on a compact Brillouin zone. 
However, unlike previous definitions of topological invariants in the continuum, which rely on either 
compactification of reciprocal space~\cite{silveirinha_chern_2015, delplace_topological_2017, souslov_topological_2019, bal_topological_2022, fossati_odd_2024} or non-Hermiticity~\cite{scheibner_non-hermitian_2020}, we define the invariants \eqref{eq:continuum_topological_invariants} for a Hermitian system
using an effective compatibility matrix.

We bridge our continuum formulation and discrete lattice-based realizations of topological mechanics using a systematic homogenization procedure. This procedure shows how our theory arises naturally from the continuum limit of a generic ball-and-spring lattice. We interpret the additional fields in our theory as local soft modes of an underlying mechanical lattice in \eqref{eq:additional_fields}. In this way, our homogenization procedure provides design principles for topological metamaterials by identifying the minimal number of local soft modes necessary for a desired behavior. Our formulation based on partial differential equations also provides computationally efficient tools for studying topological floppy modes, instead of relying on the full lattice structure.

In practice, 3D-printed metamaterials are not Maxwell lattices due to effects from bending stiffness at hinges~\cite{bilal_intrinsically_2017, bergne_scalable_2022, tang_fully_2024} to nonlinearities~\cite{chen_nonlinear_2014, upadhyaya_nuts_2020}. Nevertheless, asymmetric floppy mode localization can occur when the energy scale for hinge bending is much lower than for bond stretching~\cite{bilal_intrinsically_2017, bergne_scalable_2022, tang_fully_2024}. We envision future generalizations of our continuum theory that incorporate these additional energy scales, non-rectilinear constraints~\cite{lera_topological_2018}, as well as active and non-Hermitian effects~\cite{scheibner_odd_2020}.

\begin{acknowledgments}
I.T.~gratefully acknowledges support from the Cambridge Commonwealth, European \& International Trust.
\end{acknowledgments}

\appendix
\section{Determinant of compatibility matrix for $\varepsilon$-perturbed lattice}
\label{sec:detC_perturbed_derivation}
We derive \eqref{eq:detCperturbed}, showing that \emph{(i)} the terms lowest order in $\delta$ present in $\det \mathbf{C}_p(\mathbf{q}; \varepsilon)$ are of order $d+n_w$ in $\delta$ and that \emph{(ii)} these terms are of order at least $d$ in $\bar{q}_r$. We begin with \eqref{eq:perturbed_compatibility_matrix} and the statements about the null spaces of $\mathbf{C}$ and $\mathbf{C} + \varepsilon \mathbf{C}_w$ that immediately follow \eqref{eq:perturbed_compatibility_matrix}. Let $\mathbf{\bar{C}}(\mathbf{q}) = \mathbf{C}(\mathbf{q}) - \mathbf{C}$ and $\mathbf{\bar{C}}_w(\mathbf{q}) = \mathbf{C}_w(\mathbf{q}) - \mathbf{C}_w$. Since the elements of $\mathbf{C}(\mathbf{q})$ and $\mathbf{C}_w(\mathbf{q})$ depend on $\mathbf{q}$ via terms proportional to $e^{i \mathbf{n} \cdot \mathbf{q}}$ (c.f. \eqref{eq:detCexpansion}), all the non-zero elements of $\mathbf{\bar{C}}(\mathbf{q})$ and $\mathbf{\bar{C}}_w(\mathbf{q})$ depend on $\mathbf{q}$ via terms proportional to $e^{i \mathbf{n} \cdot \mathbf{q}} - 1$. Let the $k$th rows of $\mathbf{C}$, $\mathbf{\bar{C}}(\mathbf{q})$, $\mathbf{C}_w$ and $\mathbf{\bar{C}}_w (\mathbf{q})$ be denoted $c_k$, $\bar{c}_k(\mathbf{q})$, $w_k$ and $\bar{w}_k(\mathbf{q})$ respectively. These vectors are not written in bold font to distinguish them from $d$-dimensional vectors such as $\mathbf{u}$. Each of the vectors with an overbar $\bar{\cdot}$ is proportional to $(\mathbf{n} \cdot \mathbf{q})$ for some lattice vector $\mathbf{n}$, from expanding $e^{i \mathbf{n} \cdot \mathbf{q}} - 1$, so these vectors are $\mathcal{O}(\delta)$.

We write the determinant of an $m \times m$ matrix $\mathbf{A}$ as $\det \mathbf{A} = \mathcal{M}(a_1, \ldots, a_m)$, where $a_1, \ldots, a_m$ are the rows of $\mathbf{A}$ and $\mathcal{M}$ is an alternating, multilinear map from $m$-tuples of vectors to $\mathbb{C}$ that maps the rows of the $m \times m$ identity matrix to 1. This representation of the determinant is detailed in Ref.~\cite{rosen_geometric_2019}. The determinant of the perturbed compatibility matrix is then
\begin{equation}
\label{eq:multilinear_determinant}
    \det \mathbf{C}_p(\mathbf{q}; \varepsilon) = \mathcal{M}(c_1(\mathbf{q}; \varepsilon),\ldots, c_{n_b}(\mathbf{q}; \varepsilon)),
\end{equation}
where the rows of $\mathbf{C}_p(\mathbf{q}; \varepsilon)$ are
\begin{displaymath}
    c_k(\mathbf{q}; \varepsilon) = c_k + \varepsilon w_k + \bar{c}_k(\mathbf{q}) + \varepsilon \bar{w}_k(\mathbf{q})
\end{displaymath}
for $k = 1, \ldots, n_b$. Because $\mathcal{M}$ is multilinear, we can expand the right hand side of \eqref{eq:multilinear_determinant} so that it equals a sum involving terms such as $\mathcal{M}(c_1, \varepsilon w_2, \ldots, \bar{c}_{n_b}(\mathbf{q}))$. The row space of $\mathbf{C}$ has dimension $n_b - (d + n_w)$. Thus, there exists a choice of $n_b - (d + n_w)$ rows $c_k$ from $\mathbf{C}$ that form a basis for its row space~\cite{axler_linear_2024},.  Because all the other $c_k$ are linear combinations of those basis vectors, each non-zero term in the expansion must have at least $d + n_w$ terms involving $\varepsilon$ or $e^{i \mathbf{n} \cdot \mathbf{q}} - 1$, so the terms of lowest order in $\delta$ are $\mathcal{O}(\delta^{d + n_w})$. This is result \emph{(i)}. 

To establish result \emph{(ii)}, we note that because the row space of $\mathbf{C} + \varepsilon \mathbf{C}_w$ has dimension $n_b - d$, there exists a choice of $n_b - d$ rows $c_k + \varepsilon w_k$ that form a basis for this row space. Because all the other $c_k + \varepsilon w_k$ are linear combinations of those basis vectors, each non-zero term in the expansion of the right hand side of \eqref{eq:multilinear_determinant} when keeping terms of the form $c_k + \varepsilon w_k$ together must have at least $d$ terms involving $e^{i \mathbf{n} \cdot \mathbf{q}} - 1$, so the terms of lowest order in $\bar{q}_r$ have order $d$. This is result \emph{(ii)}.

These results enable us to write down \eqref{eq:detCperturbed} as the general form taken by $\det \mathbf{C}_p(\mathbf{q}; \varepsilon)$. The terms lowest order in $\delta$ are obtained as products of polynomials of degree $m$ in $\bar{q}_r$ and powers of $\varepsilon$, where the order of each term in $\delta$ equals $d + n_w$. Therefore, $\bar{P}_m(\bar{q}_1, \ldots, \bar{q}_d)$ multiplies $\varepsilon^{d + n_w - m}$. The sum over $m$ ranges from $d$ to $d + n_w$ because each term has order in $\bar{q}_r$ of at least $d$ (accounting for the lower limit of the sum) and $\varepsilon$ is never raised to a negative power (accounting for the upper limit). 

\section{Details of the homogenization procedure}
\subsection{Projecting onto subspaces}
\label{sec:projecting_subspaces}
We begin by projecting the equations of motion for the $n_s d$ degrees of freedom per lattice unit cell \eqref{eq:full_eqs_motion} onto each of the three subspaces defined by the eigendecomposition of $\mathbf{C}^{\mathrm{T}}\mathbf{K}\mathbf{C}$ in Sec.~\ref{sec:displacement_and_other}:
\begin{widetext}

    \begin{subequations}
	\label{eq:projections_eqs_motion_perturbed}
	\begin{align}
		a^2 \mathbf{P}_u (\mathbf{M} \partial_t^2 \Phi - F) &= \mathbf{P}_u \left(\mathbf{C}_{\mathbf{x}}^{\mathrm{T}} \mathbf{K} (\mathbf{C} + \varepsilon \mathbf{C}_w + \mathbf{C}_{\mathbf{x}}) + \left(\varepsilon \mathbf{C}_{\mathbf{x},w} + \mathcal{O}(\varepsilon^2)\right)^{\mathrm{T}} \mathbf{K} \left(\mathbf{C} + \varepsilon \mathbf{C}_w + \mathbf{C}_{\mathbf{x}} + \varepsilon \mathbf{C}_{\mathbf{x},w} + \mathcal{O}(\varepsilon^2)\right) \right) \Phi \nonumber \\
        & \quad - \mathbf{P}_u \mathbf{C}_{\mathbf{x}}^{\mathrm{T}} \mathbf{K} (\varepsilon \mathbf{C}_{\mathbf{x},w} + \mathcal{O}(\varepsilon^2)) \Phi \label{eq:eq_motion_projection_Px} \\
		a^2 \mathbf{P}_w (\mathbf{M} \partial_t^2 \Phi - F) &= \mathbf{P}_w \left(\mathbf{C}_{\mathbf{x}}^{\mathrm{T}} \mathbf{K} (\mathbf{C} + \varepsilon \mathbf{C}_w + \mathbf{C}_{\mathbf{x}}) + \left(\varepsilon \mathbf{C}_{\mathbf{x},w} + \mathcal{O}(\varepsilon^2)\right)^{\mathrm{T}} \mathbf{K} \left(\mathbf{C} + \varepsilon \mathbf{C}_w + \mathbf{C}_{\mathbf{x}} + \varepsilon \mathbf{C}_{\mathbf{x},w} + \mathcal{O}(\varepsilon^2) \right) \right) \! \Phi \nonumber \\
		& \quad + \mathbf{P}_w \left( \mathbf{C}_{\mathbf{x}}^{\mathrm{T}} \mathbf{K} \left(\varepsilon \mathbf{C}_{\mathbf{x},w} + \mathcal{O}(\varepsilon^2)\right) - \varepsilon\mathbf{C}_w^{\mathrm{T}} \mathbf{K} (\mathbf{C} + \varepsilon \mathbf{C}_w + \mathbf{C}_{\mathbf{x}}) -\varepsilon \mathbf{C}_w^{\mathrm{T}} \mathbf{K} \left(\varepsilon \mathbf{C}_{\mathbf{x},w} + \mathcal{O}(\varepsilon^2)\right) \right)\! \Phi \label{eq:eq_motion_projection_Pw} \\
		a^2 \mathbf{P}_v (\mathbf{M} \partial_t^2 \Phi - F) &= - \mathbf{P}_v (\mathbf{C} + \varepsilon \mathbf{C}_w)^{\mathrm{T}}\mathbf{K} (\mathbf{C} + \varepsilon \mathbf{C}_w + \mathbf{C}_{\mathbf{x}}) \Phi - \mathbf{P}_v (\mathbf{C} + \varepsilon \mathbf{C}_w)^{\mathrm{T}}\mathbf{K} (\varepsilon \mathbf{C}_{\mathbf{x},w} + \mathcal{O}(\varepsilon^2)) \Phi \nonumber \\ 
		& \qquad + \mathbf{P}_v \mathbf{C}_{\mathbf{x}}^{\mathrm{T}} \mathbf{K} (\mathbf{C} + \varepsilon \mathbf{C}_w + \mathbf{C}_{\mathbf{x}}) \Phi + \mathbf{P}_v \mathbf{C}_{\mathbf{x}}^{\mathrm{T}} \mathbf{K} (\varepsilon \mathbf{C}_{\mathbf{x},w} + \mathcal{O}(\varepsilon^2)) \Phi \nonumber \\
		& \qquad + \mathbf{P}_v (\varepsilon \mathbf{C}_{\mathbf{x},w} + \mathcal{O}(\varepsilon^2))^{\mathrm{T}} \mathbf{K} (\mathbf{C} + \varepsilon \mathbf{C}_w + \mathbf{C}_{\mathbf{x}} + \varepsilon \mathbf{C}_{\mathbf{x},w} + \mathcal{O}(\varepsilon^2)) \Phi, \label{eq:eq_motion_projection_Pv}
	\end{align}
\end{subequations}
\end{widetext}
where we drop the explicit dependence on $(\mathbf{x}, t)$ for brevity, the right hand side of \eqref{eq:full_eqs_motion} has been expanded and we used the properties of the three classes of eigenvector shown in Fig.~\ref{fig:eigendecomposition}. Using $\varepsilon \sim \mathcal{O}(\delta)$ and $a = L \delta$ in \eqref{eq:eq_motion_projection_Pv}, we obtain \eqref{eq:pre-pseudo-inversion_Phi_v} by neglecting terms $\mathcal{O}(\delta^2)$ and above. To see this, we first neglect terms that are clearly $\mathcal{O}(\delta^2)$ and above and are left with
\begin{align*}
     &\mathbf{P}_v (\mathbf{C} + \varepsilon \mathbf{C}_w)^{\mathrm{T}}\mathbf{K} (\mathbf{C} + \varepsilon \mathbf{C}_w + \mathbf{C}_{\mathbf{x}}) \Phi \\
     &\qquad \approx \mathbf{P}_v \mathbf{C}_{\mathbf{x}}^{\mathrm{T}} \mathbf{K} (\mathbf{C} + \varepsilon \mathbf{C}_w + \mathbf{C}_{\mathbf{x}}) \Phi,
\end{align*}
which simplifies to
\begin{multline}
\label{eq:pre_pre_pseudo_inversion_Phi_v}
     \mathbf{P}_v (\mathbf{C} + \varepsilon \mathbf{C}_w)^{\mathrm{T}} \mathbf{K} \mathbf{C} \Phi_v \\ 
     + \mathbf{P}_v (\mathbf{C} + \varepsilon \mathbf{C}_w)^{\mathrm{T}} \mathbf{K}(\varepsilon \mathbf{C}_w + \mathbf{C}_{\mathbf{x}}) \Phi \\
     \approx \mathbf{P}_v \mathbf{C}_{\mathbf{x}}^{\mathrm{T}} \mathbf{K} (\mathbf{C} + \varepsilon \mathbf{C}_w + \mathbf{C}_{\mathbf{x}}) \Phi.
\end{multline}
We establish that $\Phi_v \sim \mathcal{O}(\delta)$ by noting that $\mathbf{P}_v (\mathbf{C} + \varepsilon \mathbf{C}_w)^{\mathrm{T}} \mathbf{K} (\varepsilon \mathbf{C}_w + \mathbf{C}_{\mathbf{x}}) \Phi$ and $\mathbf{P}_v \mathbf{C}_{\mathbf{x}}^{\mathrm{T}} \mathbf{K} (\mathbf{C} + \varepsilon \mathbf{C}_w + \mathbf{C}_{\mathbf{x}}) \Phi$ are at least $\mathcal{O}(\delta)$, so the first term on the left hand side of the above equation is also $\mathcal{O}(\delta)$. Therefore, $\mathbf{P}_v \mathbf{C}^{\mathrm{T}} \mathbf{K}\mathbf{C} \Phi_v \sim \mathcal{O}(\delta)$, and since $\mathbf{P}_v \mathbf{C}^{\mathrm{T}} \mathbf{K}\mathbf{C}$ is invertible on the range space of $\mathbf{P}_v$ by construction, we conclude that $\Phi_v \sim \mathcal{O}(\delta)$. Returning to \eqref{eq:pre_pre_pseudo_inversion_Phi_v}, we can discard more terms now revealed to be $\mathcal{O}(\delta^2)$ to obtain \eqref{eq:pre-pseudo-inversion_Phi_v}.

We then substitute \eqref{eq:pre-pseudo-inversion_Phi_v} into \eqref{eq:eq_motion_projection_Px} and \eqref{eq:eq_motion_projection_Pw} and neglect $\mathcal{O}(\delta^3)$ terms (terms of lowest order are $\mathcal{O}(\delta^2)$) to obtain, after algebraic manipulation, \eqsref{eq:continuum_eqs_preliminary} in the main text. We note that the null space of $\mathbf{\tilde{K}}$ is the orthogonal complement of the null space of $\mathbf{C}^{\mathrm{T}}$ (proven in Appendix~\ref{sec:symmetries}), so the action of $\mathbf{\tilde{K}}$ involves projecting onto unit-cell periodic ($\mathbf{q} = \mathbf{0}$) states of self-stress of the geometrically unperturbed lattice.

\subsection{Generalized elastic moduli and other terms in the continuum equations of motion}
\label{sec:gen_elastic_moduli}
Here we express the generalized elastic moduli, inertial constants and body force/torque density in terms of microscopic quantities. By substituting these expressions into the continuum equations 
(\ref{eq:continuum_eq}--\ref{eq:constitutive}), we show that the main-text Eqs.~(\ref{eq:continuum_eq}--\ref{eq:constitutive}) are equivalent to \eqsref{eq:continuum_eqs_preliminary}.

To express the generalized elastic moduli concisely, we first define some auxiliary linear maps. $H$ is a linear map from the space of second rank tensors $\mathbf{A}$ to the space of bond extensions $\Xi = [e_1 \ldots e_{n_b}]^{\mathrm{T}}$, and $\{G_m\}_{m=1}^{n_w}$ are linear maps from vectors $\mathbf{c}$ to bond extensions $\Xi$. These linear maps $H$ and $G_m$ ($m = 1, \ldots, n_w$) are given by
\begin{subequations}
\label{eq:H_G_m}
    \begin{align}
		H[\mathbf{A}] &= \sum_{r = 1}^d (-i) (\partial_{\bar{q}_r} \mathbf{C}(\mathbf{q})|_{\mathbf{q} = \mathbf{0}}) D [\mathbf{A} \cdot a^{-1}  \mathbf{a}_r] \\
		G_m[\mathbf{c}] &=  \sum_{r = 1}^d (-i) (\partial_{\bar{q}_r} \mathbf{C}(\mathbf{q})|_{\mathbf{q} = \mathbf{0}})w_m(a^{-1} \mathbf{a}_r \cdot \mathbf{c} ),
	\end{align}
\end{subequations}
where $D$ is the linear map from vectors $\mathbf{v}$ to site displacements satisfying $D[\mathbf{e}_2] = \sqrt{n_s} u_n$ for $n = 1, \ldots, d$, recalling that $\{\mathbf{e}_j\}_{j = 1}^d$ are an orthonormal basis, and $\{u_n\}_{n = 1}^d$ are the unit cell site displacements $\Phi$ for uniform translations along $\mathbf{e}_j$. Thus, we can regard $D$ as mapping displacement vectors $\mathbf{u}$ to unit cell site displacements corresponding to uniform translations by $\mathbf{u}$.
These linear maps are used to express the elastic tensors, which satisfy
\begin{subequations}
	\label{eq:elastic_tensors}
	\begin{align}
		\mathsf{C}:\mathbf{A} &= H^{\mathrm{T}}[\mathbf{\tilde{K}}H[\mathbf{A}]] \\
		\mathfrak{B}_m \cdot \mathbf{c} &= H^{\mathrm{T}} \left[\mathbf{\tilde{K}}G_m[\mathbf{c}]\right] \\
		\mathbf{N}_m &= H^{\mathrm{T}} \left[a^{-1} \varepsilon \mathbf{\tilde{K}} \mathbf{C}_w w_m \right] \\
		\mathbf{M}_{km} \cdot \mathbf{c} & = G_k^{\mathrm{T}}[\mathbf{\tilde{K}} G_m[\mathbf{c}]] \\
		\mathbf{h}_{km} &= G_k^{\mathrm{T}}[\mathbf{\tilde{K}} a^{-1} \varepsilon \mathbf{C}_w w_m] \\
		J_{km} &= (a^{-1} \varepsilon)^2 w_k^{\mathrm{T}} \mathbf{C}_w^{\mathrm{T}} \mathbf{\tilde{K}}\mathbf{C}_w w_m.
	\end{align}
\end{subequations}
for all second rank tensors $\mathbf{A}$ and vectors $\mathbf{c}$, and where $\mathbf{\tilde{K}}$ is given in \eqref{eq:def_tilde_K}. Since $|\mathbf{a}_r| \sim \mathcal{O}(a)$ and $\varepsilon \sim \mathcal{O}(\delta) \sim \mathcal{O}(a) L^{-1}$, the expressions $a^{-1} \mathbf{a}_r$ and $a^{-1} \varepsilon$ are independent of $a$.

The inertial terms associated with the second time derivatives of the fields $\mathbf{u}(\mathbf{x},t)$ and $\{\varphi_m(\mathbf{x},t)\}_{m=1}^{n_w}$ contain the following constants for $m, k = 1, \mathellipsis, n_w$:
\begin{equation}
	\begin{aligned}
		\rho &= \sum_{h = 1}^{n_s} M_h \\
		\mathbf{p}_k &= D^{\mathrm{T}}[\mathbf{P}_u \mathbf{M} w_k] \\
		\mu_{mk} &= w_k^{\mathrm{T}}\mathbf{M}w_m,
	\end{aligned}
\end{equation}
where $M_h$ is the mass density associated with the $h$th site. We note that $\mu_{km} = \mu_{mk}$. For the special case of all masses being equal in the unit cell, $\mathbf{M} = \mu \mathbf{I}$, $\mathbf{p}_k = \mathbf{0}$ for all $k = 1, \mathellipsis, n_w$ and $\mu_{km} = 0$ for $k \neq m$.

The body force/torque densities derived from the external forces $\mathbf{f}_h(\mathbf{x},t)$ applied to lattice sites are given by
\begin{equation}
	\begin{aligned}
		\mathbf{f}(\mathbf{x},t) &= \sum_{h = 1}^{n_s} \mathbf{f}_h(\mathbf{x},t) \\
		\tau_k(\mathbf{x},t) &= w_k^{\mathrm{T}} F(\mathbf{x},t) 
	\end{aligned}
\end{equation}
for $k = 1, \mathellipsis, n_w$, which concludes the mapping between Eqs.~(\ref{eq:continuum_eq}--\ref{eq:constitutive}) and \eqref{eq:continuum_eqs_preliminary}.

\subsection{Symmetries of the generalized elastic moduli}
\label{sec:symmetries}
We derive symmetries satisfied by the components of the generalized elastic moduli from \eqref{eq:elastic_tensors} and from the result $\mathbf{\tilde{K}} H[\mathbf{W}] = 0$ for all antisymmetric second rank tensors $\mathbf{W}$, which we proceed to derive.

First, we show that the null space of $\mathbf{\tilde{K}}$ is the range space of $\mathbf{C}$. Let $\mathbf{E} = \mathbf{K}^{1/2} \mathbf{C}$, so that $\mathbf{\tilde{K}} = \mathbf{K}^{1/2}(\mathbf{I} - \mathbf{E} \mathbf{P}_v \mathbf{B} \mathbf{P}_v \mathbf{E}^{\mathrm{T}} )\mathbf{K}^{1/2}$. For any $v$ in the range space of $\mathbf{C}$, there exists $\Phi$ such that $v = \mathbf{C} \Phi = \mathbf{C} \mathbf{P}_v \Phi$, where the last equality holds because $\mathbf{P}_v$ is the projection onto the orthogonal complement of the null space of $\mathbf{C}$. Then, for all $v$ in the range space of $\mathbf{C}$,
	\begin{align*}
		\mathbf{\tilde{K}} v &= \mathbf{K}^{1/2}(\mathbf{I} - \mathbf{E} \mathbf{P}_v \mathbf{B} \mathbf{P}_v \mathbf{E}^{\mathrm{T}} )\mathbf{K}^{1/2} \mathbf{C} \mathbf{P}_v \Phi \\
		& = \mathbf{K}^{1/2}(\mathbf{I} - \mathbf{E} \mathbf{P}_v \mathbf{B} \mathbf{P}_v \mathbf{E}^{\mathrm{T}} ) \mathbf{E} \mathbf{P}_v \Phi \\
		& = \mathbf{K}^{1/2}(\mathbf{E} \mathbf{P}_v - \mathbf{E} \mathbf{P}_v \mathbf{B} \mathbf{P}_v \mathbf{E}^{\mathrm{T}}\mathbf{E} \mathbf{P}_v) \Phi \\
		& = \mathbf{K}^{1/2}(\mathbf{E} \mathbf{P}_v - \mathbf{E} \mathbf{P}_v) \Phi = 0,
	\end{align*}
where we have used the definition of $\mathbf{B}$ in \eqref{eq:def_B} as the pseudoinverse of $\mathbf{E}^\mathrm{T} \mathbf{E}$. Thus, $\mathrm{range} \, \mathbf{C} \subseteq \mathrm{null} \, \mathbf{\tilde{K}}$.
Suppose now that $u \in \mathrm{null} \, \mathbf{\tilde{K}}$. Then,
\begin{displaymath}
	\mathbf{K}( u - \mathbf{C} \mathbf{P}_v \mathbf{B} \mathbf{P}_v \mathbf{C}^{\mathrm{T}} \mathbf{K} u) = 0,
\end{displaymath}
which implies that $u \in \mathrm{range} \, \mathbf{C}$ because $\mathbf{K}$ is invertible. This shows that $\mathrm{null} \, \mathbf{\tilde{K}} \subseteq \mathrm{range} \, \mathbf{C}$, completing the proof that $\mathrm{null} \, \mathbf{\tilde{K}} = \mathrm{range} \, \mathbf{C}$.

Next, we show that $H[\mathbf{W}]$ is in the range space of $\mathbf{C}$ for all antisymmetric second rank tensors $\mathbf{W}$. Combining this result with the previous statement about the null space of $\mathbf{\tilde{K}}$ establishes that $\mathbf{\tilde{K}} H[\mathbf{W}] = 0$ for all antisymmetric $\mathbf{W}$. We use the notation of Sec.~\ref{sec:site_displacements_and_compatibility_matrix} in this derivation. Let $\mathbf{r}_h$ be the initial position relative to the unit cell origin of site $h$. Then, the $k$th bond vector is given by $\mathbf{s}_k = \mathbf{r}_{h_2(k)} + \mathbf{n}_k - \mathbf{r}_{h_1(k)}$, with $\mathbf{\hat{s}}_k = |\mathbf{s}_k|^{-1} \mathbf{s}_k$. We observe that the $k$th entry in the vector $(\partial_{\bar{q}_r} \mathbf{C}(\mathbf{q})|_{\mathbf{q} = \mathbf{0}}) D [\mathbf{c}]$ is $n_r^{(k)} \mathbf{\hat{s}}_k \cdot \mathbf{c}$, where $n_r^{(k)}$ is the $r$th component of $\mathbf{n}_k$, for all vectors $\mathbf{c}$. Therefore, the $k$th component of $H[\mathbf{W}]$ is
	\begin{align*}
		(H[\mathbf{W}])_k &=\left( \sum_{r = 1}^d (-i) (\partial_{\bar{q}_r} \mathbf{C}(\mathbf{q})|_{\mathbf{q} = \mathbf{0}}) D [\mathbf{W} \cdot a^{-1}  \mathbf{a}_r] \right)_k \\
		&= \sum_{r = 1}^d n_r^{(k)} \mathbf{\hat{s}}_k \cdot ( \mathbf{W} \cdot a^{-1}  \mathbf{a}_r) \\
		& =  a^{-1} \mathbf{\hat{s}}_k \cdot ( \mathbf{W} \cdot  \mathbf{n}_k) \\
		& =  a^{-1} |\mathbf{s}_k|^{-1} (\mathbf{r}_{h_2(k)} + \mathbf{n}_k - \mathbf{r}_{h_1(k)}) \cdot ( \mathbf{W} \cdot  \mathbf{n}_k) \\
		& =  a^{-1} |\mathbf{s}_k|^{-1} ( \mathbf{r}_{h_2(k)} - \mathbf{r}_{h_1(k)}) \cdot ( \mathbf{W} \cdot  \mathbf{n}_k),
	\end{align*}
where in the last equality, the antisymmetry of $\mathbf{W}$ is used. To complete the proof, we show that $H[\mathbf{W}]$ is equal to the bond extensions resulting from unit cell-periodic displacements associated with infinitesimally rotating the sites about the centroid of the unit cell by $-\mathbf{W}$. Let $\mathbf{\bar{r}} = \frac{1}{n_s} \sum_{h = 1}^{n_s} \mathbf{r}_h$ be the centroid of the unit cell with respect to site positions. We consider site displacements $\mathbf{u}_h$ given by rotating each site about the centroid using $-\mathbf{W}$, so that
	\begin{displaymath}
		\mathbf{u}_h = -\mathbf{W} \cdot (\mathbf{r}_h - \mathbf{\bar{r}}).
	\end{displaymath}
Assembling these site displacements in the vector $\Phi = [\mathbf{u}_1 \mathellipsis \mathbf{u}_{n_s}]^{\mathrm{T}}$, and using $\mathbf{u}_{h_2(k)} - \mathbf{u}_{h_1(k)} = -\mathbf{W} \cdot ( \mathbf{r}_{h_2(k)} - \mathbf{r}_{h_1(k)})$, the $k$th component (bond extension) of $\mathbf{C} \Phi$ is
\begin{align*}
	(\mathbf{C} \Phi)_k &= \mathbf{\hat{s}}_k \cdot (\mathbf{u}_{h_2(k)} - \mathbf{u}_{h_1(k)}) \\
	&= |\mathbf{s}_k|^{-1} ( \mathbf{r}_{h_1(k)} + \mathbf{n}_k - \mathbf{r}_{h_2(k)}) \cdot (\mathbf{u}_{h_2(k)} - \mathbf{u}_{h_1(k)}) \\
	& = |\mathbf{s}_k|^{-1} \mathbf{n}_k  \cdot (-\mathbf{W} \cdot ( \mathbf{r}_{h'(k)} - \mathbf{r}_{h(k)})) \\
	& = |\mathbf{s}_k|^{-1} ( \mathbf{r}_{h'(k)} - \mathbf{r}_{h(k)}) \cdot ( \mathbf{W} \cdot  \mathbf{n}_k) = a (H[\mathbf{W}])_k.
\end{align*}
Therefore, $H[\mathbf{W}]$ is in the range space of $\mathbf{C}$, as claimed.

We now list the symmetries of the generalized elastic moduli that follow from $\mathbf{\tilde{K}} H[\mathbf{W}] = 0$ for all antisymmetric $\mathbf{W}$ and from the expressions in \eqref{eq:elastic_tensors}:
\begin{subequations}
\label{eq:relations}
    \begin{align}
        & (\mathsf{C})_{ijkl} = (\mathsf{C})_{klij}, \quad (\mathsf{C})_{ijkl} = (\mathsf{C})_{jikl} = (\mathsf{C})_{ijlk} \\
        & (\mathfrak{B}_m)_{ijk} = (\mathfrak{B}_m)_{jik} \\
        & (\mathbf{N}_m)_{ij} = (\mathbf{N}_m)_{ji} \\
        & (\mathbf{M}_{pm})_{ij} = (\mathbf{M}_{mp})_{ji} \\
        & J_{pm} = J_{mp}
    \end{align}
\end{subequations}
for $m, p = 1, \ldots n_w$, which are the symmetries we aimed to show. 

These can be expressed in index-free form as $\mathbf{A}: (\mathsf{C}:\mathbf{B}) = \mathbf{B}: (\mathsf{C}:\mathbf{A})$ for all second rank tensors $\mathbf{A}, \mathbf{B}$ (major symmetry), $\mathsf{C}:\mathbf{W} = \mathbf{0}$ for all antisymmetric tensors $\mathbf{W}$ (minor symmetry), $\mathbf{N}_m = \mathbf{N}_m^{\mathrm{T}}$, $(\mathfrak{B}_m \cdot \mathbf{c}) = (\mathfrak{B}_m \cdot \mathbf{c})^{\mathrm{T}}$ for all vectors $\mathbf{c}$, and $\mathbf{M}_{km}^{\mathrm{T}} = \mathbf{M}_{mk}$.

\subsection{The condition $\nu = 1$ is necessary to capture topological polarization}
\label{sec:necessity}
We establish that $\varepsilon \sim \mathcal{O}(\delta)$ is necessary to capture topological polarization by showing that when $\nu \neq 1$ and $\nu \geq 0$ in $\varepsilon \sim \mathcal{O}(\delta^{\nu})$, the continuum equations of motion obtained from retaining only terms lowest order in $\delta$ present in \eqref{eq:full_eqs_motion} are spatial inversion symmetric. We consider the following cases, which exhaust the possible values of $\nu$.
\begin{enumerate}[leftmargin=\parindent,labelindent=\parindent]
    \item $\nu = 0$ \\
    We repeat the procedure used to obtain the continuum equations of motion in Appendix~\ref{sec:projecting_subspaces}, but with $\mathbf{C}$ replaced by $\mathbf{C} + \varepsilon \mathbf{C}_w + \mathcal{O}(\varepsilon^2)$ and $\Phi_v$ replaced by $\Phi_w + \Phi_v$. These steps are equivalent to setting $n_w = 0$, i.e., considering a lattice with no local soft modes. Thus $\Phi_w + \Phi_v$ is eliminated from the equations of motion, resulting in standard linear elasticity. This case is therefore unable to capture topological polarization.
    \item $\nu > 0$ \\
    To account for possible scaling of $|\Phi_w|$ with $\delta$, we write $\Phi_w \sim \mathcal{O}(\delta^{\sigma - \nu})$ where $\sigma \geq \nu$, so that $\varepsilon \Phi_w \sim \mathcal{O}(\delta^{\sigma})$.
    \begin{enumerate}[leftmargin=12pt,labelindent=12pt]
        \item $0 < \sigma < 1$ \\
        Retaining terms lowest order in $\delta$ present in \eqref{eq:eq_motion_projection_Pv} gives
        \begin{displaymath}
            \Phi_v = - \mathbf{B} \mathbf{C}^{\mathrm{T}} \mathbf{K} \mathbf{C}_w \varepsilon \Phi_w,
        \end{displaymath}
        where $\mathbf{B}$ is the pseudoinverse introduced in \eqref{eq:def_B}. Substituting this expression into \eqref{eq:eq_motion_projection_Px} and \eqref{eq:eq_motion_projection_Pw}, and retaining terms lowest order in $\delta$ yields
        \begin{align*}
            \mathbf{P}_u \mathbf{C}_{\mathbf{x}}^{\mathrm{T}} \mathbf{\tilde{K}} \varepsilon \mathbf{C}_w \Phi_w &= 0 \\
            \mathbf{P}_w \varepsilon \mathbf{C}_{w}^{\mathrm{T}} \mathbf{\tilde{K}} \varepsilon \mathbf{C}_w \Phi_w &= 0
        \end{align*}
        as the effective continuum equations of motion. These are not only spatial inversion symmetric but also do not include any dependence on derivatives of $\Phi_u$ and therefore are independent of strain.
        \item $\sigma > 1$
        \begin{enumerate}[leftmargin=\parindent,labelindent=\parindent]
            \item $\sigma = \nu$ \\
            Retaining terms lowest order in $\delta$ in \eqref{eq:eq_motion_projection_Pv} yields
            \begin{equation*}
                \Phi_v = - \mathbf{B} \mathbf{C}^{\mathrm{T}} \mathbf{K} \mathbf{C}_{\mathbf{x}}( \Phi_u + \Phi_w ),
            \end{equation*}
            which is substituted into \eqref{eq:eq_motion_projection_Px} and \eqref{eq:eq_motion_projection_Pw} to obtain the continuum equations of motion (after discarding terms not of lowest order in $\delta$)
            \begin{align*}
                \qquad a^2 \mathbf{P}_u (\mathbf{M} \partial_t^2 \Phi - F) 
                &= \mathbf{P}_u \mathbf{C}_{\mathbf{x}}^{\mathrm{T}} \mathbf{\tilde{K}} \mathbf{C}_{\mathbf{x}}( \Phi_u + \Phi_w )
                 \\
                a^2 \mathbf{P}_w (\mathbf{M} \partial_t^2 \Phi - F)
                &= \mathbf{P}_w \mathbf{C}_{\mathbf{x}}^{\mathrm{T}} \mathbf{\tilde{K}} \mathbf{C}_{\mathbf{x}}( \Phi_u + \Phi_w ).
            \end{align*}
            These equations are spatial inversion symmetric.
            \item $\sigma > \nu$ \\
            Retaining terms lowest order in $\delta$ in \eqref{eq:eq_motion_projection_Pv} yields
            \begin{equation*}
                \Phi_v = - \mathbf{B} \mathbf{C}^{\mathrm{T}} \mathbf{K} \mathbf{C}_{\mathbf{x}} \Phi_u,
            \end{equation*}
            which is substituted into \eqref{eq:eq_motion_projection_Px} and \eqref{eq:eq_motion_projection_Pw} to obtain the continuum equations of motion (after discarding terms not of lowest order in $\delta$)
            \begin{align*}
                a^2 \mathbf{P}_u (\mathbf{M} \partial_t^2 \Phi - F)
                &= \mathbf{P}_u \mathbf{C}_{\mathbf{x}}^{\mathrm{T}} \mathbf{\tilde{K}} \mathbf{C}_{\mathbf{x}} \Phi_u \\
                \qquad \quad a^2 \mathbf{P}_w (\mathbf{M} \partial_t^2 \Phi - F)
                & = \mathbf{P}_w (\mathbf{C}_{\mathbf{x}} - \varepsilon \mathbf{C}_w)^{\mathrm{T}} \mathbf{\tilde{K}} \mathbf{C}_{\mathbf{x}} \Phi_u.
            \end{align*}
            If $\nu = 1$, the equations can break spatial inversion symmetry, and are special cases of \eqref{eq:continuum_eqs_preliminary} obtained by discarding terms involving $\Phi_w$ because they are higher order in $\delta$ than $\Phi_u$.
            To proceed, we rule out separately $\nu < 1$ and $\nu > 1$. If $\nu < 1$, then neglecting terms higher order in $\delta$ in the second equation above gives
            \begin{equation*}
                 \mathbf{P}_w \varepsilon \mathbf{C}_w^{\mathrm{T}} \mathbf{\tilde{K}} \mathbf{C}_{\mathbf{x}} \Phi_u = 0,
            \end{equation*}
            so the continuum equations of motion are spatial inversion symmetric. 
            If $\nu > 1$, then the same equation becomes
            \begin{equation*}
                a^2 \mathbf{P}_w (\mathbf{M} \partial_t^2 \Phi - F) = \mathbf{P}_w \mathbf{C}_{\mathbf{x}}^{\mathrm{T}} \mathbf{\tilde{K}} \mathbf{C}_{\mathbf{x}} \Phi_u
            \end{equation*}
            after neglecting terms higher order in $\delta$, so the continuum equations of motion are again spatial inversion symmetric. 
        \end{enumerate}
        \item $\sigma = 1$
        \begin{enumerate}[leftmargin=\parindent,labelindent=\parindent]
            \item $0 < \nu < 1$ \\
            Retaining terms lowest order in $\delta$ present in \eqref{eq:eq_motion_projection_Pv} gives \eqref{eq:pseudo-inversion_Phi_v}, which is substituted into \eqref{eq:eq_motion_projection_Px} and \eqref{eq:eq_motion_projection_Pw} to obtain the continuum equations of motion (after discarding terms not of lowest order in $\delta$)
            \begin{align*}
                \quad \qquad a^2 \mathbf{P}_u (\mathbf{M} \partial_t^2 \Phi - F)
                &= \mathbf{P}_u \mathbf{C}_{\mathbf{x}}^{\mathrm{T}} \mathbf{\tilde{K}} (\mathbf{C}_{\mathbf{x}} \Phi_u + \varepsilon \mathbf{C}_w \Phi_w) \\
                \varepsilon \mathbf{P}_w \mathbf{C}_w^{\mathrm{T}} \mathbf{\tilde{K}} \mathbf{C}_w \Phi_w & = - \mathbf{P}_w \mathbf{C}_w^{\mathrm{T}} \mathbf{\tilde{K}} \mathbf{C}_{\mathbf{x}} \Phi_u.
            \end{align*}
            Let $\mathbf{P}_1$ be the orthogonal projection onto the orthogonal complement of the null space of $\mathbf{\tilde{K}} \mathbf{C}_w$, so that $\mathbf{\tilde{K}} \mathbf{C}_w \Phi_w = \mathbf{\tilde{K}} \mathbf{C}_w \mathbf{P}_1 \Phi_w$. The restriction of $\mathbf{P}_w \mathbf{C}_w^{\mathrm{T}} \mathbf{\tilde{K}} \mathbf{C}_w \mathbf{P}_w$ to the range space of $\mathbf{P}_1$ is invertible. We recall that the range space of $\mathbf{C}_w^{\mathrm{T}} \mathbf{\tilde{K}}$ is equal to the orthogonal complement of the null space of $\mathbf{\tilde{K}} \mathbf{C}_w$ \cite{axler_linear_2024}, so the restriction of $\mathbf{P}_w \mathbf{C}_w^{\mathrm{T}} \mathbf{\tilde{K}} \mathbf{C}_w \mathbf{P}_w$ to the range space of $\mathbf{P}_1$ is an operator on that subspace. Multiplying the second equation by the inverse of this restriction of $\mathbf{P}_w \mathbf{C}_w^{\mathrm{T}} \mathbf{\tilde{K}} \mathbf{C}_w \mathbf{P}_w$ gives
            \begin{equation*}
                \mathbf{P}_1 \Phi_w = - \mathbf{Y} \mathbf{P}_w \mathbf{C}_w^{\mathrm{T}} \mathbf{\tilde{K}} \mathbf{C}_{\mathbf{x}} \Phi_u,
            \end{equation*}
            where $\mathbf{Y}$ is the aforementioned inverse. This equation is substituted into the $\mathbf{P}_u$ equation of motion to eliminate $\Phi_w$, leading to
            \begin{align*}
                \qquad &a^2 \mathbf{P}_u (\mathbf{M} \partial_t^2 \Phi - F) \\
                & \qquad = \mathbf{P}_u \mathbf{C}_{\mathbf{x}}^{\mathrm{T}} (\mathbf{\tilde{K}} - \mathbf{\tilde{K}}\mathbf{C}_w \mathbf{Y} \mathbf{C}_w^{\mathrm{T}} \mathbf{\tilde{K}}) \mathbf{C}_{\mathbf{x}} \Phi_u.
            \end{align*}
            Thus, we have eliminated $\Phi_w$ using similar reasoning to that used in eliminating $\Phi_v$. The resulting continuum equations of motion are those of standard linear elasticity with modified effective spring constants.  These equations of motion are spatial inversion symmetric.
            \item $\nu = 1$ \\
            The continuum equations of motion are those given in \eqref{eq:continuum_eqs_preliminary} and so are able to break spatial inversion symmetry.
        \end{enumerate}
    \end{enumerate}
\end{enumerate}
We have shown that the continuum equations of motion obtained by retaining lowest order terms in $\delta$ are spatial inversion symmetric whenever $\nu \neq 1$. Therefore, $\nu = 1$ is necessary for the equations to capture topological polarization.

\subsection{Recovering the compatibility matrix determinant from generalized elastic moduli}
\label{sec:recovering_determinant}
We show that the generalized elastic moduli of the homogenized theory contain enough information to recover $\det \mathbf{C}_p(\mathbf{q}; \varepsilon)$ to lowest order in $\delta$. We use this to prove that $\det \mathbf{P}_K^{\mathrm{T}} \mathbf{\hat{C}}(\mathbf{q})$ in Sec.~\ref{sec:maxwell_condition} is proportional to the terms lowest order in $\delta$ in $\det \mathbf{C}_p(\mathbf{q}; \varepsilon)$ when the continuum theory arises from a periodic lattice satisfying the Maxwell criterion, provided the continuum Maxwell criterion is also satisfied by the generalized elastic moduli.

First, we define
\begin{equation}
	\label{eq:def_mat_A}
	\mathbf{A}(\mathbf{\check{q}}, \mathbf{q}; \varepsilon) = \mathbf{C}_p(\mathbf{\check{q}}; \varepsilon)^{\mathrm{T}} \mathbf{K} \mathbf{C}_p(\mathbf{q}; \varepsilon)
\end{equation}
and we let $\mathbf{S}_u$, $\mathbf{S}_w$ and $\mathbf{S}_v$ be matrices whose columns are given by $\{\sqrt{n_s} u_n\}_{n =1}^d$, $\{w_m\}_{m = 1}^{n_w}$, and $\{v_k\}_{k = 1}^{n_s d - (d + n_w)}$ respectively. These three sets of vectors are the eigenvectors of $\mathbf{C}^{\mathrm{T}} \mathbf{K} \mathbf{C}$ introduced in Sec.~\ref{sec:displacement_and_other}. Let $\mathbf{S} = [\mathbf{S}_u \, \mathbf{S}_w \, \mathbf{S}_v]$ be the matrix containing columns from $\mathbf{S}_u$, $\mathbf{S}_w$ and $\mathbf{S}_v$. We see that
\begin{equation}
\label{eq:detC_propto}
\begin{aligned}
    \det \mathbf{C}_p(\mathbf{\check{q}}; \varepsilon) \,  \det \mathbf{C}_p(\mathbf{q}; \varepsilon) &\varpropto \det \mathbf{A}(\mathbf{\check{q}}, \mathbf{q}; \varepsilon) \\
    &\varpropto \det \mathbf{S}^{\mathrm{T}} \mathbf{A}(\mathbf{\check{q}}, \mathbf{q}; \varepsilon)\mathbf{S},
\end{aligned}
\end{equation}
since $\det \mathbf{K} > 0$ and $\det \mathbf{S} > 0$ are constants independent of $\mathbf{q}$. We express $\det \mathbf{S}^{\mathrm{T}} \mathbf{A}(\mathbf{\check{q}}, \mathbf{q}; \varepsilon) \mathbf{S}$ in terms of the generalized elastic constants. For notational convenience, we write $\mathbf{S} = [\mathbf{S}_{u,w} \, \mathbf{S}_v]$, where $\mathbf{S}_{u,w}$ combines the columns of $\mathbf{S}_{x}$ and $\mathbf{S}_{w}$. Then,
\begin{equation}
\label{eq:detC_reduction_matrix}
\begin{aligned}
    &\det \mathbf{S}^{\mathrm{T}} \mathbf{A}(\mathbf{\check{q}}, \mathbf{q}; \varepsilon) \mathbf{S} \\
    &= \det \begin{bmatrix}
        \mathbf{S}_{u,w}^{\mathrm{T}} \mathbf{A}(\mathbf{\check{q}}, \mathbf{q}; \varepsilon) \mathbf{S}_{u,w} & \mathbf{S}_{u,w}^{\mathrm{T}} \mathbf{A}(\mathbf{\check{q}}, \mathbf{q}; \varepsilon) \mathbf{S}_{v} \\
        \mathbf{S}_{v}^{\mathrm{T}} \mathbf{A}(\mathbf{\check{q}}, \mathbf{q}; \varepsilon) \mathbf{S}_{u,w} & \mathbf{S}_{v}^{\mathrm{T}} \mathbf{A}(\mathbf{\check{q}}, \mathbf{q}; \varepsilon) \mathbf{S}_{v}
    \end{bmatrix} \\
    &= \det \mathbf{S}_{v}^{\mathrm{T}} \mathbf{A}(\mathbf{\check{q}}, \mathbf{q}; \varepsilon) \mathbf{S}_{v} \, \det \mathbf{\tilde{A}}(\mathbf{\check{q}}, \mathbf{q}; \varepsilon),
\end{aligned}
\end{equation}
for sufficiently small $q_r$, $\varepsilon$, where
\begin{align*}
    \mathbf{\tilde{A}}(\mathbf{\check{q}}, \mathbf{q}; \varepsilon) = \mathbf{S}_{u,w}^{\mathrm{T}}( \mathbf{A} - \mathbf{A} \mathbf{S}_{v} (\mathbf{S}_{v}^{\mathrm{T}} \mathbf{A} \mathbf{S}_{v})^{-1}\mathbf{S}_{v}^{\mathrm{T}} \mathbf{A}) \mathbf{S}_{u,w},
\end{align*}
writing $\mathbf{A} = \mathbf{A}(\mathbf{\check{q}}, \mathbf{q}; \varepsilon)$, and we have used the matrix identity
\begin{equation}
\label{eq:det_identity}
	\det \begin{bmatrix}
		\mathbf{B} & \mathbf{C} \\
		\mathbf{D} & \mathbf{E}
	\end{bmatrix} = \det \mathbf{E} \, \det(\mathbf{B} - \mathbf{C} \mathbf{E}^{-1} \mathbf{D})
\end{equation}
for invertible $\mathbf{E}$. The requirement that $\mathbf{S}_{v}^{\mathrm{T}} \mathbf{A}(\mathbf{\check{q}}, \mathbf{q}; \varepsilon) \mathbf{S}_{v}$ be invertible is satisfied for sufficiently small $\check{q}_r$, $q_r$, $\varepsilon$ because the columns of $\mathbf{S}_{v}$ are eigenvectors of $\mathbf{A}(\mathbf{0},\mathbf{0}; 0) = \mathbf{C}^{\mathrm{T}} \mathbf{K} \mathbf{C}$ with strictly positive eigenvalues.

Using the definition of $\mathbf{A}(\mathbf{\check{q}}, \mathbf{q}; \varepsilon)$, algebraic manipulations give
\begin{align*}
    \mathbf{\tilde{A}}(\mathbf{\check{q}},\mathbf{q}; \varepsilon)
    &= \mathbf{S}_{u,w}^{\mathrm{T}}\mathbf{C}_p(\mathbf{\check{q}}; \varepsilon)^{\mathrm{T}} (\mathbf{\tilde{K}} + \mathcal{O}(\delta)) \mathbf{C}_p(\mathbf{q}; \varepsilon) \mathbf{S}_{u,w} \\
    &= \begin{bmatrix}
        \mathbf{S}_{x}^{\mathrm{T}} \mathbf{R}(\mathbf{\check{q}}, \mathbf{q}; \varepsilon) \mathbf{S}_{x} & \mathbf{S}_{x}^{\mathrm{T}} \mathbf{R}(\mathbf{\check{q}}, \mathbf{q}; \varepsilon) \mathbf{S}_{w} \\
        \mathbf{S}_{w}^{\mathrm{T}} \mathbf{R}(\mathbf{\check{q}}, \mathbf{q}; \varepsilon) \mathbf{S}_{x} & \mathbf{S}_{w}^{\mathrm{T}} \mathbf{R}(\mathbf{\check{q}}, \mathbf{q}; \varepsilon) \mathbf{S}_{w}
    \end{bmatrix},
\end{align*}
where $\mathbf{R}(\mathbf{\check{q}}, \mathbf{q}; \varepsilon) = \mathbf{C}_p(\mathbf{\check{q}}; \varepsilon)^{\mathrm{T}} (\mathbf{\tilde{K}} + \mathcal{O}(\delta)) \mathbf{C}_p(\mathbf{q}; \varepsilon)$. We relate $\mathbf{\tilde{A}}(\mathbf{\check{q}}, \mathbf{q}; \varepsilon)$ to the generalized elastic moduli using \eqref{eq:H_G_m}, giving
\begin{subequations}
    \begin{align}
        (\mathbf{C}_p(\mathbf{q}; \varepsilon) \mathbf{S}_{x})_{jk} &= a(H[\mathbf{e}_k \otimes (i \mathbf{q})])_{j} + \mathcal{O}(\delta^2) \\
        (\mathbf{C}_p(\mathbf{q}; \varepsilon) \mathbf{S}_{w})_{jk} &= a(G_k[(i \mathbf{q})])_{j} + \varepsilon \mathbf{C}_w w_k +\mathcal{O}(\delta^2).
    \end{align}
\end{subequations}
Since the values of $\mathbf{\check{q}}$, $\mathbf{q}$ are complex, the inner products $\cdot$ and $:$ must be complexified so that they are antilinear in their first argument: $(c \mathbf{a}) \cdot \mathbf{b} = c^* \mathbf{a} \cdot \mathbf{b}$ and $(c \mathbf{A}) : \mathbf{B} = c^* \mathbf{A} : \mathbf{B}$ for $c \in \mathbb{C}$, where $^*$ denotes complex conjugation. For consistency with the complexified inner products, the tensor product is complexified such that $\mathbf{a} \otimes (c \mathbf{b}) = c^* \mathbf{a} \otimes \mathbf{b}$. The operators $H$ and $G_m$ introduced in \eqref{eq:H_G_m} must also be complexified to be compatible with the complexified inner products. This modification involves replacing the transposes $^\mathrm{T}$ in \eqref{eq:elastic_tensors} by conjugate transposes $^{\dagger}$ (Hermitian transposes). 

We introduce the generalized elastic moduli $\mathbf{\hat{K}}$ by computing the matrix (with respect to some fixed orthonormal basis $\{\mathbf{e}_j\}_{j=1}^d$ of $d$-dimensional space)
\begin{equation}
\label{eq:CKC_elastic_moduli}
    \mathbf{\hat{C}}(\mathbf{\check{q}})^{\dagger} \mathbf{\hat{K}} \mathbf{\hat{C}}(\mathbf{q}) = \begin{bmatrix}
        \mathbf{F}(\mathbf{\check{q}}, \mathbf{q}) & \mathbf{G}(\mathbf{q}, \mathbf{\check{q}})^{\dagger} \\
        \mathbf{G}(\mathbf{\check{q}}, \mathbf{q}) & \mathbf{H}(\mathbf{\check{q}}, \mathbf{q})
    \end{bmatrix},
\end{equation}
where the blocks are given by the matrices
\begin{subequations}
\label{eq:F_G_H}
    \begin{align}
    (\mathbf{F}(\mathbf{\check{q}}, \mathbf{q}))_{jk} &= (\mathbf{e}_j \otimes \mathbf{\check{q}}): \mathsf{C} : (\mathbf{e}_k \otimes \mathbf{q}) \\
    (\mathbf{G}(\mathbf{\check{q}}, \mathbf{q}))_{jk} &= (-\mathfrak{B}_j \cdot \mathbf{\check{q}} +i \mathbf{N}_j):(\mathbf{e}_k \otimes \mathbf{q}) \\
    (\mathbf{H}(\mathbf{\check{q}}, \mathbf{q}))_{jk} &= \mathbf{\check{q}} \cdot \mathbf{M}_{jk} \cdot  \mathbf{q} + i(\mathbf{h}_{kj} \cdot \mathbf{q} - \mathbf{\check{q}} \cdot \mathbf{h}_{jk}) + J_{jk}. \label{eq:H}
    \end{align}
\end{subequations}
Since the elements of $\mathbf{\hat{C}}(\mathbf{q})$ have polynomial dependence on $i\mathbf{q}$, we see that $\mathbf{\hat{C}}(\mathbf{q})^{\mathrm{T}} = \mathbf{\hat{C}}(-\mathbf{q}^*)^{\dagger}$, where the complex conjugate of a complex vector $\mathbf{q} = \mathbf{q}^R + i \mathbf{q}^I$ expressed in terms of real vectors $\mathbf{q}^R, \mathbf{q}^I$ is given by $\mathbf{q}^* = \mathbf{q}^R - i \mathbf{q}^I$. Similarly, since the elements of $\mathbf{C}_p(\mathbf{q}; \varepsilon)$ have polynomial dependence on $e^{i \mathbf{n} \cdot \mathbf{q}}$ for some vectors $\mathbf{n}$, $\mathbf{C}_p(\mathbf{q}; \varepsilon)^{\mathrm{T}} = \mathbf{C}_p(-\mathbf{q}^*; \varepsilon)^{\dagger}$.

We can now compute the following blocks in $\mathbf{\tilde{A}}(\mathbf{\check{q}}, \mathbf{q}; \varepsilon)$:
\begin{subequations}
\begin{equation}
        \begin{aligned}
        &(\mathbf{S}_{x}^{\mathrm{T}} \mathbf{C}_p(\mathbf{\check{q}}; \varepsilon)^{\mathrm{T}} \mathbf{\tilde{K}} \mathbf{C}_p(\mathbf{q}; \varepsilon) \mathbf{S}_{x})_{jk} \\
        & = (\mathbf{S}_{x}^{\mathrm{T}} \mathbf{C}_p(\mathbf{-\check{q}^*}; \varepsilon)^{\dagger} \mathbf{\tilde{K}} \mathbf{C}_p(\mathbf{q}; \varepsilon) \mathbf{S}_{x})_{jk} \\
        &= a^2 (\mathbf{e}_j \otimes (-i \mathbf{\check{q}}^*)): \mathsf{C} : (\mathbf{e}_k \otimes (i \mathbf{q})) + \mathcal{O}(\delta^3) \\
        &= -a^2 (\mathbf{e}_j \otimes \mathbf{\check{q}}^*): \mathsf{C} : (\mathbf{e}_k \otimes \mathbf{q}) + \mathcal{O}(\delta^3) \\
        &= a^2 (\mathbf{F}(-\mathbf{\check{q}}^*, \mathbf{q}))_{jk} + \mathcal{O}(\delta^3)
    \end{aligned}
\end{equation}
\begin{equation}
        \begin{aligned}
        &(\mathbf{S}_{w}^{\mathrm{T}} \mathbf{C}_p(\mathbf{\check{q}}; \varepsilon)^{\mathrm{T}} \mathbf{\tilde{K}} \mathbf{C}_p(\mathbf{q}; \varepsilon) \mathbf{S}_{x})_{jk} \\
        &= (\mathbf{S}_{w}^{\mathrm{T}} \mathbf{C}_p(-\mathbf{\check{q}}^*; \varepsilon)^{\dagger} \mathbf{\tilde{K}} \mathbf{C}_p(\mathbf{q}; \varepsilon) \mathbf{S}_{x})_{jk} \\
        &= a^2 (\mathfrak{B}_j \cdot (-i\mathbf{\check{q}}^*) + \mathbf{N}_j) :(\mathbf{e}_k \otimes (i \mathbf{q})) + \mathcal{O}(\delta^3) \\
        &= a^2 ( -\mathfrak{B}_j \cdot (-\mathbf{\check{q}}^*) + i\mathbf{N}_j) :(\mathbf{e}_k \otimes \mathbf{q}) + \mathcal{O}(\delta^3) \\
        &= a^2 (\mathbf{G}(-\mathbf{\check{q}}^*, \mathbf{q}))_{jk} + \mathcal{O}(\delta^3)
    \end{aligned}
\end{equation}
\begin{equation}
        \begin{aligned}
        &(\mathbf{S}_{x}^{\mathrm{T}} \mathbf{C}_p(\mathbf{\check{q}}; \varepsilon)^{\mathrm{T}} \mathbf{\tilde{K}} \mathbf{C}_p(\mathbf{q}; \varepsilon) \mathbf{S}_{w})_{jk} \\
        & = (\mathbf{S}_{x}^{\mathrm{T}} \mathbf{C}_p(\mathbf{-\check{q}^*}; \varepsilon)^{\dagger} \mathbf{\tilde{K}} \mathbf{C}_p(\mathbf{q}; \varepsilon) \mathbf{S}_{w})_{jk} \\
        & = a^2 (\mathbf{e}_j \otimes (-i \mathbf{\check{q}}^*)): (\mathfrak{B}_k \cdot (i \mathbf{q}) + \mathbf{N}_k) + \mathcal{O}(\delta^3) \\
        & = a^2 (\mathbf{e}_j \otimes (- \mathbf{\check{q}}^*)): (-\mathfrak{B}_k \cdot \mathbf{q} + i\mathbf{N}_k) + \mathcal{O}(\delta^3) \\
        & = a^2 ((-\mathfrak{B}_k \cdot \mathbf{q} + i\mathbf{N}_k) : (\mathbf{e}_j \otimes (- \mathbf{\check{q}}^*)))^* + \mathcal{O}(\delta^3) \\
        &= a^2 (\mathbf{G}(\mathbf{q}, -\mathbf{\check{q}}^*)^{\dagger})_{jk} + \mathcal{O}(\delta^3)
    \end{aligned}
\end{equation}
\begin{equation}
        \begin{aligned}
        &(\mathbf{S}_{w}^{\mathrm{T}} \mathbf{C}_p(\mathbf{\check{q}}; \varepsilon)^{\mathrm{T}} \mathbf{\tilde{K}} \mathbf{C}_p(\mathbf{q}; \varepsilon) \mathbf{S}_{w})_{jk} \\
        & = (\mathbf{S}_{w}^{\mathrm{T}} \mathbf{C}_p(-\mathbf{\check{q}}^*; \varepsilon)^{\dagger} \mathbf{\tilde{K}} \mathbf{C}_p(\mathbf{q}; \varepsilon) \mathbf{S}_{w})_{jk} \\
        &= a^2 (-i\mathbf{\check{q}}^*) \cdot \mathbf{M}_{jk} \cdot (i \mathbf{q}) \\
        & \qquad + (-i\mathbf{\check{q}}^*) \cdot \mathbf{h}_{jk} + \mathbf{h}_{kj} \cdot (i\mathbf{q}) + J_{jk}) + \mathcal{O}(\delta^3) \\
        &= a^2 ((-\mathbf{\check{q}}^*) \cdot \mathbf{M}_{jk} \cdot  \mathbf{q}  \\
        &  \qquad + i(\mathbf{h}_{kj} \cdot \mathbf{q} - (-\mathbf{\check{q}}^*) \cdot \mathbf{h}_{jk}) - J_{jk}) + \mathcal{O}(\delta^3) \\
        &= a^2 (\mathbf{H}(-\mathbf{\check{q}}^*, \mathbf{q}))_{jk} + \mathcal{O}(\delta^3),
    \end{aligned}
\end{equation}
\end{subequations}
where $\mathbf{F}(\mathbf{\check{q}}, \mathbf{q})$, $\mathbf{G}(\mathbf{\check{q}}, \mathbf{q})$ and $\mathbf{H}(\mathbf{\check{q}}, \mathbf{q})$ are introduced in \eqref{eq:F_G_H}.

Therefore, we can use \eqref{eq:CKC_elastic_moduli} to obtain
\begin{align*}
    \mathbf{\tilde{A}}(\mathbf{\check{q}},\mathbf{q}; \varepsilon)
    &= a^2 \begin{bmatrix}
        \mathbf{F}(-\mathbf{\check{q}}^*, \mathbf{q}) & \mathbf{G}(\mathbf{q}, -\mathbf{\check{q}}^*)^{\dagger} \\
        \mathbf{G}(-\mathbf{\check{q}}^*, \mathbf{q}) & \mathbf{H}(-\mathbf{\check{q}}^*, \mathbf{q})
    \end{bmatrix} + \mathcal{O}(\delta^3) \\
    &= a^2 \mathbf{\hat{C}}(\mathbf{-\check{q}}^*)^{\dagger} \mathbf{\hat{K}} \mathbf{\hat{C}}(\mathbf{q}) + \mathcal{O}(\delta^3)\\
    &= a^2 \mathbf{\hat{C}}(\mathbf{\check{q}})^{\mathrm{T}} \mathbf{\hat{K}} \mathbf{\hat{C}}(\mathbf{q}) + \mathcal{O}(\delta^3).
\end{align*}
Combining this result with Eqs.~(\ref{eq:detC_propto}, \ref{eq:detC_reduction_matrix}), and using $\det \mathbf{S}_{v}^{\mathrm{T}} \mathbf{A}(\mathbf{\check{q}}, \mathbf{q}; \varepsilon) \mathbf{S}_{v} = \det \mathbf{S}_{v}^{\mathrm{T}} \mathbf{A}(\mathbf{0}, \mathbf{0}; 0) \mathbf{S}_{v} + \mathcal{O}(\delta)$, we obtain
\begin{align}
\label{eq:detC_double_propto}
    &\det \mathbf{C}_p(\mathbf{\check{q}}; \varepsilon) \,  \det \mathbf{C}_p(\mathbf{q}; \varepsilon) \\
    & \qquad \varpropto a^{2(d + n_w)} \det \mathbf{\hat{C}}(\mathbf{\check{q}})^{\mathrm{T}} \mathbf{\hat{K}} \mathbf{\hat{C}}(\mathbf{q}) + \mathcal{O}\left(\delta^{2(d + n_w) + 1}\right) \nonumber
\end{align}
where the constant of proportionality is positive and we recall that $a = \delta L$. 

Recalling the definition of $\mathbf{P}_K$ in Sec.~\ref{sec:maxwell_condition}, 
\begin{align}
\label{eq:detChat_propto}
    &\det \mathbf{\hat{C}}(\mathbf{q})^{\mathrm{T}} \mathbf{\hat{K}} \mathbf{\hat{C}}(\mathbf{q}) \\
    & \qquad = \det \mathbf{\hat{C}}(\mathbf{q})^{\mathrm{T}} \mathbf{P}_K \mathbf{P}_K^{\mathrm{T}} \mathbf{\hat{K}}^{\frac{1}{2}} \mathbf{P}_K \mathbf{P}_K^{\mathrm{T}} \mathbf{\hat{K}}^{\frac{1}{2}} \mathbf{P}_K \mathbf{P}_K^{\mathrm{T}} \mathbf{\hat{C}}(\mathbf{q}) \nonumber \\
    & \qquad = (\det \mathbf{P}_K^{\mathrm{T}} \mathbf{\hat{K}}^{\frac{1}{2}} \mathbf{P}_K \mathbf{P}_K^{\mathrm{T}} \mathbf{\hat{C}}(\mathbf{q}))^2 \nonumber \\
    & \qquad = (\det \mathbf{P}_K^{\mathrm{T}} \mathbf{\hat{K}}^{\frac{1}{2}} \mathbf{P}_K)^2 (\det \mathbf{P}_K^{\mathrm{T}} \mathbf{\hat{C}}(\mathbf{q}))^2 \nonumber
\end{align}
when the Maxwell criterion \eqref{eq:maxwell_continuum} is satisfied, and where $\mathbf{\hat{K}}^{\frac{1}{2}}$ is the unique positive semi-definite square root of the positive semi-definite matrix $\mathbf{\hat{K}}$. Therefore, setting $\mathbf{\check{q}} = \mathbf{q}$ in \eqref{eq:detC_double_propto} and using \eqref{eq:detChat_propto} gives
\begin{align}
    (\det \mathbf{C}_p(\mathbf{q}; \varepsilon))^2 &\varpropto (-a^2)^{d + n_w} (\det \mathbf{P}_K^{\mathrm{T}} \mathbf{\hat{C}}(\mathbf{q}))^2 \\
    & \qquad + \mathcal{O}\left(\delta^{2(d + n_w) + 1}\right).\nonumber
\end{align}
We have shown that the terms lowest order in $\delta$ in $\det \mathbf{C}_p(\mathbf{q}; \varepsilon)$ are proportional to $\det \mathbf{P}_K^{\mathrm{T}} \mathbf{\hat{C}}(\mathbf{q})$. Our result here is also consistent with our derivation in Appendix~\ref{sec:detC_perturbed_derivation} showing that the terms of lowest order in $\delta$ present in $\det \mathbf{C}_p(\mathbf{q}; \varepsilon)$ have order $d + n_w$. Since $\det \mathbf{P}_K^{\mathrm{T}} \mathbf{\hat{C}}(\mathbf{q})$ depends only on the generalized elastic moduli, we have shown that the generalized elastic moduli contain enough information to recover $\det \mathbf{C}_p(\mathbf{q}; \varepsilon)$ to lowest order in $\delta$, up to a constant of proportionality. This constant is irrelevant to the study of zero modes.

\subsection{Elastic energy density in the continuum limit}
\label{sec:elastic_energy}
In this subsection, we provide a physical interpretation for the scaling relations $\varepsilon \sim \mathcal{O}(\delta)$ and $|\Phi_v| \sim \mathcal{O}(\delta)$ in terms of the elastic energy density in the continuum limit.

The elastic potential energy per unit cell, in terms of the continuum fields, is given by $V_a (\mathbf{x}, t) = \frac{1}{2} \sum_{k = 1}^{n_b} K_k e_k(\mathbf{x}, t)^2$, where $e_k(\mathbf{x}, t) = \mathbf{C}_{p, \mathbf{x}} \Phi(\mathbf{x},t)$ is the extension of the $k$th bond, and $K_k$ is its spring constant. The elastic energy density is then
\begin{equation*}
    \begin{aligned}
        \frac{V_a (\mathbf{x}, t)}{a^d} \! &= \! \frac{1}{2 a^d} |\mathbf{K}_a^{\frac{1}{2}} \mathbf{C}_{p, \mathbf{x}} \Phi(\mathbf{x}, t)|^2 = \! \frac{1}{2} \! \left|\mathbf{K}^{\frac{1}{2}} \mathbf{C}_{p, \mathbf{x}} \frac{1}{a}\Phi(\mathbf{x}, t)\right|^2 \!,
    \end{aligned}
\end{equation*}
where we used the scaling $\mathbf{K}_a = a^{d - 2} \mathbf{K}$ introduced in Sec.~\ref{sec:site_displacements_and_compatibility_matrix}. Using the decomposition of $\Phi$ in \eqref{eq:phi_decomposition}, the term inside the vector norm is
\begin{equation*}
    \begin{aligned}
        & \mathbf{K}^{\frac{1}{2}}\mathbf{C}_{p, \mathbf{x}} \frac{1}{a}\Phi(\mathbf{x}, t) \\
        &= \frac{1}{a} \mathbf{K}^{\frac{1}{2}}(\mathbf{C}\Phi_v + \varepsilon \mathbf{C}_w (\Phi_w + \Phi_v) + \mathcal{O}(\varepsilon^2)) + \frac{1}{a} \mathcal{O}(\delta) \\
        &= \frac{1}{L \delta} \mathbf{K}^{\frac{1}{2}}(\mathbf{C}\Phi_v + \varepsilon \mathbf{C}_w (\Phi_w + \Phi_v) + \mathcal{O}(\varepsilon^2)) + \frac{1}{L} \mathcal{O}(1),
    \end{aligned}
\end{equation*}
where $\frac{1}{a} \mathcal{O}(\delta)$ in the first equality arises from $(\mathbf{C}_{\mathbf{x}} + \varepsilon \mathbf{C}_{\mathbf{x}, w}) \Phi(\mathbf{x},t)$, because terms involving these differential operators are $\mathcal{O}(\delta)$. If \emph{(i)} $|\Phi_v| \sim \mathcal{O}(\delta)$ and \emph{(ii)} $\varepsilon \sim \mathcal{O}(\delta)$, then we see that the elastic energy density remains bounded as $\delta \to 0$.

\section{Counting continuum edge modes with a topological invariant}
\label{sec:counting_edge_modes}
We derive \eqref{eq:edge_modes_invariant}. Let $g_{\alpha}(\tau) = \det \mathbf{P}_K^{\mathrm{T}} \mathbf{\hat{C}}(\mathbf{q}_{\alpha}^k(\tau))$. Since $g_{\alpha}(\tau)$ is a polynomial of degree $d + n_w$ in $\tau$ with $\alpha$-dependent coefficients, we can write the roots of $g_{\alpha}(\tau) = 0$ as $\tau = f_1(\alpha), \ldots, f_{d+n_w}(\alpha)$, where $f_j$ is a continuous function. Using the theory of Puiseux series~\cite{knopp1996theory}, the set of functions $f_j$ is partitioned into cycles, and $f_j$ can be expressed as a fractional power series
\begin{equation}
    f_j(\alpha) = \sum_{m = 0}^{\infty} \gamma_{k,m} \alpha^{\frac{m}{p_j}},
\end{equation}
where $p_j$ is the number of elements in the cycle to which $f_k$ belongs. Since $1 \leq p_j \leq d + n_w$, we see that
\begin{equation}
    \label{eq:f_j_limit}
    \lim_{\alpha \to 0^+} \frac{f_j(\alpha)}{\alpha^c} = 0
\end{equation}
whenever $f_j(0) = 0$ and $0 < c < \frac{1}{d + n_w}$. This result, which we use below, is the reason for the upper bound on $c$ in \eqref{eq:edge_modes_invariant}. If all the functions $f_j$ satisfying $f_j(0) = 0$ are analytic (i.e., $p_k = 1$), then $0 < c < 1$ is sufficient for \eqref{eq:f_j_limit} to hold for all functions $f_j$ satisfying $f_j(0) = 0$.

Since $g_{\alpha}(\tau) = Z \prod_{k = 1}^{d + n_w} (\tau - f_j(\alpha))$ for some constant $Z$, the integrand in \eqref{eq:edge_modes_invariant} is
\begin{equation}
\label{eq:sum_over_roots}
    \frac{\frac{\mathrm{d}}{\mathrm{d} \tau} \det \mathbf{P}_K^{\mathrm{T}} \mathbf{\hat{C}}(\mathbf{q}_{\alpha}^k(\tau))}{\det \mathbf{P}_K^{\mathrm{T}} \mathbf{\hat{C}}(\mathbf{q}_{\alpha}^k(\tau))} = \sum_{j = 1}^{d + n_w} \frac{1}{\tau - f_j(\alpha)}.
\end{equation}
We restrict $\alpha$ to real values in the integral. Each root $f_j(\alpha)$ contributes to the integral
\begin{equation}
\label{eq:integral_single_root}
	\begin{aligned}
		&\int_{-\alpha^c}^{\alpha^c} \frac{1}{\tau - f_j(\alpha)} \, \mathrm{d}\tau \\
        &= \log(\alpha^c - f_j(\alpha)) - \log(-\alpha^c - f_j(\alpha)) \\
		&= \log \left| \frac{\alpha^c - f_j(\alpha)}{-\alpha^c - f_j(\alpha)} \right| \\
        & \qquad + i(\arg (\alpha^c - f_j(\alpha)) - \arg (-\alpha^c - f_j(\alpha))).
	\end{aligned}
\end{equation}
To evaluate this expression, we choose a branch of the natural logarithm. Since branches of the logarithm differ by integer multiples of $2 \pi i$, the integral is independent of our branch choice as it is a difference between logarithmic terms. We choose the principal branch of the logarithm (with branch cut along the negative real axis) because the straight line in the complex plane joining $-\alpha^c - f_j(\alpha)$ and $\alpha^c - f_j(\alpha)$ is parallel to the real axis. This line lies either in the positive or negative imaginary half-plane, since $\mathrm{Im} \, f_j(\alpha) \neq 0$ for $\alpha > 0$, by the assumption that there are no bulk modes. Thus, this line does not cross the branch cut. With this branch choice, we have $\arg z \in (-\pi, \pi]$ and when $z \in \mathbb{C} \setminus (-\infty, 0]$,
\begin{equation}
\label{eq:arg_def}
	\begin{aligned}
	    \arg z &=  |\mathrm{sgn}(\mathrm{Re}(z))| \arctan \left(\frac{\mathrm{Im}(z)}{\mathrm{Re}(z)}\right) \\
     & \qquad \qquad+ H(-\mathrm{Re}(z)) \, \mathrm{sgn}(\mathrm{Im} (z)) \,\pi,
	\end{aligned}
\end{equation}
where
\begin{displaymath}
	\mathrm{sgn}(x) = 
    \begin{cases}
	   -1 & x < 0 \\
	   0 & x = 0\\
	   1 & x > 0,
    \end{cases}, \quad
    H(x) = \begin{cases}
		      0 & x < 0 \\
			\frac{1}{2} & x = 0 \\
			1 & x > 0
		\end{cases}.
\end{displaymath}
By considering separately $f_j(0) = 0$ and $f_j(0) \neq 0$, we see that
\begin{equation}
    \lim_{\alpha \to 0^+} \log \left| \frac{\alpha^c - f_j(\alpha)}{-\alpha^c - f_j(\alpha)} \right| = \log 1 = 0,
\end{equation}
in both cases, where we have used \eqref{eq:f_j_limit} for the case $f_j(0) = 0$. Therefore, the real part of the integral vanishes in the limit $\alpha \to 0^+$.

To evaluate the imaginary part of the integral, we use \eqref{eq:arg_def} to obtain
\begin{equation}
\label{eq:general_arg_expression}
    \begin{aligned}
        & \arg (\pm \alpha^c - f_j(\alpha)) \\
        & = |\mathrm{sgn}(\mathrm{Re}(\pm \alpha^c - f_j(\alpha)))| \arctan \left(\frac{\mathrm{Im}(\pm \alpha^c - f_j(\alpha))}{\mathrm{Re}(\pm \alpha^c - f_j(\alpha))}\right) \\
     & \qquad + H(-\mathrm{Re}(\pm \alpha^c - f_j(\alpha))) \, \mathrm{sgn}(\mathrm{Im} (\pm \alpha^c - f_j(\alpha))) \,\pi, \\
     & = -\arctan \left(\frac{\mathrm{Im}(f_j(\alpha))}{\pm \alpha^c - \mathrm{Re}(f_j(\alpha))}\right) \\
     & \qquad - H(\mp \alpha^c + \mathrm{Re}(f_j(\alpha))) \, \mathrm{sgn}(\mathrm{Im} (f_j(\alpha))) \,\pi,
    \end{aligned}
\end{equation}
for values of $\alpha > 0$ sufficiently close to zero, since $\lim_{\alpha \to 0^+} |\mathrm{sgn}(\mathrm{Re}(\pm \alpha^c - f_j(\alpha)))| = 1$.

When $j$ corresponds to $f_j(0) \neq 0$, we consider two cases: $\mathrm{Re}(f_j(0)) \neq 0$ and $\mathrm{Re}(f_j(0)) = 0$. When $\mathrm{Re}(f_j(0)) \neq 0$,
\begin{equation}
    \begin{aligned}
        & \lim_{\alpha \to 0^+} \arg (\pm \alpha^c - f_j(\alpha)) \\
        & = \lim_{\alpha \to 0^+} \left[ \arctan \left(\frac{\mathrm{Im}(f_j(\alpha))}{\mathrm{Re}(f_j(\alpha))}\right) \right. \\
        & \qquad \qquad \quad \left. - H(\mathrm{Re}(f_j(\alpha))) \, \mathrm{sgn}(\mathrm{Im} (f_j(\alpha))) \,\pi \vphantom{\frac{1}{2}}\right],
    \end{aligned}
\end{equation}
so the limit is independent of the sign in $\pm \alpha^c$, leading to
\begin{equation}
\label{eq:zero_contribution_arg}
    \lim_{\alpha \to 0^+} (\arg (\alpha^c - f_j(\alpha)) - \arg (-\alpha^c - f_j(\alpha))) = 0.
\end{equation}
When $\mathrm{Re}(f_j(0)) = 0$, \eqref{eq:f_j_limit} holds with $f_j(\alpha)$ replaced by $\mathrm{Re}(f_j(\alpha))$, and by assumption $\mathrm{Im}(f_j(0)) \neq 0$, so we obtain
\begin{equation}
    \begin{aligned}
        & \lim_{\alpha \to 0^+} \arg (\pm \alpha^c - f_j(\alpha)) \\
        & = \lim_{\alpha \to 0^+} \left[ \arctan \left(\frac{\mathrm{Im}(f_j(\alpha))}{\mp \alpha^c}\right) \right.\\
        & \qquad \qquad \qquad \left. - H(\mp 1) \, \mathrm{sgn}(\mathrm{Im} (f_j(\alpha))) \,\pi \vphantom{\frac{1}{2}} \right], \\
        & = \left(\lim_{\alpha \to 0^+} \mathrm{sgn}(\mathrm{Im} (f_j(\alpha)))\right)\left( \mp \frac{1}{2} - H(\mp 1)\right) \pi.
    \end{aligned}
\end{equation}
We again find that this case leads to \eqref{eq:zero_contribution_arg} because the limit is independent of the sign in $\pm \alpha^c$. In summary, when $j$ corresponds to $f_j(0) \neq 0$, the limit of the imaginary part of the integral in \eqref{eq:integral_single_root} vanishes. Therefore, in view of \eqref{eq:sum_over_roots}, roots $f_j(\alpha)$ such that $f_j(0) \neq 0$ do not contribute to the topological invariant \eqref{eq:edge_modes_invariant}.

Finally, when $f_j(0) = 0$, we obtain from \eqref{eq:general_arg_expression}
\begin{equation}
    \begin{aligned}
        & \lim_{\alpha \to 0^+} \arg (\pm \alpha^c - f_j(\alpha)) \\
        & = \lim_{\alpha \to 0^+} \left[ - H(\mp 1) \, \mathrm{sgn}(\mathrm{Im} (f_j(\alpha))) \,\pi \right],
    \end{aligned}
\end{equation}
which leads to
\begin{equation}
\label{eq:pi_contribution_arg}
    \begin{aligned}
        &\lim_{\alpha \to 0^+} (\arg (\alpha^c - f_j(\alpha)) - \arg (-\alpha^c - f_j(\alpha))) \\
        &= \left(\lim_{\alpha \to 0^+} \mathrm{sgn}(\mathrm{Im} (f_j(\alpha)))\right) \pi.
    \end{aligned}
\end{equation}
Bringing all our results of this section together gives
\begin{equation*}
    \begin{aligned}
        & \lim_{\alpha \to 0^{+}} \frac{1}{\pi i} \int_{-\alpha^c}^{\alpha^c} \frac{\frac{\mathrm{d}}{\mathrm{d} \tau} \det \mathbf{P}_K^{\mathrm{T}} \mathbf{\hat{C}}(\mathbf{q}_{\alpha}^k(\tau))}{\det \mathbf{P}_K^{\mathrm{T}} \mathbf{\hat{C}}(\mathbf{q}_{\alpha}^k(\tau))} \, \mathrm{d}\tau \\
        & = \lim_{\alpha \to 0^{+}} \frac{1}{\pi i} \sum_{j = 1}^{d + n_w} \int_{-\alpha^c}^{\alpha^c} \frac{1}{\tau - f_j(\alpha)} \, \mathrm{d}\tau \\
        & = \frac{1}{\pi i} \sum_{\substack{j = 1: \\ f_j(0) = 0}}^{d + n_w} i\left(\lim_{\alpha \to 0^+} \mathrm{sgn}(\mathrm{Im} (f_j(\alpha)))\right) \pi \\
        & = \sum_{\substack{j = 1: \\ f_j(0) = 0}}^{d + n_w} \left(\lim_{\alpha \to 0^+} \mathrm{sgn}(\mathrm{Im} (f_j(\alpha)))\right) = N_{+,k} - N_{-,k},
    \end{aligned}
\end{equation*}
which proves \eqref{eq:edge_modes_invariant} because $N_{+,k}$ is the number of roots with $\lim_{\alpha \to 0^+} \mathrm{sgn}(\mathrm{Im} (f_j(\alpha)) = 1$ and $N_{-,k}$ is the number of roots with $\lim_{\alpha \to 0^+} \mathrm{sgn}(\mathrm{Im} (f_j(\alpha)) = -1$.

\section{Guest-Hutchinson modes and $\Delta$}
\label{sec:guest-hutchinson}
Guest-Hutchinson modes of two-dimensional Maxwell lattices as studied in Refs.~\cite{lubensky_phonons_2015, mao_maxwell_2018, rocklin_transformable_2017} are defined using standard linear elasticity, as strains in the null space of the elasticity tensor. To define Guest-Hutchinson modes in our generalized elasticity theory, we first obtain an effective standard linear elastic continuum from our augmented theory. We then prove that $\Delta > 0$ is equivalent to the Guest-Hutchinson mode being shear-dominant and that $\Delta < 0$ is equivalent to the Guest-Hutchinson mode being dilation-dominant.

We consider spatially uniform generalized strains $\Lambda$, which implies that $\nabla \varphi_m = \mathbf{0}$ because the strain measures $\varphi_m$ are constants. For a fixed linearized strain $\nabla^s \mathbf{u}$, we solve for the values of $\varphi_1, \ldots, \varphi_{n_w}$ that minimize the elastic energy density. Physically, we are imposing a spatially uniform strain $\nabla^s \mathbf{u}$ and allowing the the other kinematic fields to relax to a spatially uniform equilibrium. Using \eqref{eq:elastic_energy_density_general} and the elastic moduli in \eqref{eq:constitutive}, the elastic energy density in this spatially uniform state is given by
\begin{equation}
    \label{eq:spatially_uniform_elastic_energy}
    V = \frac{1}{2} \left( \nabla^s \mathbf{u}: \mathsf{C} : \nabla^s \mathbf{u} + 2 U^{\mathrm{T}}\Lambda_{\varphi} + \Lambda_{\varphi}^{\mathrm{T}} \mathbf{J} \Lambda_{\varphi} \right),
\end{equation}
where $U \in \mathbb{R}^{n_w}$ has components $U_m = \mathbf{N}_m : \nabla^s \mathbf{u}$, the vector $\Lambda_{\varphi} = [\varphi_1 \ldots \varphi_{n_w}]^{\mathrm{T}}$ contains the additional fields, and $\mathbf{J}$ is the symmetric $n_w \times n_w$ matrix with components given by the generalized elastic moduli $J_{km}$. Since the matrix of generalized elastic moduli $\mathbf{\hat{K}}$ is positive semi-definite, $\mathbf{J}$ is also positive semi-definite, implying that $V$ is a convex function of $\Lambda_{\varphi}$. Thus, stationary points of $V$ correspond to global minima. Differentiating $V$ with respect to $\Lambda_{\varphi}$ yields the stationary point condition
\begin{equation}
\label{eq:stationary_condition}
    \mathbf{J} \Lambda_{\varphi} = - U.
\end{equation}
The minimum elastic energy density $V$ for given $\nabla^s \mathbf{u}$ is therefore
\begin{align}
    \label{eq:spatially_uniform_elastic_energy_min}
    V &= \frac{1}{2} \left( \nabla^s \mathbf{u}: \mathsf{C} : \nabla^s \mathbf{u} - U^{\mathrm{T}} \mathbf{J}^{-1} U \right) \\
    & = \frac{1}{2} \nabla^s \mathbf{u} : \left( \mathsf{C}   - \sum_{k,m = 1}^{n_w} (\mathbf{J}^{-1})_{km} \mathbf{N}_k \otimes \mathbf{N}_m \right) : \nabla^s \mathbf{u}, \nonumber
\end{align}
where $\mathbf{J}^{-1}$ is the inverse of $\mathbf{J}$ if it exists, and its Moore-Penrose pseudoinverse if not. When $\mathbf{J}$ is not invertible, \eqref{eq:stationary_condition} does not have unique solutions $\Lambda_{\varphi}$, but since any stationary point of $V$ is a \emph{global} minimum, the minimum $V$ is independent of the choice of $\Lambda_{\varphi}$ satisfying \eqref{eq:stationary_condition} and is given by \eqref{eq:spatially_uniform_elastic_energy_min}.

We can therefore define an effective elasticity tensor
\begin{equation}
    \label{eq:effective_elasticity_tensor}
    \mathsf{C}_{\mathrm{eff}} = \mathsf{C}   - \sum_{k,m = 1}^{n_w} (\mathbf{J}^{-1})_{km} \mathbf{N}_k \otimes \mathbf{N}_m.
\end{equation}
Thus, when we impose a spatially uniform strain $\nabla^s \mathbf{u}$ and allow the system to relax to a spatially uniform equilibrium state, the system has elastic energy density equal to that of a standard linear continuum with elasticity tensor $\mathsf{C}_{\mathrm{eff}}$. The Guest-Hutchinson modes in our generalized elasticity theory are defined to be strains in the null space of $\mathsf{C}_{\mathrm{eff}}$. Physically, these Guest-Hutchinson modes are uniform zero-energy deformations when the additional fields $\varphi_1, \ldots \varphi_{n_w}$ are free to relax to spatially uniform equilibrium values.

Since $\Delta$ is defined in terms of the coefficients of the $\mathbf{q}$-expansion of $\det \mathbf{P}_K^{\mathrm{T}} \mathbf{\hat{C}}(\mathbf{q})$, we need to relate $\mathsf{C}_{\mathrm{eff}}$ to this expansion in order to link $\Delta$ to Guest-Hutchinson modes in our theory. We take the determinant of the matrix in \eqref{eq:CKC_elastic_moduli} and use the matrix identity \eqref{eq:det_identity} to obtain
\begin{align}
\label{eq:application_det_identity}
    & \det \mathbf{\hat{C}}(\mathbf{\check{q}})^{\dagger} \mathbf{\hat{K}} \mathbf{\hat{C}}(\mathbf{q}) \\
    &= \det \mathbf{H}(\mathbf{\check{q}}, \mathbf{q}) \nonumber \\
    & \qquad \times \det (\mathbf{F}(\mathbf{\check{q}}, \mathbf{q}) - \mathbf{G}(-\mathbf{q}^*, \mathbf{\check{q}})^{\dagger} \mathbf{H}(\mathbf{\check{q}}, \mathbf{q})^{-1} \mathbf{G}(\mathbf{\check{q}}, \mathbf{q})) \nonumber
\end{align}
when $\mathbf{H}(\mathbf{\check{q}}, \mathbf{q})$ is invertible. If $\mathbf{J}$ is invertible, then \eqref{eq:H} implies that $\mathbf{H}(\mathbf{\check{q}}, \mathbf{q})$ is invertible for sufficiently small components $q_j, \check{q}_j$. Setting $\mathbf{\check{q}} = \mathbf{q}$, and using \eqref{eq:F_G_H}, the terms lowest order in $q_j$ present in $\mathbf{F}(\mathbf{q}, \mathbf{q}) - \mathbf{G}(-\mathbf{q}^*, \mathbf{q})^{\dagger} \mathbf{H}(\mathbf{q}, \mathbf{q})^{-1} \mathbf{G}(\mathbf{q}, \mathbf{q})$ are proportional to the $d \times d$ matrix $\mathbf{Z}(\mathbf{q})$ whose elements are given by
\begin{equation}
\label{eq:Z_C_eff}
    (\mathbf{Z}(\mathbf{q}))_{jk} = (\mathbf{e}_j \otimes \mathbf{q}): \mathsf{C}_{\mathrm{eff}} : (\mathbf{e}_k \otimes \mathbf{q}).
\end{equation}
Therefore, the terms lowest order in $q_j$ present in $\det \mathbf{\hat{C}}(\mathbf{\check{q}})^{\dagger} \mathbf{\hat{K}} \mathbf{\hat{C}}(\mathbf{q})$ are proportional to $\det \mathbf{Z}(\mathbf{q})$. Since $| \det\mathbf{P}_K^{\mathrm{T}} \mathbf{\hat{C}}(\mathbf{q}) |^2 \varpropto \det \mathbf{\hat{C}}(\mathbf{\check{q}})^{\dagger} \mathbf{\hat{K}} \mathbf{\hat{C}}(\mathbf{q})$ and the terms lowest order in $q_j$ present in $| \det\mathbf{P}_K^{\mathrm{T}} \mathbf{\hat{C}}(\mathbf{q}) |^2$ are proportional to $|P_d(\mathbf{q})|^2$ in \eqref{eq:general_form_continuum_det_C}, we see that
\begin{equation}
\label{eq:P_d_propto_detZ}
    |P_d(\mathbf{q})|^2 \varpropto \det \mathbf{Z}(\mathbf{q}).
\end{equation}

The results we have derived so far in this section apply in $d$ spatial dimensions. Because $\Delta$ is defined for $d = 2$, we consider only this case here. Since $\mathbf{Z}(\mathbf{q})$ is a $2 \times 2$ matrix with elements quadratic in $q_j$, it is a homogeneous polynomial of degree 4 in $q_j$. We conclude that $\det \mathbf{Z}(\mathbf{q})$ is proportional to $|P_2(\mathbf{q})|^2$ in \eqref{eq:general_form_continuum_det_C}. From this, we show that when $\Delta \neq 0$, there is exactly one Guest-Hutchinson mode. To begin, $\Delta \neq 0$ implies that $P_2(\mathbf{q})$ is not identically zero. Since $\mathsf{C}_{\mathrm{eff}}$ has both major and minor symmetry, it is a symmetric linear operator on the space of symmetric second rank tensors, which is three-dimensional when $d = 2$. Clearly, $\mathsf{C}_{\mathrm{eff}} \neq 0$ because $P_2(\mathbf{q})$ is not identically zero. Therefore, the number of non-zero eigenvalues of $\mathsf{C}_{\mathrm{eff}}$ is at least 1 and at most 3. Suppose that there are two Guest-Hutchinson modes. Then by the rank-nullity theorem, $\mathsf{C}_{\mathrm{eff}}$ has $3 - 2 = 1$ non-zero eigenvalues and takes the form $\mathsf{C}_{\mathrm{eff}} = \lambda \mathbf{E} \otimes \mathbf{E}$ for $\lambda > 0$ and some symmetric second rank tensor $\mathbf{E}$. Then, using $(\mathbf{a} \otimes \mathbf{b}): \mathbf{E} = \mathbf{a} \cdot (\mathbf{E} \cdot \mathbf{b})$, 
\begin{align*}
    (\mathbf{Z}(\mathbf{q}))_{jk} &= (\mathbf{e}_j \otimes \mathbf{q}): \lambda \mathbf{E} \otimes \mathbf{E} : (\mathbf{e}_k \otimes \mathbf{q}) \\
    & = \lambda (\mathbf{e}_j \cdot (\mathbf{E} \cdot \mathbf{q}))((\mathbf{e}_k \cdot (\mathbf{E} \cdot \mathbf{q}))^* \\
    & = \lambda (\mathbf{e}_j \cdot (\mathbf{E} \cdot \mathbf{q})) ((\mathbf{E} \cdot \mathbf{q}) \cdot \mathbf{e}_k) \\
    & = \mathbf{e}_j \cdot (\lambda (\mathbf{E} \cdot \mathbf{q}) \otimes (\mathbf{E} \cdot \mathbf{q}) ) \cdot \mathbf{e}_k.
\end{align*}
So, $\det \mathbf{Z}(\mathbf{q}) = \det (\lambda (\mathbf{E} \cdot \mathbf{q}) \otimes (\mathbf{E} \cdot \mathbf{q})) = 0$ identically, as the determinant of a second-rank tensor with a non-trivial null space, which is equal to the set of vectors orthogonal to $\mathbf{E} \cdot \mathbf{q}$. This implies that $P_2(\mathbf{q})$ is identically zero, which is a contradiction. Therefore, $\Delta \neq 0$ implies that there is at most one Guest-Hutchinson mode. 

Finally, we show that $\Delta > 0$ and $\Delta < 0$ are equivalent to  shear-dominant and dilation-dominant Guest-Hutchinson modes respectively. We recall that a Guest-Hutchinson mode $\mathbf{E}_{\mathrm{GH}}$ in two dimensions is shear-dominant when $\det \mathbf{E}_{\mathrm{GH}} < 0$ and dilation-dominant when $\det \mathbf{E}_{\mathrm{GH}} > 0$, corresponding to principal strains of opposite signs and of the same sign, respectively~\cite{rocklin_transformable_2017, mao_maxwell_2018}. Since all the expressions we work with are homogeneous polynomials in $q_1, q_2$, we can set $q_1 = 1$ so that solving for $q_2$ is equivalent to solving for the ratio $q_2/q_1$, which determines the direction of $\mathbf{q}$ in 2D. Thus, we set $\mathbf{q} = \mathbf{e}_1 + q_2 \mathbf{e}_2$, where $q_2 \in \mathbb{C}$. So $q_2$ satisfies $P_2(\mathbf{q}) = 0$ if and only if $\det \mathbf{Z}(\mathbf{q}) = 0$. When this last equality holds, there exists a vector $\mathbf{d}$ whose components $d_k$ satisfy
\begin{displaymath}
    (\mathbf{Z}(\mathbf{q}))_{jk} d_k = 0,
\end{displaymath}
where we can scale $\mathbf{d}$ so that $d_1 = 1$. Since $\mathsf{C}_{\mathrm{eff}}$ is symmetric, it can be diagonalized to obtain
\begin{equation}
    \mathsf{C}_{\mathrm{eff}} = \lambda_1 \mathbf{E}_1 \otimes \mathbf{E}_1 + \lambda_2 \mathbf{E}_2 \otimes \mathbf{E}_2,
\end{equation}
where $\lambda_1, \lambda_2 > 0$ are the two non-zero eigenvalues of $\mathsf{C}_{\mathrm{eff}}$ and $\mathbf{E}_1, \mathbf{E}_2$ are eigenstrains orthogonal to the Guest-Hutchinson mode. We find that
\begin{align}
\label{eq:is_G-H_mode}
    0 &= d_j^*(\mathbf{Z}(\mathbf{q}))_{jk} d_k \\
    &= d_j^*(\mathbf{e}_j \otimes \mathbf{q}): \mathsf{C}_{\mathrm{eff}} : (\mathbf{e}_k \otimes \mathbf{q}) \, d_k \nonumber \\
    & = (\mathbf{d} \otimes \mathbf{q}): \mathsf{C}_{\mathrm{eff}} : (\mathbf{d} \otimes \mathbf{q}) \nonumber \\
    &= \sum_{j = 1}^2 \lambda_j  ((\mathbf{d} \otimes \mathbf{q}):\mathbf{E}_j)(\mathbf{E}_j:(\mathbf{d} \otimes \mathbf{q})) \nonumber \\
    &= \lambda_1 |\mathbf{E}_1:(\mathbf{d} \otimes \mathbf{q}) |^2 + \lambda_2 |\mathbf{E}_2:(\mathbf{d} \otimes \mathbf{q}) |^2, \nonumber
\end{align}
from which we conclude that $\mathbf{E}_j:(\mathbf{d} \otimes \mathbf{q}) = 0$ for $j = 1,2$. Thus, $\mathsf{C}_{\mathrm{eff}} : \mathrm{sym}(\mathbf{d} \otimes \mathbf{q}) = \mathbf{0}$. We show now that $\mathbf{d}$ satisfies $P_2(\mathbf{d}^*) = 0$. Since $\mathbf{E}_j$ are symmetric second rank tensors, $\mathbf{E}_j:(\mathbf{d} \otimes \mathbf{q}) = \mathbf{E}_j:(\mathbf{q}^* \otimes \mathbf{d}^*)$. Therefore, $\mathsf{C}_{\mathrm{eff}} : (\mathbf{q}^* \otimes \mathbf{d}^*) = \mathbf{0}$, implying that
\begin{displaymath}
    (\mathbf{e}_j \otimes \mathbf{d}^*) : \mathsf{C}_{\mathrm{eff}} : (\mathbf{e}_k \otimes \mathbf{d}^*) q_k^*= 0
\end{displaymath}
for $j = 1,2$. We see that $\det \mathbf{Z}(\mathbf{d}^*) = 0$ from \eqref{eq:Z_C_eff}, so $P_2(\mathbf{d}^*) = 0$ by \eqref{eq:P_d_propto_detZ}.

When $\Delta > 0$, the values $q_2^A, q_2^B$ satisfying $P_2(\mathbf{e}_1 + q_2 \mathbf{e}_2) = 0$ are real and distinct, so $\mathbf{q}$ and $\mathbf{d}$ in \eqref{eq:is_G-H_mode} are vectors with real components. Therefore, $\mathrm{sym}(\mathbf{d} \otimes \mathbf{q})$ is the Guest-Hutchinson mode. We show that if $\mathbf{q} = \mathbf{e}_1 + q_2^A \mathbf{e}_2$, then $\mathbf{d} = \mathbf{e}_1 + q_2^B \mathbf{e}_2$. Suppose instead that $\mathbf{d} = \mathbf{q}$. Then, employing the reasoning in the previous paragraph, we find a Guest-Hutchinson mode corresponding to the other solution $q_2^B$, given by $\mathrm{sym}((\mathbf{e}_1 + q_2^B \mathbf{e}_2) \otimes \mathbf{\check{q}})$, where $\mathbf{\check{q}}$ satisfies $\mathbf{P}_2(\mathbf{\check{q}}) = 0$. Therefore, we have two independent Guest-Hutchinson modes: $\mathbf{q} \otimes \mathbf{q}$, and $\mathrm{sym}((\mathbf{e}_1 + q_2^B \mathbf{e}_2) \otimes \mathbf{\check{q}})$, since $\mathbf{q} = \mathbf{e}_1 + q_2^A \mathbf{e}_2$ and $\mathbf{e}_1 + q_2^B \mathbf{e}_2$ are linearly independent. We have arrived at a contradiction with our previously established result that there is at most one Guest-Hutchinson mode when $P_2(\mathbf{q})$ is not identically zero. Therefore, when $\Delta > 0$, the Guest-Hutchinson mode is
\begin{equation}
    \mathbf{E}_{\mathrm{GH}} = \mathrm{sym}((\mathbf{e}_1 + q_2^A \mathbf{e}_2) \otimes (\mathbf{e}_1 + q_2^B \mathbf{e}_2)),
\end{equation}
with
\begin{align}
    \det \mathbf{E}_{\mathrm{GH}} &= \det \begin{bmatrix}
        1 & \frac{1}{2}(q_2^B + q_2^A) \\
        \frac{1}{2}(q_2^A + q_2^B) & q_2^A q_2^B
    \end{bmatrix} \\
    & = q_2^A q_2^B - \frac{1}{4}\left((q_2^A)^2 + 2 q_2^A q_2^B +(q_2^A)^2\right) \nonumber \\
    & = - \frac{1}{4}\left(q_2^A - q_2^B \right)^2 < 0. \nonumber
\end{align}
Thus, $\Delta > 0$ implies that the Guest-Hutchinson mode is shear-dominant.

When $\Delta < 0$, the values $q_2^A, q_2^B$ satisfying $P_2(\mathbf{e}_1 + q_2 \mathbf{e}_2) = 0$ are a complex conjugate pair $q_2^{A,B} = \zeta \pm i \xi$ for $\zeta, \xi \in \mathbb{R}$. The symmetrization operation when the vectors $\mathbf{a}$ and $\mathbf{b}$ are complex is given by
\begin{displaymath}
    \mathrm{sym}(\mathbf{a} \otimes \mathbf{b}) = \frac{1}{2}(\mathbf{a} \otimes \mathbf{b} + \mathbf{b}^* \otimes \mathbf{a}^*).
\end{displaymath}
Letting $\mathbf{q} = \mathbf{e}_1 + (\zeta + i \xi) \mathbf{e}_2$, our goal is to express the Guest-Hutchinson mode $\mathbf{E}_{\mathrm{GH}}$ in terms of $\mathbf{q}$. The candidates for a (complex) second rank tensor $\mathbf{E}$ that satisfies $\mathsf{C}_{\mathrm{eff}}:\mathbf{E} = \mathbf{0}$ are $\mathrm{sym}(\mathbf{q} \otimes \mathbf{q}) = \mathrm{sym}(\mathbf{q}^* \otimes \mathbf{q}^*)$, $\mathrm{sym}(\mathbf{q} \otimes \mathbf{q}^*)$ and $\mathrm{sym}(\mathbf{q}^* \otimes \mathbf{q})$. If $\mathsf{C}_{\mathrm{eff}}:\mathbf{E} = \mathbf{0}$ and $\mathbf{E} = \mathbf{E}_R + i \mathbf{E}_I$, where $\mathbf{E}_R, \mathbf{E}_I$ are real second rank tensors, then $\mathsf{C}_{\mathrm{eff}}:\mathbf{E}_R = \mathbf{0}$ and $\mathsf{C}_{\mathrm{eff}}:\mathbf{E}_I = \mathbf{0}$. Therefore, $\mathbf{E}_R$ and $\mathbf{E}_I$ are both proportional to the Guest-Hutchinson mode $\mathbf{E}_{\mathrm{GH}}$. Suppose $\mathsf{C}_{\mathrm{eff}}:\mathrm{sym}(\mathbf{q} \otimes \mathbf{q}^*)$. Expanding $\mathrm{sym}(\mathbf{q} \otimes \mathbf{q}^*)$ gives
\begin{align}
\label{eq:candidate_not_G-H_mode}
    \mathrm{sym}(\mathbf{q} \otimes \mathbf{q}^*) &= \frac{1}{2}(\mathbf{q} \otimes \mathbf{q}^* + (\mathbf{q}^*)^* \otimes \mathbf{q}^*) \\
    &= \mathbf{q} \otimes \mathbf{q}^* \nonumber \\
    &= (\mathbf{e}_1 + \zeta \mathbf{e}_2) \otimes (\mathbf{e}_1 + \zeta \mathbf{e}_2) - \xi^2 \mathbf{e}_2 \otimes \mathbf{e}_2 \nonumber \\
    & \quad + i \xi ( \mathbf{e}_2 \otimes (\mathbf{e}_1 + \zeta \mathbf{e}_2) + (\mathbf{e}_1 + \zeta \mathbf{e}_2) \otimes \mathbf{e}_2). \nonumber
\end{align}
We see that the real and imaginary parts of $\mathrm{sym}(\mathbf{q} \otimes \mathbf{q}^*)$ are linearly independent second rank tensors, as the real part has a non-zero $\mathbf{e}_1 \otimes \mathbf{e}_1$ component whereas the imaginary part does not. Therefore, $\mathsf{C}_{\mathrm{eff}}:\mathrm{sym}(\mathbf{q} \otimes \mathbf{q}^*) \neq \mathbf{0}$, otherwise there would be two independent Guest-Hutchinson modes given by the real and imaginary parts of $\mathrm{sym}(\mathbf{q} \otimes \mathbf{q}^*)$. Similarly, $\mathsf{C}_{\mathrm{eff}}:\mathrm{sym}(\mathbf{q}^* \otimes \mathbf{q}) \neq \mathbf{0}$, which we can see by replacing $\xi$ with $-\xi$ in \eqref{eq:candidate_not_G-H_mode}. The last candidate for $\mathbf{E}$ is
\begin{align}
    \mathrm{sym}(\mathbf{q} \otimes \mathbf{q}) &= \frac{1}{2}(\mathbf{q} \otimes \mathbf{q} + \mathbf{q}^* \otimes \mathbf{q}^*) \\
    &= (\mathbf{e}_1 + \zeta \mathbf{e}_2) \otimes (\mathbf{e}_1 + \zeta \mathbf{e}_2) + \xi^2 \mathbf{e}_2 \otimes \mathbf{e}_2, \nonumber
\end{align}
where the imaginary parts cancel to give a purely real second rank tensor. Since there exists $\mathbf{q}$ satisfying $\det \mathbf{Z}(\mathbf{q}) = 0$, and \eqref{eq:is_G-H_mode} implies that for this $\mathbf{q}$ there exists a vector $\mathbf{d}$ such that $\mathsf{C}_{\mathrm{eff}} : \mathrm{sym}(\mathbf{d} \otimes \mathbf{q}) = \mathbf{0}$, a Guest-Hutchinson mode $\mathbf{E}_{\mathrm{GH}}$ must exist. Because we have ruled out the other candidates, $\mathbf{E}_{\mathrm{GH}} = \mathrm{sym}(\mathbf{q} \otimes \mathbf{q})$, with
\begin{align}
    \det \mathbf{E}_{\mathrm{GH}} &= \det \begin{bmatrix}
        1 & \zeta \\
        \zeta & \zeta^2 + \xi^2
    \end{bmatrix} = \xi^2 > 0.
\end{align}
Thus, we have shown that $\Delta < 0$ implies that the Guest-Hutchinson mode is dilation-dominant.

Bringing our results together, we see that the converses to the previous implications also hold: when $\Delta \neq 0$, the presence of a shear-dominant Guest-Hutchinson mode implies that $\Delta > 0$ and the presence of a dilation-dominant Guest-Hutchinson mode implies that $\Delta < 0$. Therefore, we have established that $\Delta > 0$ is equivalent to the presence of a shear-dominant Guest-Hutchinson mode and $\Delta < 0$ is equivalent to the presence of a dilation-dominant Guest-Hutchinson mode.

\section{Analyticity of $q_2$ as a function of $q_1$ corresponding to continuum zero modes}
\label{sec:analyticity}
Our goal in this section is to prove that when solutions to $\det \mathbf{P}_K^{\mathrm{T}} \mathbf{\hat{C}}(q_1 \mathbf{e}_1 + q_2 \mathbf{e}_2) = 0$ are expressed in the form $q_2 = F(q_1)$, where $\lim_{q_1 \to 0} F(q_1) = 0$ (to satisfy \eqref{eq:def_continuum_edge_mode}), $F$ is an analytic function on some neighborhood of zero.

In this derivation, we assume $A_{0, 2} \neq 0$, which is valid as long as the unit vector $\mathbf{e}_2$ corresponding to $q_2$ is not pointing along a soft direction (as defined in Sec.~\ref{sec:polarization_2D}). Therefore, as we showed in the main text, there are two continuum zero modes, corresponding to functions $F_1, F_2$ such that the wavevector components of each zero mode are given by $q_2 = F_k(q_1)$, $k = 1, 2$. We apply the theory of Puiseux series~\cite{knopp1996theory}: $F_k$ can always be written as a Puiseux series (fractional power series)
\begin{equation}
    F_k(q_1) = \sum_{m = 1}^{\infty} \gamma_{k,m} q_1^{\frac{m}{m_k}}
\end{equation}
for sufficiently small $|q_1|$, where $m_k$ is some positive integer. To prove that $F_k$ is analytic, we need to show that the lowest order non-zero term in this expansion is proportional to $q_1^a$ where $a \geq 1$. In the theory of Puiseux series, the set of functions $F_k$ is partitioned into cycles, and $m_k$ is the number of elements in the cycle to which $F_k$ belongs. Therefore, since there are only two functions $F_k$ representing solutions to $\det \mathbf{P}_K^{\mathrm{T}} \mathbf{\hat{C}}(\mathbf{q}) = 0$ that satisfy $F_k(0) = 0$, we have $m_k = 1$ or $m_k = 2$. We show that $F_k$ is analytic by contradiction. Suppose $F_k$ is not analytic. Then, $m_k = 2$ because $m_k = 1$ implies that $F_k$ is analytic. Therefore, $F_k$ belongs to a cycle of two elements, implying that this cycle contains both $F_1$ and $F_2$. Since at least one of the $F_k$ is not analytic, $\gamma_{k,1} \neq 0$ for some $k \in \{1,2\}$. Otherwise, the lowest order non-zero term in the Puiseux expansion of $F_k$ would be proportional to $q_1^a$ where $a \geq 1$, implying that $F_k$ is analytic. The coefficients $\gamma_{1,1}$ and $\gamma_{2,1}$ are related by $\gamma_{1,1} = -\gamma_{2,1}$ because $F_1$ and $F_2$ belong to the same cycle (in general, coefficients of Puiseux series belonging to the same cycle of $p$ elements are related by the $p$th roots of unity)~\cite{knopp1996theory}. Therefore, both $F_1$ and $F_2$ are not analytic and the lowest order non-zero terms in their Puiseux series are proportional to $q_1^{\frac{1}{2}}$. The product of roots of $\det \mathbf{P}_K^{\mathrm{T}} \mathbf{\hat{C}}(\mathbf{q})$, taken as a polynomial in $q_2$ with $q_1$-dependent coefficients, has lowest order non-zero term proportional to the product of the lowest order non-zero terms in $F_1$ and $F_2$, because the other functions corresponding to solutions of $\det \mathbf{P}_K^{\mathrm{T}} \mathbf{\hat{C}}(\mathbf{q}) = 0$ have non-zero constant terms in their expansions. Therefore, this product of roots has lowest order non-zero term proportional to $q_1$, from considering the Puiseux expansions of $F_1$ and $F_2$. However, the lowest order term possible in the product of roots is proportional to $A_{2,0} q_1^2$, leading to a contradiction. We therefore conclude that both $F_1$ and $F_2$ are analytic whenever the unit vector $\mathbf{e}_2$ corresponding to $q_2$ is not pointing along a soft direction.

\section{Orthogonal soft directions}
\label{sec:orthogonal_soft_directions}
Here we consider the case of orthogonal soft directions. We show that the results previously derived for the non-orthogonal case also apply here. We demonstrate that the wavevector of the floppy mode associated with a soft direction $\mathbf{e}_1$ can be parametrized as $\mathbf{q} = q_1 \mathbf{e}_1 + F(q_1)\mathbf{e}_2$, where $F$ is an analytic function. When soft directions are orthogonal, choosing an orthonormal basis $\{\mathbf{e}_1, \mathbf{e}_2\}$ such that $\mathbf{e}_1$ is parallel to a soft direction necessarily means that $\mathbf{e}_2$ is parallel to the other soft direction. Therefore, the coefficients $A_{j,k}$ of the polynomials $P_m(q_1 \mathbf{e}_1 + q_2 \mathbf{e}_2) = \sum_{j=1}^m A_{j,m-j} q_1^j q_2^{m-j}$ in \eqref{eq:general_form_continuum_det_C} satisfy $A_{2,0} = A_{0,2} = 0$. In particular $A_{0,2} = 0$ means that the results proven in Appendix~\ref{sec:analyticity} do not apply in this case. However, when $1 \leq n_w \leq 2$, we can still show that there is at least one solution $\mathbf{q}$ with $q_2 = F(q_1)$ to $\det \mathbf{P}_K^{\mathrm{T}} \mathbf{\hat{C}}(\mathbf{q}) = 0$ such that $F$ is analytic, $F(0) = 0$ and $F'(0) = 0$. The last property concerning $F'(0)$ implies that this floppy mode has wavevector $\mathbf{q} = q_1 \mathbf{e}_1 + i \frac{1}{2} \mathrm{Im}(F''(0)) q_1^2 \mathbf{e}_2 + \mathcal{O}(q_1^3)$. Therefore, this floppy mode can be associated with the soft direction parallel to $\mathbf{e}_1$, as explained following \eqref{eq:zero_mode_wavevector}. This association is sufficient for the results proven in Appendix~\ref{sec:zero_modes_polarization_directions} to hold, in particular \eqref{eq:continuum_zero_mode_soft_directions}. Thus, the characterization of topological polarization in Sec.~\ref{sec:polarization_2D} holds when the soft directions are orthogonal, if $1 \leq n_w \leq 2$. This is a sufficient, not necessary condition. Weaker conditions are possible but we give this condition because it is easiest to state and covers the cases in which an elasticity theory can exhibit topological polarization and Weyl modes.

We have assumed that there are no bulk floppy modes, so there are no solutions $\mathbf{q}$ with purely real components to $\det \mathbf{P}_K^{\mathrm{T}} \mathbf{\hat{C}}(\mathbf{q}) = 0$. Therefore, there exists at least one $3 \leq m' \leq 2 + n_w$ such that $A_{0,m'} \neq 0$, otherwise $q_1 = 0, q_2 \neq 0$ would be a bulk zero mode. The lower bound on $m'$ comes from $A_{0,2} = 0$. Let $m$ be the least such $m'$. Treating $\det \mathbf{P}_K^{\mathrm{T}} \mathbf{\hat{C}}(\mathbf{q})$ as a polynomial in $q_2$ with $q_1$-dependent coefficients, we see that there are $m$ roots in $q_2$ that tend to zero as $q_1 \to 0$. Similarly, there is at least one $3 \leq s' \leq 2 + n_w$ such that $A_{s',0} \neq 0$, otherwise $q_1 \neq 0, q_2 = 0$ would be a bulk floppy mode. The lower bound on $s'$ comes from $A_{2,0} = 0$. Let $s$ be the least such $s'$. If $1 \leq n_w \leq 2$, then $m - 1 \leq 3$. Suppose (for contradiction) that none of the functions $F_j$ corresponding to zero modes are analytic. Let $p > 1$ be the number of Puiseux cycles among the roots (in $q_2$) of $\det \mathbf{P}_K^{\mathrm{T}} \mathbf{\hat{C}}(\mathbf{q}) = 0$, where $m_k$ is the number of elements in the $k$th cycle, $1 \leq k \leq p$. If none of the roots correspond to analytic functions, then the lowest order terms in $q_1$ present in the Puiseux expansions in the $k$th cycle have $q_1$ raised to $\frac{m_k - 1}{m_k}$ at most (c.f. Appendix~\ref{sec:analyticity}). Therefore, the lowest order term in $q_1$ present in the product of roots has order of at most $\sum_{k = 1}^p \frac{m_k - 1}{m_k} m_k = m - p$, since $\sum_{k = 1}^{p} m_k = m$. However, $A_{s,0} \neq 0$, so the the lowest order term in the product of roots has order of $s \geq 3$ in $q_1$. But the hypothesis $1 \leq n_w \leq 2$ implies that $m - p < m - 1 \leq 3$, which is a contradiction. Therefore, there is at least one root in $q_2$ corresponding to an analytic function $F$. The value of $F'(0)$ is determined by the equation $P_2(1, F'(0)) = A_{1,1} F'(0) = 0$ (c.f. discussion after \eqref{eq:power_series}), which gives the stated result $F'(0) = 0$. Thus, we have shown that when the soft directions are orthogonal and $\mathbf{e}_1$ is parallel to a soft direction, the floppy mode associated with that soft direction corresponds to an analytic function.

\section{Characterizing 2D edge floppy modes in terms of polarization directions}
\label{sec:zero_modes_polarization_directions}
Here, we derive \eqref{eq:continuum_zero_mode_soft_directions}. Let $\{\mathbf{e}_1^S, \mathbf{e}_2^S\}$ be an orthonormal basis with $\mathbf{e}_1^S$ parallel to a soft direction, and $\{\mathbf{e}_1, \mathbf{e}_2\}$ be an orthonormal basis obtained by rotating $\{\mathbf{e}_1^S, \mathbf{e}_2^S\}$ counterclockwise through an angle $\theta$. We showed in Sec.~\ref{sec:polarization_2D} that there is a floppy mode associated with the soft direction parallel to $\mathbf{e}_1^S$, with wavevector $\mathbf{q} = q_1^S \mathbf{e}_1 + F_S(q_1^S) \mathbf{e}_2$. The function $F_S(q_1^S)$ is analytic and satisfies $F_S(0) = 0$ and $F_S'(0) = 0$. We express the wavevector of this floppy mode in the rotated basis $\{\mathbf{e}_1, \mathbf{e}_2\}$ as $\mathbf{q} = q_1 \mathbf{e}_1 + F_{\theta}(q_1) \mathbf{e}_2$, where $F_{\theta}$ is a continuous function satisfying $\det \mathbf{P}_K^{\mathrm{T}}\mathbf{\hat{C}}(q_1 \mathbf{e}_1 + F_{\theta}(q_1) \mathbf{e}_2) = 0$ for $q_1$ in some neighborhood of zero. The components of the floppy mode wavevector in the two bases considered are related by
\begin{equation}
		\begin{bmatrix}
			q_1 \\ F_{\theta}(q_1)
		\end{bmatrix} = \begin{bmatrix}
			\cos \theta & \sin \theta \\
			-\sin \theta & \cos \theta
		\end{bmatrix}
		\begin{bmatrix}
			q_1^S \\ F_S(q_1^S)
		\end{bmatrix},
\end{equation}
so by substituting the expression for $q_1$ from the first row into the second row, we obtain
\begin{equation}
\label{eq:F_theta}
	F_{\theta}(q_1^S \cos \theta + F_S(q_1^S) \sin \theta) = -q_1^S \sin \theta + F_S (q_1^S) \cos \theta.
\end{equation}
To show that $F_{\theta}$ is analytic, let $h_{\theta}(q_1^S) = q_1^S \cos \theta + F_S(q_1^S) \sin \theta$. Then the preceding equation shows that $F_{\theta} \circ h_{\theta}$ is an analytic function, since the right-hand side of \eqref{eq:F_theta} is analytic in $q_1^S$ because $F_S$ has been shown to be analytic. Using $F_S'(0) = 0$, we see that $h_{\theta}'(0) = \cos \theta \neq 0$ for $\theta \neq \pm \pi / 2$. By Theorem 5.7.17 in Ref.~\cite{asmar2018complex}, $h_{\theta}$ is invertible on some neighbourhood of zero with an analytic inverse. Restricted to this neighbourhood, $h_{\theta}$ can be taken to be invertible with an analytic inverse for all $q_1$ under consideration. Then, $F_{\theta} = (F_{\theta} \circ h_{\theta}) \circ h_{\theta}^{-1}$ so $F_{\theta}$ is the composition of analytic functions and is therefore analytic on some neighbourhood of zero.

Differentiating \eqref{eq:F_theta} with respect to $q_1^S$ and using the analyticity of $F_S$ and $F_{\theta}$ for $\theta \neq \pm \pi / 2$ gives
\begin{multline}
\label{eq:F_theta_first_derivative}
	F_{\theta}'(q_1^S \cos \theta + F_S(q_1^S) \sin \theta) (\cos \theta + F_S'(q_1^S) \sin \theta) \\
    = -\sin \theta + F_S' (q_1^S) \cos \theta.
\end{multline}
Setting $q_1^S = 0$ gives
\begin{equation}
    F_{\theta}'(0) = - \tan \theta.
\end{equation}
Differentiating \eqref{eq:F_theta_first_derivative} with respect to $q_1$ results in
\begin{multline}
\label{eq:F_theta_second_derivative}
	F_{\theta}''(q_1^S \cos \theta + F_S(q_1^S) \sin \theta) (\cos \theta + F_S'(q_1^S) \sin \theta)^2 \\
    +  F_{\theta}'(q_1^S \cos \theta + F_S(q_1^S) \sin \theta) \sin \theta F_S '' (q_1^S) \\
    = F_S'' (q_1^S) \cos \theta,
\end{multline}
and setting $q_1^S = 0$ gives
\begin{equation}
	F_{\theta}''(0) = \frac{1}{\cos^3 \theta} F_S''(0).
\end{equation}
Therefore, the floppy mode wavevector $\mathbf{q} = q_1^S \mathbf{e}_1^{S} + \frac{1}{2} F_S''(0)(q_1^S)^2 \mathbf{e}_2^{S} + \mathcal{O}((q_1^S)^3)$ associated with the soft direction parallel to $\mathbf{e}_1^S$ can be expressed in terms of an orthonormal basis rotated by $\theta$ as
\begin{equation}
\label{eq:rotated_q_preliminary}
    \mathbf{q} = q_1 (\mathbf{e}_1 - \tan \theta \, \mathbf{e}_2) + \frac{1}{2} \frac{F_S''(0)}{\cos^3 \theta} q_1^2 \mathbf{e}_2 + \mathcal{O}(q_1^3).
\end{equation}
We see that $\mathbf{e}_1^S = \cos \theta \, \mathbf{e}_1 - \sin \theta \, \mathbf{e}_2 = \cos \theta \, (\mathbf{e}_1 - \tan \theta \, \mathbf{e}_2)$, and $\cos \theta$ can be expressed in terms of the polarization direction $\mathbf{p}$ associated with the soft direction parallel to $\mathbf{e}_1^S$ using $\cos \theta = \mathbf{e}_2 \cdot \mathbf{e}_2^S = -\mathrm{sgn}(\mathrm{Im}(F_S''(0))) \, \mathbf{e}_2 \cdot \mathbf{p}$ using \eqref{eq:polarization_direction}. Therefore, \eqref{eq:rotated_q_preliminary} becomes
\begin{equation}
\label{eq:rotated_q}
    \mathbf{q} = -\frac{\mathrm{sgn}(\mathrm{Im}(F_S''(0)))}{\mathbf{e}_2 \cdot \mathbf{p}}q_1 \mathbf{e}_1^S -i \frac{|\mathrm{Im}(F_S''(0))|}{2(\mathbf{e}_2 \cdot \mathbf{p})^3} q_1^2 \mathbf{e}_2 + \mathcal{O}(q_1^3),
\end{equation}
which is \eqref{eq:continuum_zero_mode_soft_directions}.

\section{Implications for topological polarization in discrete lattices}
\label{sec:polarization_discrete}
Here we connect our continuum definition of topological polarization with the definition for discrete lattices.
Given a lattice with a shear-dominant Guest-Hutchinson mode, if the primitive lattice vectors $\{\mathbf{a}_1, \mathbf{a}_2\}$ are chosen so that their dual reciprocal lattice vectors $\{\mathbf{b}_1, \mathbf{b}_2\}$ satisfy $\mathrm{sgn}(\mathbf{b}_j \cdot \mathbf{p}_1) = -\mathrm{sgn}(\mathbf{b}_j \cdot \mathbf{p}_2)$ for $j = 1, 2$, then $n_1 = n_2 = 0$ in the definition of the polarization lattice vector $\mathbf{R}_\mathrm{T} = n_1 \mathbf{a}_1 + n_2 \mathbf{a}_2$~\cite{kane_topological_2014}, once appropriate unit cell choices are made. In other words, if each reciprocal lattice vector $\mathbf{b}_j$ makes an acute angle with one polarization direction and an obtuse angle with the other polarization direction, then $\mathbf{R}_{\mathrm{T}} = \mathbf{0}$ for an appropriate gauge choice of unit cell. However, $\mathbf{R}_{\mathrm{T}} = \mathbf{0}$ does not mean that the lattice is unpolarized, because we can choose a different set of primitive lattice vectors so that $\mathbf{R}_{\mathrm{T}} \neq \mathbf{0}$.

Given a set of primitive reciprocal lattice vectors $\{\mathbf{b}_1, \mathbf{b}_2\}$ and a lattice with a shear-dominant Guest-Hutchinson mode, we construct a new set $\{\mathbf{\tilde{b}}_1, \mathbf{\tilde{b}}_2\}$ of reciprocal lattice vectors such that at least one vector makes an acute angle with both polarization directions. Whenever the Guest-Hutchinson mode of the lattice is shear-dominant, we can choose a direction $\mathbf{n}$ that makes an acute angle with both polarization directions $\mathbf{p}_1, \mathbf{p}_2$. We take an integer linear combination of the reciprocal lattice vectors $m_1 \mathbf{b}_1 + m_2 \mathbf{b}_2$, $m_1, m_2 \in \mathbb{Z}$, that is sufficiently aligned with $\mathbf{n}$ to make an acute angle with both polarization directions. Let $\tilde{b}_{1,j} = m_j / \mathrm{gcd}(m_1, m_2)$ for $j = 1,2$, where $\mathrm{gcd}(m_1, m_2)$ is the greatest common divisor of $(m_1, m_2)$. Therefore, $\mathrm{gcd}(\tilde{b}_{1,1}, \tilde{b}_{1,2}) = 1$. We set
\begin{equation}
    \tilde{\mathbf{b}}_1 = \tilde{b}_{1,1} \mathbf{b}_1 + \tilde{b}_{1,2} \mathbf{b}_2.
\end{equation} 
By construction, $\tilde{\mathbf{b}}_1  \cdot \mathbf{p}_m > 0$ for $m = 1,2$. We must now choose $\tilde{\mathbf{b}}_2$ such that $\{\tilde{\mathbf{b}}_1, \tilde{\mathbf{b}}_2\}$ generates the same reciprocal lattice as $\{\mathbf{b}_1, \mathbf{b}_2\}$. By a result in the theory of lattices~\cite{Micciancio_lattices_2002}, $\{\tilde{\mathbf{b}}_1, \tilde{\mathbf{b}}_2\}$ and $\{\mathbf{b}_1, \mathbf{b}_2\}$ generate the same lattice if and only if the (integer) coefficients $\tilde{b}_{2,1}, \tilde{b}_{2,2}$ in $\tilde{\mathbf{b}}_2 = \tilde{b}_{2,1} \mathbf{b}_1 + \tilde{b}_{2,2} \mathbf{b}_2$ satisfy
\begin{equation}
	\det \begin{bmatrix}
		\tilde{b}_{1,1} & \tilde{b}_{1,2} \\
		\tilde{b}_{2,1} & \tilde{b}_{2,2}
	\end{bmatrix} =	\tilde{b}_{1,1} \tilde{b}_{2,2} - \tilde{b}_{1,2} \tilde{b}_{2,1} = \pm 1.
\end{equation}
Using a result from abstract algebra (the remarks following Definition 6.8 in Ref.~\cite{fraleigh2003first}) that guarantees the existence of integers $r, s$ that satisfy $\mathrm{gcd}(\tilde{b}_{1,1}, \tilde{b}_{1,2}) = \tilde{b}_{1,1} r + \tilde{b}_{1,2} s$, we conclude that there exist $\tilde{b}_{2,1}, \tilde{b}_{2,2}$ that satisfy $\tilde{b}_{1,1} \tilde{b}_{2,2} - \tilde{b}_{1,2} \tilde{b}_{2,1} = 1$. Therefore, we have constructed a new set of reciprocal lattice vectors $\{\mathbf{\tilde{b}}_1, \mathbf{\tilde{b}}_2\}$ such that $\mathbf{\tilde{b}}_1$ makes an acute angle with both polarization directions. As a result, the polarization lattice vector $\mathbf{\tilde{R}}_{\mathrm{T}}$ computed with respect to the primitive lattice vectors $\{\mathbf{\tilde{a}}_1, \mathbf{\tilde{a}}_2\}$ dual to the new set of reciprocal lattice vectors is non-zero because the winding number $\tilde{n}_1$ accounts for the asymmetric floppy mode localization between the edges orthogonal to $\mathbf{\tilde{b}}_1$. Thus, in the context of discrete lattices whose edge floppy modes coincide with the continuum edge modes of \eqref{eq:def_continuum_edge_mode}, the presence of a shear-dominant Guest-Hutchinson mode implies that we can always choose a set of primitive lattice vectors such that $\mathbf{R}_{\mathrm{T}} \neq \mathbf{0}$, consistent with our continuum results, as we aimed to show.

\section{Solving partial differential equations for topological floppy modes}
\label{sec:numerical}
To obtain numerical solutions for floppy modes from the the system of partial differential equations \eqref{eq:constitutive}, we first compute a basis for the range space of $\mathbf{\hat{K}}$. We use the corresponding projection matrix $\mathbf{P}_K^{\mathrm{T}}$ and the compact representation \eqref{eq:constitutive_compact} to obtain a system of $n_K = \textrm{rank } \mathbf{\hat{K}}$ partial differential equations in $d + n_w$ dependent variables $u_1, \ldots, u_d, \varphi_1, \ldots, \varphi_{n_w}$:
\begin{equation}
    \mathbf{P}_K^{\mathrm{T}} \mathbf{\hat{K}} \Lambda = 0,
\end{equation}
where we set the stress measures $\Sigma = 0$ and recall from the main text that $\Lambda = [\nabla^s \mathbf{u} \; \nabla \varphi_1 \ldots \nabla \varphi_{n_w} \varphi_1 \ldots \varphi_{n_w}]^{\mathrm{T}}$.
We see that the continuum Maxwell criterion \eqref{eq:maxwell_continuum} guarantees that the number of equations $n_K$ equals the number of dependent variables $d + n_w$. We add a small diffusive term for stability, so the system of equations solved numerically is
\begin{equation}
\label{eq:to_solve_numerically}
    \mathbf{P}_K^{\mathrm{T}} \mathbf{\hat{K}} \Lambda + c \nabla^2 \Psi = 0,
\end{equation}
where $\Psi = [\mathbf{u} \; \varphi_1 \ldots \varphi_{n_w}]^{\mathrm{T}}$ are the dependent variables, $\nabla^2$ is the Laplacian, and $c$ is a constant with numerical value much smaller than the elements of $\mathbf{P}_K^{\mathrm{T}} \mathbf{\hat{K}}$. In our numerical computations, the elements of $\mathbf{P}_K^{\mathrm{T}} \mathbf{\hat{K}}$ have numerical values in the range $10^{-2}$ to $1$, and we use $0 < c \leq 10^{-10}$. We use Mathematica to solve \eqref{eq:to_solve_numerically} and specify the Dirichlet boundary conditions below.
\vspace*{-\baselineskip}
\subsection{Edge modes}
We solve \eqref{eq:to_solve_numerically} with $n_w = 1$ on the square-shaped domain $\Omega = \{ (x_1, x_2) \in \mathbb{R}^2: |x_1|, |x_2| \leq 0.5 \}$, and set $\lambda = 0.5$ to impose Dirichlet boundary conditions
\begin{align*}
    &\Psi(\mathbf{x}) = \mathbf{0} \textrm{ on } x_1 = -0.5, \\
    &\mathbf{u}(\mathbf{x}) \cdot \mathbf{e}_2 = \sin (4 \pi x_1) \textrm{ on } |x_2| = 0.5.
\end{align*}
\vspace*{-\baselineskip}
\subsection{Weyl modes}
We solve \eqref{eq:to_solve_numerically} with $n_w = 2$ on the square-shaped domain $\Omega = \{ (x_1, x_2) \in \mathbb{R}^2: |x_1|, |x_2| \leq 1 \}$, with Dirichlet boundary conditions
\begin{align*}
    &\Psi(\mathbf{x}) = \mathbf{0} \textrm{ on } x_1 = -1, \\
    & \Psi(\mathbf{x}) = \mathrm{Re}(\Psi^W e^{i \mathbf{q}^W \cdot \mathbf{x}}) \textrm{ on } x_2 = -1,
\end{align*}
where $\mathbf{q}^W$ is the Weyl wavevector whose coordinates are given by \eqref{eq:weyl_coordinates}, and the Weyl mode shape $\Psi^W$ satisfies $\mathbf{P}_K^{\mathrm{T}} \mathbf{\hat{C}}(\mathbf{q}^W) \Psi^W = 0$.

\end{document}